# Visualizing surface-reaction-driven chloride migration and cascading failure in sulfide solid electrolytes

Xiaoya He[1,2†], Haoran Zhang[2†], Xinyang Yue[3†], Hang Liu[1], Zian Mao[1], Jialong Wang[4], Min Liu[4], Shufen Chu[1], Liwei Chen[3]*, Zhaoxin Yu[4]*, Xi Liu[3,5]*, Chuying Ouyang[4]*, Ziyang Ning[4]*, Yujun Xie[1]*

[1]Global Institute of Future Technology, Shanghai Jiao Tong University, Shanghai 200240, P.R. China

[2]Shanghai Jiao Tong University Global College, Shanghai Jiao Tong University, Shanghai 200240, P.R. China

[3]School of Chemistry and Chemical Engineering, In-situ Center for Physical Science, Frontiers Science Center for Transformative Molecules, Shanghai Jiao Tong University, Shanghai 200240, P.R. China

[4]Contemporary Amperex Technology Co., Ltd., Ningde, P.R. China

[5]School of Chemistry and Chemical Engineering, Ningxia University, Yinchuan 750021, P.R. China

[†]These authors contributed equally to this work.

*Corresponding author. Email: lwchen2018@sjtu.edu.cn; yuzx@catl-21c.com; liuxi@sjtu.edu.cn; ouyangcy@catl.com; ningzy02@catl-21c.com; yujun.xie@sjtu.edu.cn

Moisture instability of sulfide solid electrolytes limits practical use in all-solid-state batteries. Spatially resolving gas–solid reactions during moisture exposure remains challenging. Here, cryogenic four-dimensional scanning transmission electron microscopy (cryo-4D-STEM) maps nanoscale degradation in low- and high-Cl argyrodite electrolytes, $Li_6PS_5Cl$ and $Li_{5.5}PS_{4.5}Cl_{1.5}$. $Li_6PS_5Cl$ undergoes predominantly surface-localized degradation that progressively extends inward, whereas $Li_{5.5}PS_{4.5}Cl_{1.5}$ exhibits rapid interior amorphization accompanied by outward chloride redistribution. Density functional theory calculations indicate a lower chloride-migration barrier along the modeled pathway in the higher-Cl composition, while chloride removal and vacancy formation distort the $PS_4^{3-}$ framework and may destabilize the lattice. Surface treatment suppresses detectable large-scale chloride segregation and improves moisture tolerance, retaining 99.9% of the capacity measured at the start of 0.33 C cycling after 150 cycles. These results identify chloride redistribution as a key process in degradation propagation and suggest a route toward moisture-tolerant solid-state batteries.

All-solid-state batteries utilizing sulfide solid electrolytes are highly promising for next-generation energy-storage systems due to their potential to deliver superior safety and high energy density[1-3]. Among candidate electrolytes, argyrodite-type $Li_{7-x}PS_{6-x}Cl_x$ solid-state electrolytes (SSEs) offer a compelling platform by leveraging their high ionic conductivity alongside favorable mechanical processability[4,5]. However, these electrolytes face substantial stability challenges [6–10]. In particular, the extreme moisture sensitivity of sulfide SSEs demands ultra-dry manufacturing environments at dew points below −60 °C that are difficult to implement within existing battery production infrastructure[4,11,12]. Furthermore, moisture-induced hydrolysis triggers structural degradation, electrochemical decay, and toxic $H_2S$ release, presenting formidable safety and handling hurdles[13–15].

To date, mechanistic understanding of moisture degradation in $Li_{7-x}PS_{6-x}Cl_x$ has been largely guided by the hard and soft acids and bases (HSAB) principle[14,16], which attributes hydrolysis primarily to the unfavorable hard-acid/soft-base interaction between $P^{5+}$ and $S^{2-}$ in thiophosphate units and the tendency of $H_2O$ to convert P–S bonds into more stable P–O motifs[15,17]. Accordingly, strategies to improve moisture tolerance have focused on stabilizing the thiophosphate framework through partial O-for-S substitution[18,19] and tuning chemistry by partially replacing $P^{5+}$ with softer or more polarizable cations[20,21]. Previous studies also identified free $S^{2-}$ and bridging sulfur motifs as the reactive species that strongly promote hydrolysis and $H_2S$ evolution[17,22]. Nevertheless, this thiophosphate-centered model does not fully account for the role of halogen chemistry in shaping degradation pathways. Given that chlorination has been shown to substantially alter the structural and interfacial reactivity of lithium argyrodites, the structural instability of $Li_{7-x}PS_{6-x}Cl_x$ is unlikely to be governed solely by the population of sulfide sites[21–23]. These observations point to a halogen-dependent trade-off between transport optimization and moisture tolerance.

Specifically, $Li_{5.5}PS_{4.5}Cl_{1.5}$ (Cl1.5) exhibits an ionic conductivity of 4.5 mS cm$^{-1}$, compared with 2.3 mS cm$^{-1}$ for $Li_6PS_5Cl$ (Cl1.0) (Supplementary Fig. 1). This improvement is consistent with previous reports that increasing Cl content enhances $Li^+$ transport in $Li_{7-x}PS_{6-x}Cl_x$ by modifying anion distribution and Li-site connectivity[24,25]. However, Cl enrichment also changes the moisture reactivity of argyrodite sulfides. Recent studies have shown that Cl content influences $H_2O$ uptake and subsequent thiophosphate hydrolysis, with LiCl formation being closely involved in the degradation process[22]. However, how Cl content governs where degradation initiates and how it propagates within individual particles remains largely unresolved.

To elucidate these degradation pathways, it is necessary to map local structural heterogeneities at high spatial resolution during moisture exposure. However, conventional ensemble-averaged techniques, such as X-ray diffraction (XRD), Raman spectroscopy, and solid-state nuclear magnetic resonance (NMR), provide only averaged information on structural changes[26–28], because they lack the spatial resolution required to determine where degradation initiates and how it propagates within individual particles. Although *in situ* transmission electron microscopy (TEM) has been widely used to investigate structural dynamics in battery materials under working conditions[29–32], its application to sulfide electrolytes is severely constrained by their high electron-beam sensitivity, which complicates reliable imaging of early-stage degradation[33–35]. Furthermore, selected-area electron diffraction (SAED) provides crystallographic information from selected regions, but its selected-area diffraction signal averages over nanoscale heterogeneities and therefore cannot readily resolve local phase distributions[36]. Although cryogenic and low-dose protocols can mitigate beam damage during structural and chemical analysis of battery materials[33,36], direct visualization of moisture-induced structural evolution in $Li_{7-x}PS_{6-x}Cl_x$ remains challenging because hydrolysis may involve coupled $H_2O$ uptake, thiophosphate conversion, ion redistribution, and progressive crystallinity loss, rather than simple bond cleavage[22,35,37]. These limitations highlight the need for characterization methods capable of mapping local phase evolution in beam- and moisture-sensitive solid electrolytes[36,38,39].

Here, we report a data-driven cryo-4D-STEM workflow for examining moisture-induced structural evolution in sulfide solid electrolytes. By integrating cryogenic protection, nanobeam electron diffraction, and machine-learning-assisted phase classification, the workflow reconstructs dominant local phase distributions from a large volume of diffraction patterns. We reveal distinct Cl-content-dependent degradation pathways in $Li_{7-x}PS_{6-x}Cl_x$ under controlled moisture exposure. While Cl1.0 undergoes predominantly surface-localized hydrolysis with an inward-progressing reaction front, Cl1.5 exhibits rapid amorphization extending into the particle interior, accompanied by outward chloride redistribution. Density functional theory calculations and spectroscopic analyses further suggest that the local anion configuration in Cl1.5 enables a more accessible chloride-migration pathway and that vacancy-mediated chloride redistribution can destabilize the thiophosphate framework and contribute to lattice collapse. The spatial association among surface reaction products, outward chloride redistribution, and internal amorphization suggests that chloride redistribution contributes to the propagation of structural disorder beyond the initially reacted surface region. These results identify chloride redistribution as a key process associated with the propagation of moisture-induced degradation in Cl-rich sulfide argyrodites and establish a generalizable framework for studying degradation in beam- and moisture-sensitive energy materials.

## Structural evolution revealed by data-driven cryo-4D-STEM

Given that Cl1.0 and Cl1.5 represent the most commercially promising systems and currently dominate industrial R&D, they were selected as the model systems for this study. Unless otherwise specified, the controlled moisture-exposure experiments were conducted in an argon atmosphere containing 1000 ppm $H_2O$, corresponding to a dew point of −20 °C. *In situ* XRD and Raman spectroscopy were used to monitor bulk structural evolution under this condition (Fig. 1a,b, and Supplementary Fig. 2)[27] , whereas complementary XRD measurements under higher-humidity conditions are shown in Supplementary Fig. 3. Cl1.0 exhibited a stable (222) diffraction peak and retained the Raman signal of $PS_4^{3-}$ during the initial stage of moisture exposure (<1 h), indicating the relative stability of the pristine argyrodite framework[22,40]. In contrast, the (222) diffraction peak of Cl1.5 shifted to higher angles after 20 minutes of exposure, accompanied by quenching of low-frequency Raman lattice modes, suggesting rapid lattice contraction and incipient structural destabilization[18]. The pronounced redshift of the symmetric stretching mode of $PS_4^{3-}$ in Cl1.5 was consistent with perturbation of the local thiophosphate environment and possibly with the formation of partially oxygen-substituted $PS_{4-x}O_x$ species (Supplementary Fig. 4)[15,22,27]. Because increasing the Cl content replaces labile, non-tetrahedral $S^{2-}$ anions on the *4a/4d* sites rather than the sulfur atoms within $PS_4^{3-}$ tetrahedra, it would be expected to reduce the concentration of highly reactive free sulfide sites. Nevertheless, the Cl-rich composition unexpectedly exhibited faster structural degradation, indicating that this process cannot be explained solely by the population of reactive sites. However, the bulk-averaged data from XRD and Raman spectroscopy do not provide sufficient spatial resolution to clarify the underlying mechanism.

To resolve the heterogeneous phase evolution of sulfide electrolytes during moisture-induced degradation, we developed a data-driven cryo-4D-STEM workflow using an electron probe with a diameter of ≤10 nm to collect diffraction patterns across electrolyte particles (Fig. 1c). Because sulfide electrolytes are highly sensitive to electron irradiation (Supplementary Fig. 5)[41], a key challenge was to acquire diffraction patterns with a sufficient signal-to-noise ratio while minimizing beam-induced structural changes. We optimized the beam current and exposure time under cryogenic conditions at −177 °C, achieving an electron dose as low as 55 $e^{-}Å^{-2}$ while still providing sufficient diffraction signal for phase identification. Under these conditions, no detectable additional amorphization or phase transformation was observed within the sensitivity of the diffraction measurements (Supplementary Fig. 6), allowing spatially resolved phase mapping with minimized beam-induced artifacts.

Another challenge lies in extracting spatially resolved phase information from a large volume of diffraction patterns. Overlapping diffraction signals from coexisting phases, together with variations in crystallographic orientation, often hinder the differentiation of degradation products with similar lattice parameters. To address this challenge, we developed a simulation-trained, domain-adapted machine-learning workflow for automated phase identification (Fig. 1c; Methods). Briefly, experimental diffraction patterns were corrected for elliptical distortion, center-aligned, and denoised using wavelet filtering. Electron diffraction patterns were then simulated from the crystallographic structures of candidate hydrolysis products over a broad range of orientations and aligned with the experimental data in reciprocal-space coordinates and scale. A simulation-to-real module was subsequently used to introduce experimentally representative noise, background, and intensity variations while preserving the underlying Bragg geometry and phase labels. The resulting domain-adapted synthetic patterns were used to train a convolutional neural-network classifier, which was applied to the experimental datasets for spatially resolved phase identification. For patterns containing contributions from multiple crystalline phases, the assignment represents the dominant crystalline phase associated with the most prominent Bragg features. Regions dominated by diffuse scattering were evaluated separately using the local crystallinity index. Accordingly, the reconstructed phase maps represent dominant crystalline-phase distributions and are complemented by independent crystallinity-index maps.

We demonstrate how 4D-STEM can resolve structural details of the extensively degraded Cl1.5 particle that were previously difficult to access. While conventional high-angle annular dark-field scanning transmission electron microscopy (HAADF-STEM) imaging provides minimal information regarding the structural phases, our phase-identification workflow reveals a highly heterogeneous nanoscale spatial distribution (Fig. 1e,f). Specifically, crystalline LiCl and $LiCl \cdot H_2O$ accounted for projected pixel-area fractions of approximately 3.8% and 9.1%, respectively, and were enriched near the particle exterior, producing a projected surface layer dominated by chloride-containing phases. The subsurface region was dominated by $Li_3PS_4$ (35%) and was accompanied by ring-like distributions of $Li_2S$ (4.6%) and additional $Li_3PO_4$ domains (16%), indicating a compositionally distinct decomposition zone[42]. In contrast, $LiOH \cdot H_2O$ (20%) was predominantly located within the projected particle core rather than near the surface. LiOH accounted for an additional 3.5% of the projected particle area, while residual crystalline Cl1.5 constituted 8%, indicating that limited domains of the parent argyrodite structure remained after extensive moisture-induced degradation. Analysis of multiple particles revealed qualitatively similar spatial distributions, indicating that the observed degradation pattern was not unique to a single particle (Supplementary Fig. 7). These observations are also consistent with previous studies of extensively degraded, moisture-exposed sulfide electrolyte particles, which identified

LiCl/LiCl hydrates, LiOH-containing species, $Li_2S$, and thiophosphate-derived decomposition products as major degradation products[26,43]. However, static end-state phase mapping alone cannot resolve how these phases form and evolve or how the associated species redistribute and spatially segregate during moisture exposure.

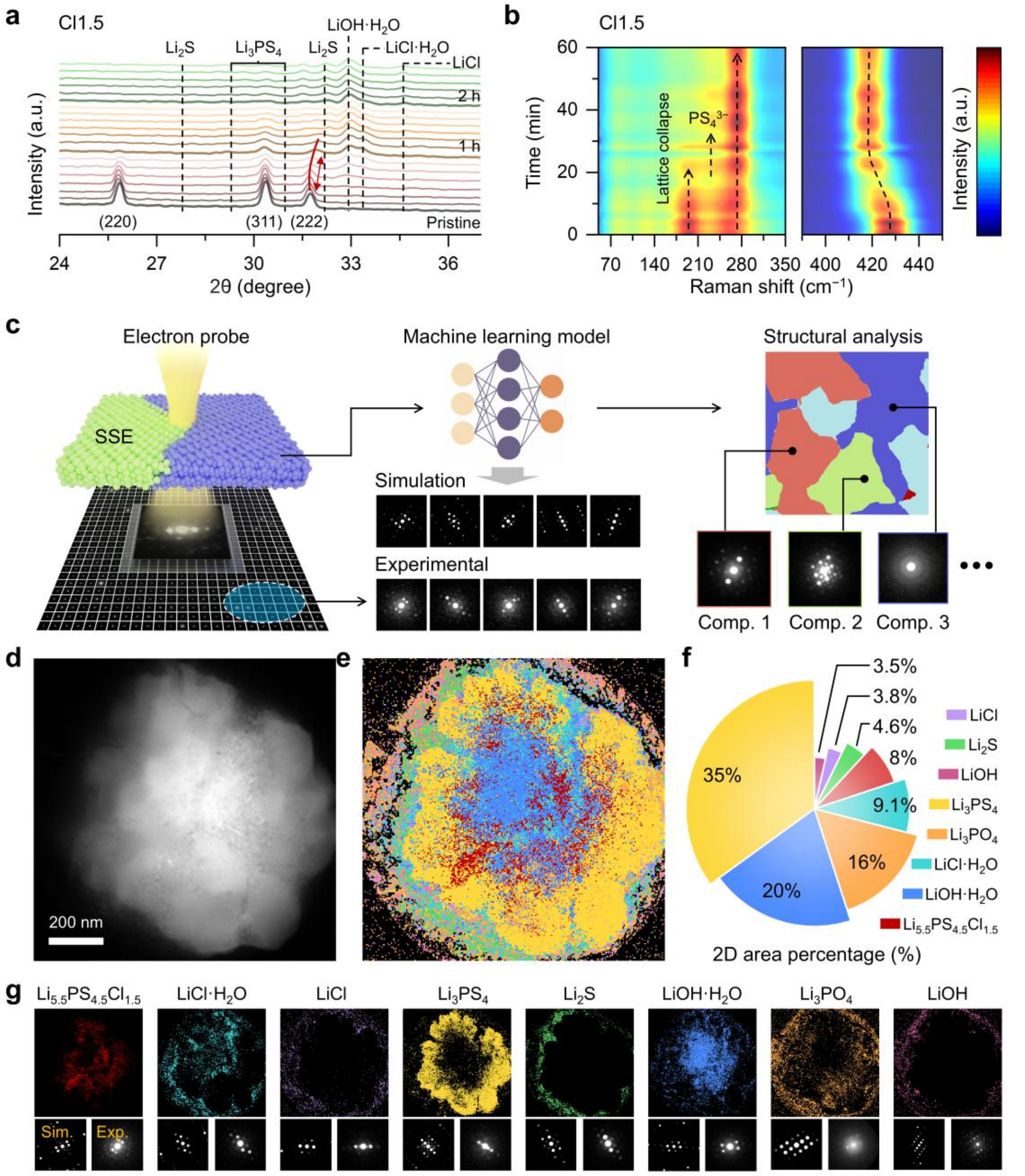

**Fig. 1 | Moisture-induced degradation of sulfide SSEs investigated using data-driven cryo-4D-STEM and complementary bulk in situ techniques.** (**a**) *In situ* XRD patterns and (**b**) *in situ* Raman spectra of $Li_{5.5}PS_{4.5}Cl_{1.5}$ (Cl1.5) during exposure to Ar containing 1000 ppm $H_2O$. (**c**) Schematic of cryo-4D-STEM data acquisition,

simulation-trained diffraction-pattern classification, and dominant-phase-map reconstruction. Representative HAADF-STEM image (**d**) and corresponding color-coded dominant-phase map (**e**) of an extensively degraded Cl1.5 particle. (**f**) Projected pixel-area fractions derived from the dominant-phase map in (**e**). (**g**) Phase-resolved maps and representative comparisons between experimental diffraction patterns and the corresponding simulated diffraction patterns.

**Spatiotemporal analysis of moisture-induced degradation**

To compare the Cl-content-dependent degradation kinetics, we reconstructed the exposure-time-dependent phase distributions of Cl1.0 and Cl1.5. These experiments were conducted under flowing Ar containing 1000 ppm $H_2O$, corresponding to a dew point of −20 °C (Fig. 2). By comparing the projected pixel-area fractions of pristine samples with those of samples exposed to moisture for 30 minutes and 3 hours, we obtained a quasi-in situ view of degradation at discrete time points. A normalized crystallinity index, ranging from 0 to 1, was calculated for each pixel to generate a spatial crystallinity map. During moisture-induced structural degradation, each local diffraction pattern may contain both Bragg-associated and non-Bragg contributions rather than arising solely from a well-ordered crystalline phase. Therefore, the crystallinity index serves as a complementary indicator of local structural order beyond dominant-phase classification alone. In these maps, brighter contrast corresponds to a higher crystallinity index and greater local structural order, whereas darker regions indicate a lower crystallinity index and greater structural disorder. Before moisture exposure, both Cl1.0 and Cl1.5 were dominated by the highly crystalline $Li_{7-x}PS_{6-x}Cl_x$ argyrodite phase, exhibited high crystallinity indices, and contained no detectable degradation products (Fig. 2a and 2d, and Supplementary Fig. 8). Bulk XRD characterization further confirmed the long-range crystalline order of the synthesized materials (Supplementary Fig. 9 and Supplementary Table 1), while STEM imaging documented their pristine particle morphologies (Supplementary Fig. 10).

After 30 min of exposure (Fig. 2b), the Cl1.0 particles retained largely intact interiors dominated by the pristine $Li_{7-x}PS_{6-x}Cl_x$ argyrodite phase. Quantitatively, this pristine phase accounted for approximately 75% of the projected particle area based on the segmented regions in the phase map. Degradation was primarily confined to the particle periphery, which contained $Li_3PO_4$ (6.5%), regions classified as amorphous according to the criterion defined in Methods (18%), and a trace amount of $Li_3PS_4$ (0.3%). These amorphous-classified regions were consistent with localized lattice disorder associated with near-surface hydrolysis (Supplementary Fig. 11)[44]. Notably, the presence of near-surface $Li_3PO_4$ after 30 minutes, alongside the minor fraction of $Li_3PS_4$, suggests that thiophosphate-derived species may undergo rapid local conversion toward oxygen-rich phosphate products upon interaction with $H_2O$. However, because this interpretation relies on the phase distribution at a single exposure time point, it should be regarded as being consistent with rapid local conversion rather than as direct evidence of the reaction sequence.

After 3 hours of exposure, Cl1.0 underwent nearly complete degradation and developed an approximately layered spatial distribution of degradation products from the particle exterior toward the interior. The degradation products comprised LiOH (7.9%) and $LiOH{\cdot}H_2O$ (3.9%), LiCl (15.7%) and $LiCl{\cdot}H_2O$ (18.8%), $Li_2S$ (4.9%), $Li_3PS_4$ (23.6%), and $Li_3PO_4$ (17.7%) (Fig. 2c and Supplementary Fig. 12). These percentages represent projected pixel-area fractions assigned according to the dominant local diffraction features and should not be interpreted as bulk phase fractions. Some phase-assigned pixels also contained substantial diffuse scattering, indicating the coexistence of crystalline and structurally disordered contributions. Together with the separately identified amorphous-classified projected pixel-area fraction of 7.1%, these findings indicate that prolonged moisture exposure produced a mixed crystalline–amorphous degradation state. Thus, the degradation state of Cl1.0 after 3 hours reflected both phase decomposition and structural disorder.

In contrast, Cl1.5 followed a markedly different structural degradation trajectory during the first 30 minutes of exposure (Fig. 2e and Supplementary Fig. 13). The crystallinity map revealed substantially lower crystallinity-index values in both the near-surface and interior regions than in the corresponding regions of Cl1.0, indicating a rapid loss of long-range crystalline order. Dominant-phase mapping showed that the projected pixel-area fraction assigned to pristine Cl1.5 was only 20.8% after 30 minutes of exposure, whereas the separately identified amorphous-classified regions accounted for 44.9%. Compared with the apparent surface recession observed for Cl1.0, the degraded Cl1.5 particles exhibited an apparent increase in projected dimensions and pronounced surface roughening, as revealed by HAADF-STEM and secondary-electron STEM imaging (Supplementary Fig. 14). The degradation products included LiCl (7.0%), $Li_3PO_4$ (7.8%), LiOH (5.8%), and $Li_3PS_4$ (13.3%), with LiCl-containing regions preferentially located near the particle exterior. We propose that the surface localization of these chloride-rich regions should not be interpreted as evidence of increased chloride occupancy within the residual surface argyrodite lattice. Instead, this behavior is consistent with chloride redistribution from the argyrodite framework, leading to chloride accumulation as LiCl-containing precipitates within or adjacent to the reacted surface layer.

After 3 hours of exposure, the pristine Cl1.5 argyrodite phase was no longer detected, and the particles exhibited a mixed amorphous–crystalline degradation state (Fig. 2f and Supplementary Fig. 15). The projected pixel-area fraction assigned to amorphous-classified regions reached 52.2%, with these disordered domains extending from near-surface regions deep into the particle interior. Meanwhile, the projected $Li_3PS_4$ fraction was higher at 3 hours (33.8%) than at 30 minutes (13.3%), accompanied by locally enhanced crystallinity in selected regions (Supplementary Fig. 16). This increase is consistent with reorganization and partial crystallization of disordered thiophosphate-containing material into $Li_3PS_4$ during prolonged exposure.

These observations indicate that the degradation pathway of Cl1.5 differs markedly from that of Cl1.0. Whereas Cl1.0 developed a more clearly layered spatial distribution of crystalline degradation products, Cl1.5 exhibited more extensive amorphization during moisture exposure. At corresponding exposure times, LiCl, $Li_3PO_4$, and $Li_3PS_4$ were preferentially located near the particle exterior in Cl1.5, whereas these phases were mainly observed in the subsurface regions of the Cl1.0 particles. Additional particles exhibited qualitatively similar degradation features, indicating that the observed behavior was not unique to the representative particles (Supplementary Figs. 17–22). Furthermore, internal void formation and structural fragmentation observed after 3 h of moisture exposure may have provided additional pathways for moisture ingress and thereby promoted further structural disorder in Cl1.5. Collectively, these findings indicate that the moisture-induced degradation pathway depends strongly on electrolyte composition.

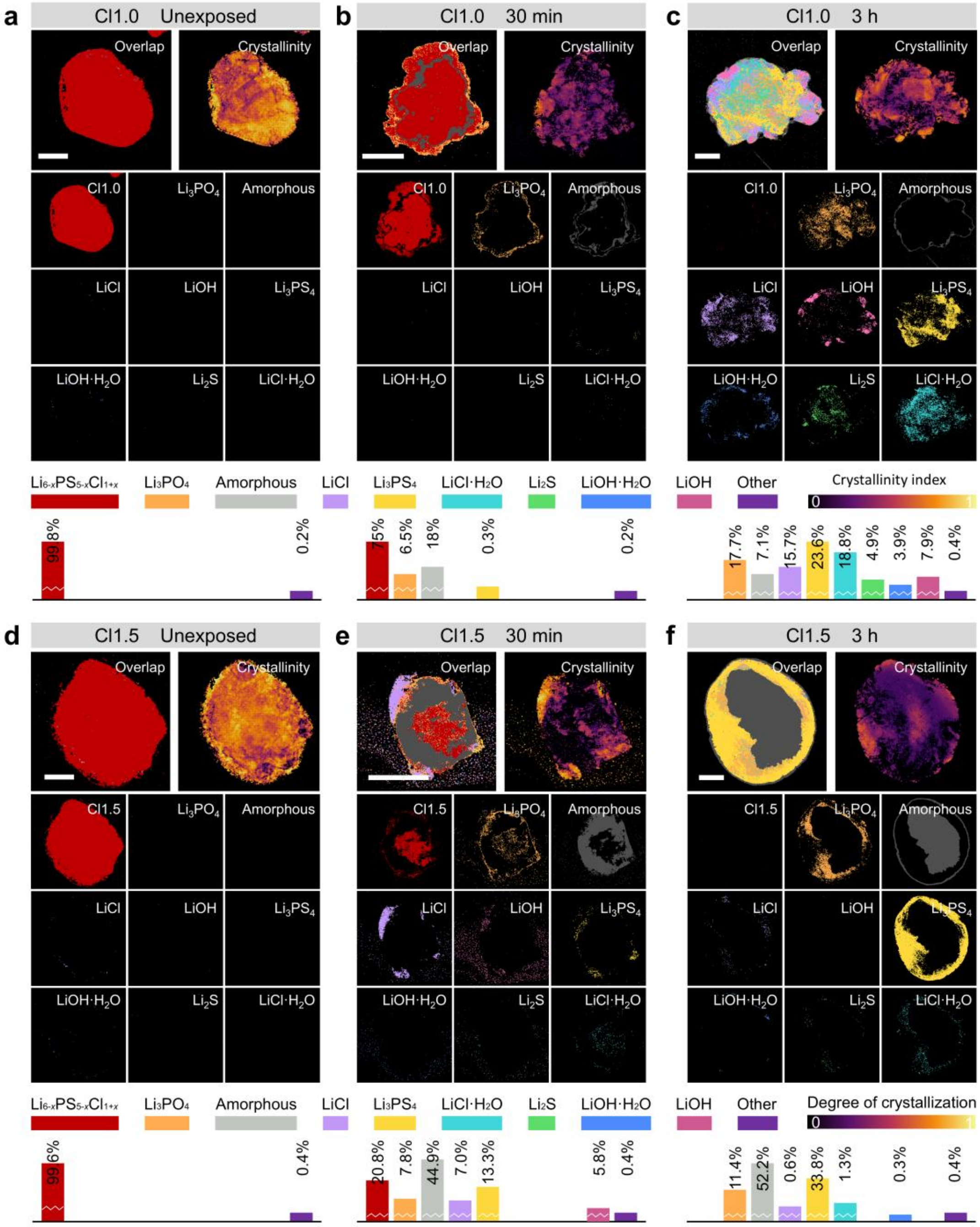

**Fig. 2 | Spatiotemporal evolution of moisture-induced degradation in Cl1.0 and Cl1.5.** Projected pixel-area fraction bars, crystallinity maps, color-coded dominant-phase maps, and phase-resolved maps are shown for Cl1.0 after exposure to Ar containing 1000 ppm $H_2O$ for (**a**) 0 minutes (unexposed), (**b**) 30 minutes, and (**c**) 3 hours, and for Cl1.5 after the same exposure durations of (**d**) 0 minutes (unexposed), (**e**) 30 minutes, and (**f**) 3 hours. Scale bars, 200 nm. The widths of the schematic phase-composition bars are not proportional to the corresponding projected pixel-area

fractions.

**Moisture-induced degradation pathways**

We next combined scanning transmission electron microscopy with energy-dispersive X-ray spectroscopy (STEM-EDS) and electron energy-loss spectroscopy (STEM-EELS) to map elemental redistribution and examine the degradation pathway in Cl1.5 (Supplementary Fig. 23). Cl enrichment near the particle surface was more pronounced with increasing moisture-exposure time, producing a projected shell-like distribution of Cl with a Cl-depleted interior after 3 hours (Fig. 3a–c). Regional quantitative EDS analysis of degraded Cl1.5 particles revealed a lower S/P ratio in the particle cores than in the near-surface regions (Supplementary Figs. 24 and 25), consistent with relative sulfur depletion and compositional reorganization associated with the amorphous interior regions. These observations suggest that degradation involves two spatially distinct and potentially coupled processes. At the particle surface, moisture-driven decomposition is associated with structural disorder and the formation of lithium–phosphorus–sulfur–oxygen (LPSO)-like or lithium–phosphorus–sulfur (LPS)-like species[45]. In the particle interior, the loss of long-range order is spatially associated with chloride depletion and may involve the accumulation of chloride vacancies in argyrodite-derived regions. Distinct diffraction signatures acquired from the particle edge and interior further support the presence of different local structural states (Supplementary Fig. 26).

To examine the spatial relationship between chloride redistribution and structural amorphization, we analyzed a representative partially degraded single-crystalline particle that exhibited pronounced Cl segregation (Fig. 3d and 3e). An EDS line profile and local diffraction patterns were acquired along a surface-to-core trajectory at six representative positions (Fig. 3e). Correlative cryo-4D-STEM and STEM-EDS mapping enabled the local structural states to be directly compared with the elemental distributions across the same particle. The six positions extended from the Cl-enriched surface layer (#1), through the near-surface region (#2) and a localized Cl-deficient zone (#3), to internal domains with varying Cl concentrations (#4–#6). The representative diffraction pattern extracted from the Cl-deficient zone at position #3 exhibited diffuse scattering without discernible Bragg reflections, consistent with a localized loss of long-range crystalline order (Fig. 3f). Although the temporal sequence was not resolved, the spatial coexistence of a surface layer enriched in LiCl-containing phases, a Cl-deficient amorphous region, and a more crystalline interior is consistent with surface-initiated outward chloride redistribution. On the basis of these observations, we propose that moisture-assisted Li/Cl extraction may generate chloride vacancies near the particle surface, thereby promoting outward chloride redistribution and producing a vacancy-enriched region extending into the particle interior that is associated with structural amorphization (Supplementary Fig. 27).

STEM-EELS line scans and mapping were further used to examine the relationship among chloride redistribution, oxygen penetration, and structural amorphization (Fig. 3g). After normalization to account for thickness variations, the Li-K and Cl-L edge intensities peaked at the particle surface, consistent with the surface enrichment of Li- and Cl-containing species and with the LiCl and $LiCl \cdot H_2O$ phases independently identified by phase mapping. The O-K edge signal likewise peaked at the surface and decayed sharply toward the particle interior, indicating that moisture-derived oxygen-containing products were concentrated primarily within the reacted exterior region. Within the detection limits of EELS, the amorphous particle interior exhibited markedly weaker oxygen-related signals than the outer reaction layer. This spatial decoupling suggests that oxygen incorporation to the extent observed in the outer reaction layer is not necessary for the loss of long-range order in the particle interior. It is therefore consistent with a surface-initiated process in which chloride depletion and associated defect accumulation may contribute to the inward propagation of structural disorder.

Despite elemental redistribution and structural reorganization, characteristic Li-K, S-L, and P-L edge features persisted in the interior, indicating the retention of Li–P–S-related local chemical environments[46]. Combined with the loss of Bragg reflections revealed by 4D-STEM, these observations suggest that the internal amorphous region retained substantial thiophosphate-related local coordination despite the loss of long-range argyrodite order. This structural disorder may involve chloride depletion, defect accumulation, and local lattice distortion rather than extensive decomposition of the thiophosphate units. This interpretation is further supported by $^{31}P$ magic-angle spinning nuclear magnetic resonance spectroscopy (Supplementary Fig. 28). After moisture exposure, Cl1.5 exhibited a resonance assigned to $P_2S_7^{4-}$ that was not detected in Cl1.0. Because $P_2S_7^{4-}$ retains P–S–P-linked thiophosphate connectivity, its formation is consistent with partial retention and reorganization of the thiophosphate network, rather than complete decomposition of the thiophosphate framework during amorphization. These $P_2S_7^{4-}$ units may form through the condensation of partially desulfurized $PS_4^{3-}$ units during reorganization of the amorphous phase (Supplementary Fig. 29)[22].

Gas-evolution measurements provided additional evidence for composition-dependent degradation behavior. Cl1.0 exhibited a pronounced initial $H_2S$ release peak that decayed rapidly, whereas Cl1.5 exhibited a lower initial $H_2S$ evolution rate followed by more persistent $H_2S$ evolution (Supplementary Figs. 30 and 31). These results are consistent with a rapid initial hydrolysis response in Cl1.0 and a prolonged degradation process in Cl1.5, although gas evolution alone cannot identify the rate-limiting process or fully represent the overall degradation kinetics. No gaseous HCl was detected during degradation of Cl1.5 under the conditions examined, within the detection limits of our measurements. Together with the identification of LiCl-containing surface phases by phase mapping, this observation is consistent with redistributed chloride remaining predominantly in condensed chloride-containing products rather than being released as detectable gaseous HCl.

Together, these analyses suggest distinct degradation pathways for Cl1.0 and Cl1.5 (Supplementary Fig. 32). In Cl1.0, crystalline hydrolysis products exhibited a roughly layered spatial distribution, consistent with the development of an inward-advancing degradation front. Rapid $H_2S$ evolution and the accumulation of LiOH-containing species in the outermost layer are consistent with strongly surface-localized hydrolysis. The resulting surface reaction layer may have influenced inward moisture transport and contributed to the spatially localized progression of degradation. Although chloride redistribution also occurred in Cl1.0, it proceeded more slowly and remained spatially coupled to the inward progression of the reaction front. By contrast, Cl1.5 exhibited pronounced outward chloride redistribution accompanied by structural amorphization. The limited oxygen enrichment in the particle interior suggests that early-stage amorphization was not governed solely by bulk moisture penetration but likely also involved chloride depletion, ion redistribution, and the formation of LiCl-containing surface products. The persistence of amorphous domains in phase maps acquired at different exposure times, together with prolonged $H_2S$ evolution, indicates the coexistence of sustained hydrolysis and structural disorder during extended moisture exposure.

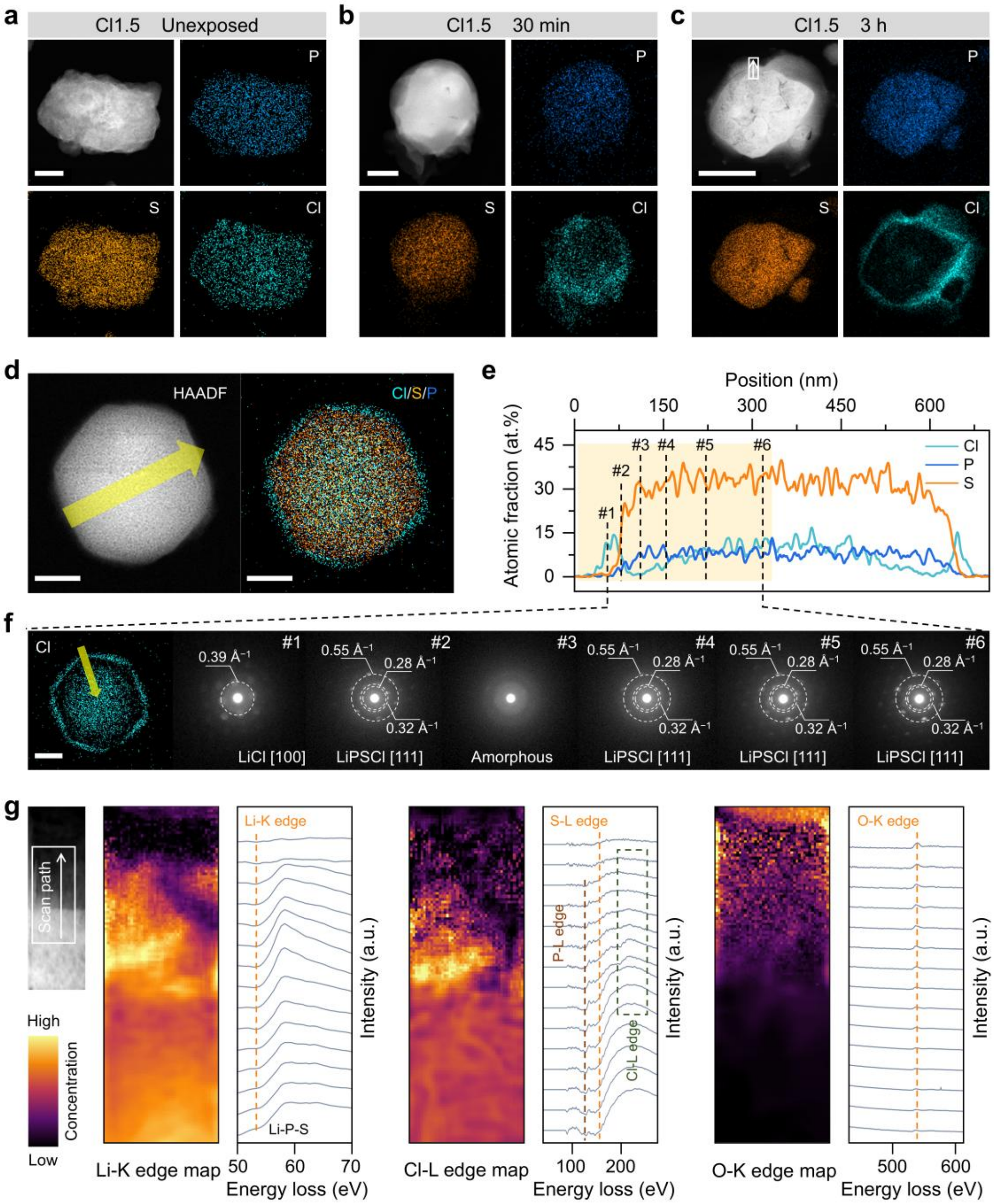

**Fig. 3 | Degradation pathway of Cl1.5 particles during moisture exposure.** (**a–c**) HAADF-STEM images and corresponding EDS maps of P, S, and Cl for Cl1.5 before exposure and after exposure to Ar containing 1000 ppm $H_2O$ for 30 minutes and 3 hours. (**d**) HAADF-STEM image and composite Cl/S/P EDS map of a partially degraded single-crystalline Cl1.5 particle; the yellow arrow indicates the line-scan direction. (**e**) EDS line-scan profiles and (**f**) Cl EDS map with local 4D-STEM diffraction patterns extracted from six positions along the surface-to-core trajectory. (**g**) STEM-EELS

elemental maps and corresponding spectra of the Li-K, Cl-L, S-L, P-L, and O-K edges after 3 hours of exposure. Scale bars, 200 nm.

**Mechanism of chloride migration**

Compared with Cl1.0, Cl1.5 is reported to exhibit preferential occupation of the *4d* sites by $Cl^-$, with partial replacement of $S^{2-}$, and redistribution of $Li^+$ from the *24g* sites to neighboring *48h* sites within the Li sublattice, together with a high concentration of Li vacancies[47]. The close proximity of neighboring *48h* sites and the large fraction of crystallographically unoccupied *48h* sites (~55%), coupled with the stoichiometric Li deficiency of Cl1.5, likely increase local configurational flexibility and susceptibility to ion redistribution. During moisture exposure, $H_2O$ interacts with surface or near-surface $Li^+$ ions and promotes the formation of LiCl-containing products, potentially generating chloride-deficient regions near the particle surface. We propose that the resulting chloride chemical-potential gradient drives outward $Cl^-$ redistribution, while local lattice distortion further facilitates vacancy-mediated hopping between anion sites (Fig. 4a and Supplementary Fig. 27). Because moisture-assisted LiCl formation can proceed via ion redistribution and phase separation without requiring extensive oxygen penetration into the particle interior, chloride-deficient regions may develop near the surface, thereby destabilizing the residual crystalline framework.

Density functional theory (DFT) calculations further revealed distinct energetics for the modeled chloride-migration pathways in the two compositions. In Cl1.0, chloride migration along the *4a–4a* pathway exhibited a calculated migration barrier of 3.2 eV, whereas the Cl1.5 structure provided an alternative *4a–4d* pathway with a lower barrier of 1.3 eV (Fig. 4b,c and Supplementary Fig. 33)[48]. The lower barrier is likely attributable to a shorter hopping distance and a less constrained local environment along the *4a–4d* pathway, suggesting that the Cl-rich structure offers a more accessible route for vacancy-mediated chloride redistribution. Nevertheless, these calculated barriers are best interpreted as relative energetic comparisons between the modeled pathways rather than a complete description of the experimental transport kinetics, as they alone cannot account for the observed redistribution timescale. Water-assisted transport, surface- or defect-mediated pathways, and other coupled processes likely contribute to the overall kinetics. Importantly, while the migration barrier governs kinetic accessibility, it does not by itself determine the net direction of chloride transport. In the proposed mechanism, outward redistribution is driven by surface chloride depletion and the resulting chloride chemical-potential gradient.

The removal of Li and Cl from the surface lattice, followed by LiCl formation, also modified the local electrostatic environment. Electrostatic-potential calculations revealed a more positive surface potential in the modeled surface configuration after Li/Cl extraction and LiCl precipitation (Fig. 4d), suggesting an electrostatic driving force that promotes the outward redistribution of negatively charged chloride ions. Together with continued chloride extraction at the surface, this potential difference likely drives the migration of chloride from less-reacted interior regions toward vacancy-containing near-surface regions. Repeated vacancy filling and chloride extraction at the surface may, in turn, allow a chloride-deficient, defect-enriched region to extend progressively from the reacted exterior into the particle interior. Structural relaxation calculations further showed that chloride removal elongated selected P–S bonds and altered the geometry of neighboring $PS_4^{3-}$ tetrahedra (Fig. 4e and Supplementary Figs. 34 and 35). These distortions indicate that chloride vacancies perturb the local argyrodite framework, consistent with their contribution to the experimentally observed loss of long-range crystalline order. The apparent particle expansion, void formation, and fragmentation associated with structural disorder may create additional pathways for moisture ingress. Enhanced moisture access can further accelerate hydrolysis, Li/Cl extraction, and chloride redistribution. These coupled processes suggest a feedback mechanism in which surface reactions with moisture initiate chloride extraction, vacancy-mediated outward chloride redistribution propagates compositional and structural disorder into the particle interior, and the resulting loss of structural integrity facilitates further moisture penetration.

Guided by this proposed mechanism, we sought to mitigate chloride redistribution by limiting the initiating surface reaction. Accordingly, Cl1.5 particles were treated with difluoromethyltrimethylsilane (DFMS) to reduce direct contact between moisture and the argyrodite surface. Upon exposure to trace water, DFMS is proposed to undergo surface reactions that generate a Si- and F-rich passivation layer (Fig. 4f and Supplementary Fig. 36). This layer restricts water–electrolyte interactions, thereby limiting moisture-assisted Li/Cl extraction from the surface lattice. By suppressing the initial formation of surface chloride vacancies and LiCl-containing products, the treatment weakens the chemical-potential gradient that drives outward chloride redistribution. Consistent with this mechanism, no pronounced chloride segregation was detected by EDS in DFMS-treated Cl1.5 particles after 7 hours of moisture exposure, within the detection limits of the measurement. Ionic-conductivity retention and thermal-recovery measurements further indicated that DFMS treatment mitigated moisture-induced performance degradation (Supplementary Fig. 37 and Supplementary Table 2).

Long-term cycling performance was evaluated in Li–In || $LiNi_{0.8}Co_{0.1}Mn_{0.1}O_2$ (NCM811) solid-state cells assembled with pristine Cl1.5, dry-room-exposed untreated Cl1.5, or dry-room-exposed DFMS-treated Cl1.5. The cell containing pristine untreated Cl1.5 delivered an initial reversible capacity of approximately 207 mAh $g^{-1}$ at 0.1 C

and retained 172 mAh $g^{-1}$ after 170 cycles at 0.33 C (Fig. 4g). This gradual capacity decay may reflect parasitic interfacial reactions between the NCM811 cathode and the sulfide electrolyte. The cell containing dry-room-exposed untreated Cl1.5 exhibited a lower reversible capacity than the pristine counterpart but a higher normalized capacity retention. This seemingly improved fractional retention may reflect partial passivation of the cathode–electrolyte interface by Cl-rich surface degradation products, although further evidence is required to verify this interpretation. By comparison, the cell containing dry-room-exposed DFMS-treated Cl1.5 retained 99.9% of the capacity measured at the beginning of 0.33 C cycling after 150 cycles. It also exhibited a higher reversible capacity and lower polarization than the corresponding cell containing dry-room-exposed untreated Cl1.5 (Supplementary Fig. 38). Together, these results indicate that DFMS surface treatment mitigates the loss of electrochemical performance associated with dry-room exposure.

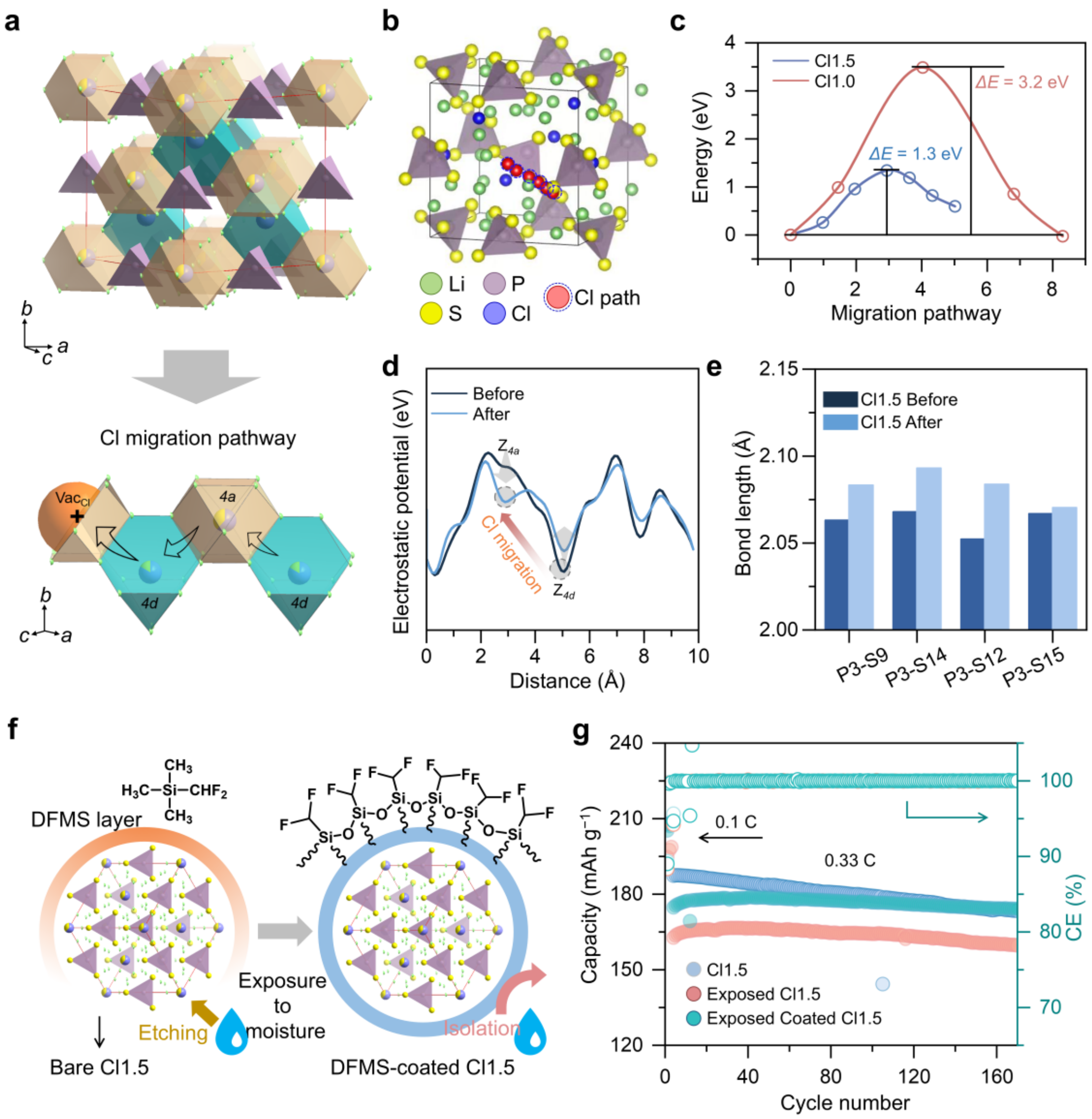

**Fig. 4 | Mechanism and mitigation of chloride-migration-driven degradation in argyrodite electrolytes.** (**a**) Schematic of the proposed chloride-migration pathway in the argyrodite lattice. (**b**) Calculated $Cl^-$ migration pathway and (**c**) corresponding energy profiles for Cl1.0 and Cl1.5. (**d**) Electrostatic potential profiles before and after chloride extraction and LiCl formation, where the horizontal axis represents the z-direction and the vertical axis denotes the *xy*-plane-averaged electrostatic potential. (**e**) Calculated local structural distortion of the $PS_4^{3-}$ framework, showing P–S bond elongation. (**f**) Schematic illustration of the surface passivation mechanism by DFMS molecules. (**g**) Cycling stability of Li–In || NCM811 full cells using bare Cl1.5, exposed Cl1.5, and exposed DFMS-coated Cl1.5.

## Conclusions

This work addresses an important methodological challenge in the characterization of solid-state battery materials by establishing a cryo-4D-STEM platform tailored for highly beam- and moisture-sensitive solid electrolytes. This approach integrates cryogenic sample preservation, low-dose convergent nanobeam electron diffraction, and machine-learning-assisted diffraction analysis to convert large four-dimensional datasets into spatially resolved maps of phase distributions and local structural disorder. Using this platform, we resolved distinct composition-dependent degradation pathways in lithium argyrodites. While Cl1.0 undergoes predominantly surface-localized hydrolysis characterized by crystalline products arranged along an inward-progressing reaction front, Cl1.5 exhibits rapid amorphization accompanied by outward chloride redistribution and the formation of LiCl-containing surface precipitates. DFT calculations provide insight into this behavior by showing that the local anion configuration in Cl1.5 enables a more accessible *4a–4d* chloride-migration pathway, which lowers the calculated migration barrier and may facilitate vacancy-mediated lattice destabilization that contributes to amorphization. Translating these mechanistic insights into a practical mitigation strategy, we demonstrated that a targeted hydrophobic surface treatment with DFMS mitigates chloride redistribution and preserves electrochemical performance under dry-room processing conditions. More broadly, this cryo-4D-STEM framework establishes a generalizable paradigm for resolving spatially heterogeneous and time-dependent degradation pathways in sensitive energy materials, providing a critical diagnostic tool to guide the rational design of next-generation energy technologies.

## Methods

**Synthesis of $Li_6PS_5Cl$ and $Li_{5.5}PS_{4.5}Cl_{1.5}$**

All starting materials were used as received without further purification. Phosphorus pentasulfide ($P_2S_5$, >99% purity), anhydrous lithium chloride (LiCl, >99.98% purity), and lithium sulfide ($Li_2S$, 99.9% purity) were purchased from Sigma-Aldrich. Given the moisture-sensitive nature of sulfide-based materials, all sample handling and storage procedures were conducted within an argon-filled glove box maintaining $H_2O$ and $O_2$ levels below 0.1 ppm. The raw material powders were first pre-mixed using an agate mortar inside the glove box for 10 minutes to ensure homogeneous distribution. The resulting mixture was then transferred to a planetary ball mill jar together with zirconia balls at a powder-to-ball mass ratio of 1:40. High-energy mechanical milling was performed at 510 rpm for 13 hours under an argon atmosphere to promote amorphization and preliminary reaction.

Following ball milling, the obtained powder was uniaxially pressed into pellets under appropriate pressure. The pellets were subsequently loaded into quartz tubes, which were evacuated, backfilled with argon, and sealed to prevent contamination during thermal treatment. Annealing was carried out by heating the sealed tubes to 550 °C at a controlled ramping rate of 5 °C $min^{-1}$ and maintaining the target temperature for 5–7 h to facilitate crystallization. After annealing, the samples were allowed to cool naturally to room temperature within the sealed quartz tubes, yielding the final crystalline products, $Li_6PS_5Cl$ (Cl1.0) and $Li_{5.5}PS_{4.5}Cl_{1.5}$ (Cl1.5).

**Surface treatment of Cl1.5 with DFMS**

All procedures were conducted in an argon-filled glovebox with $H_2O$ and $O_2$ concentrations below 0.1 ppm. Cl1.5 powder and (difluoromethyl)trimethylsilane (DFMS, ≥98%) were combined at a Cl1.5-to-DFMS mass ratio of 10:1 in a sealed 12-mL plastic vial containing six zirconia balls with a diameter of 2 mm. The mixture was homogenized using a planetary centrifugal mixer at 1800 rpm for 20 min. No additional solvent was used during the treatment. The resulting mixture was subsequently dried under vacuum at 30 °C for 3 h to remove residual unbound DFMS. The obtained powder was denoted as DFMS-treated Cl1.5 and stored in the argon-filled glovebox before subsequent moisture-exposure experiments and characterization.

**Structural and Morphological Characterization**

**Transmission Electron Microscopy (TEM)**

Microstructural imaging and energy-dispersive X-ray spectroscopy (EDS) analyses of the synthesized sulfide electrolytes were performed using a Talos F200X transmission electron microscope (FEI, USA). To prevent air exposure during sample transfer, all preparation steps were conducted in an argon-filled glovebox, where trace amounts of powder were dry-dispersed onto copper grids and subsequently loaded into the microscope column using a cryogenic vacuum transfer holder. High-angle annular dark-field scanning transmission electron microscopy (HAADF-STEM) imaging and secondary-electron imaging were carried out on a probe-corrected Hitachi HF5000 microscope operated at 200 kV.

**Four-Dimensional Scanning Transmission Electron Microscopy (4D-STEM)**

4D-STEM experiments were conducted on a Thermo Fisher Scientific Spectra 300 (S)TEM equipped with a Merlin Medipix3 direct electron detector for convergent beam electron diffraction pattern acquisition. The microscope was operated at an acceleration voltage of 300 kV with a probe convergence semi-angle of 0.6 mrad and a spot size of 7. Diffraction patterns were recorded in scanning mode across a region of interest using a dwell time of 1–2 ms per pixel and a scan step size of 2.67–10.0 nm over 256 × 256 pixels. Each dataset thus consisted of a two-dimensional real-space scan with a full diffraction pattern acquired at every probe position. The probe current was maintained at 0.044 nA. Data processing, including analysis and visualization, was carried out using the Py4DSTEM software package (see Data processing section for details).

**Electron Energy-Loss Spectroscopy (EELS)**

After cryogenic transfer of the air-sensitive sulfide samples into the Spectra 300 column, electron energy-loss spectroscopy was performed at an operating voltage of 300 kV with the probe current kept below 50 pA to minimize beam-induced damage. The Gatan Quantum ER Imaging Filter was calibrated by aligning the zero-loss peak to channel 500 using adjacent vacuum regions, providing an energy dispersion of 0.1 eV per channel and an energy resolution better than 1.0 eV as determined from the full width at half maximum of the zero-loss peak. All spectra were acquired in Dual EELS mode, enabling simultaneous recording of low-loss and core-loss signals. Core-loss spectra were collected for the S L-edge, P L-edge and Li K-edge using an exposure time of 0.5 s. Elemental maps were subsequently obtained through spectrum imaging with a step size of 0.5 nm, a dwell time of 2 ms per pixel and an energy slit width of 20 eV at a collection angle of 25 mrad.

**X-ray Diffraction (XRD)**

Phase identification and structural analysis were performed by X-ray diffraction using Cu Kα radiation ($\lambda$ = 1.5406 Å) at 45 kV and 40 mA. Diffraction patterns were collected over a 2θ range of 10° to 60° with a scanning speed of 0.5° $min^{-1}$. To prevent damage during handling and transfer, all samples were sealed on the sample holder with polyimide film inside an argon-filled glove box for testing.

**Solid-State Nuclear Magnetic Resonance (NMR) Spectroscopy**

The local chemical environments of phosphorus were examined by solid-state $^{31}P$ nuclear magnetic resonance spectroscopy. Experiments were carried out on a Bruker Avance 500 MHz spectrometer at a magnetic field of 11.75 T, corresponding to a $^{31}P$

Larmor frequency of 202.5 MHz. Powder samples were packed into 2.5-mm rotors and hermetically sealed under an argon atmosphere inside a glovebox to maintain sample stability during the measurements. Data were acquired at a magic-angle spinning rate of 25 kHz. A 3.2-μs π/2 pulse was applied with a recycle delay of 15 s to ensure complete longitudinal relaxation. Chemical shifts were externally referenced to 85% $H_3PO_4$ at 0 ppm. Spectral analysis and fitting were performed using the DMFIT software package to quantify the different phosphorus environments.

**Raman Spectroscopy**

Raman spectroscopy was carried out using a Bruker confocal micro-Raman spectrometer with a spectral resolution of 1 $cm^{-1}$. A 532 nm laser was focused through a microscope objective onto micron-sized sample regions, and Raman spectra were collected over the range of 100 to 4000 $cm^{-1}$. Analysis of the characteristic Raman shifts enabled the examination of molecular vibrational modes, chemical bonding states, and lattice dynamics. To prevent beam-induced damage, the laser power was strictly maintained at 5 mW during all measurements.

**Electrochemical Impedance Spectroscopy (EIS)**

Ionic conductivity of the synthesized solid-state electrolytes was determined via electrochemical impedance spectroscopy using a multichannel potentiostat (VMP3, Bio-Logic Science Instruments, France). Cold-pressed pellets were prepared by compacting 100 mg of electrolyte powder in a 10-mm-diameter die under 300 MPa pressure for 6 minutes. Pellets of varying thicknesses were sandwiched between two stainless-steel current collectors in a blocking electrode configuration.

EIS spectra were acquired over a frequency range of 0.1 Hz to 1 MHz with an AC perturbation amplitude of 10 mV at 298 K. The ionic conductivity (σ) was calculated according to Equation (1):

$$\sigma = \frac{t}{R \times A} \qquad (1)$$

where R represents the total electrolyte resistance derived from impedance data, t denotes pellet thickness, and A is the electrode contact area (0.785 $cm^2$ for 10 mm diameter).

**Humidity Conditions during in situ and Particle Evolution Experiments**

Hydrolysis experiments were performed in an argon atmosphere containing 1000 ppm $H_2O$ to simulate controlled moisture ingress. At a standard atmospheric pressure, this volumetric concentration yields a water vapor partial pressure of 101.325 Pa. Based on the relationship between saturation vapor pressure and temperature, this value corresponds to a dew point of approximately −20 °C. At the experimental temperature of 25 °C, this concentration represents a relative humidity of approximately 3.2% as determined by the ratio of the water partial pressure to the saturation vapor pressure of water at this temperature, which is approximately 3169 Pa.

**Density Functional Theory (DFT) Calculations**

All density functional theory calculations were performed using the Vienna Ab initio Simulation Package (VASP)[49]. The generalized gradient approximation with the Perdew-Burke-Ernzerhof (GGA-PBE) functional was selected for the exchange-correlation potential[50]. Weak van der Waals interactions were accounted for using the DFT-D3 correction method[51]. A plane-wave cutoff energy of 400 eV was employed throughout the calculations. Brillouin zone integration was performed using a 2 × 2 × 2 k-point mesh. Total energies were converged to $10^{-5}$ eV in the iterative solution of the Kohn-Sham equations. Geometry optimization was considered complete when the residual forces on each atom were reduced to below 0.05 eV $Å^{-1}$. Chloride ion diffusion pathways and corresponding energy barriers were determined using the climbing image nudged elastic band (CI-NEB) method[52].

## Air Stability and In Situ Characterization

### Controlled humidity stability testing

The atmospheric stability of sulfide electrolytes was evaluated using a custom-built humidity control system. This system generated controlled-humidity environments by bubbling dry argon through thermostatic water baths maintained at different temperatures. During testing, 20 mg of electrolyte powder was placed in a glass reaction vial and continuously exposed to the $H_2O$-containing Ar stream at a flow rate of 6 mL $min^{-1}$, controlled using a mass flow controller. Hydrogen sulfide ($H_2S$) generated from hydrolysis reactions was monitored in real-time using an electrochemical sensor, enabling quantification of decomposition behavior through continuous $H_2S$ concentration profiling.

### *In Situ* Raman Spectroscopy

*In situ* Raman spectroscopy was performed using a custom-designed sealed reaction cell featuring a gas inlet system for introducing humidity-controlled gas through a flow meter and an outlet connected to an exhaust treatment unit. Continuous gas flow conditions were maintained throughout the experiment, with spectral acquisition performed at 2-minute intervals to monitor real-time chemical evolution under controlled atmospheric conditions. All measurement parameters (532 nm laser, ≤5 mW power, 1 $cm^{-1}$ resolution) were consistent with those described in **Raman Spectroscopy**.

### *In Situ* X-ray Diffraction (XRD)

Structural evolution of the sulfide electrolytes under reactive gas environments was monitored in real time using Cu Kα radiation (λ = 1.5406 Å, 45 kV, 40 mA). Diffraction patterns were initially collected across the 10°–60° 2θ range at a scan speed of approximately 7.1° $min^{-1}$, with measurements taken at 7 min intervals during exposure to controlled-humidity argon (50% relative humidity). All samples were sealed in gas-permeable containment cells prepared inside an argon-filled glovebox to prevent degradation prior to measurement. Based on the initial full-spectrum scans, the 23°–37° 2θ region was identified as the most dynamic range. Subsequent measurements were focused on this interval under a lower humidity level (3–5% relative humidity) with improved time resolution to track the evolution of characteristic diffraction peaks in detail.

**Data Processing Methodology for 4D-STEM**

**Simulation-trained domain-adapted learning for automated phase identification in 4D-STEM**

To address the limited availability of labeled experimental datasets for 4D-STEM phase classification, we developed a multi-component neural-network framework that integrates physics-based diffraction simulations with generative domain adaptation. Multislice simulations based on the crystallographic information files of the candidate phases were first used to construct a labeled library of ideal electron-diffraction patterns spanning a broad range of crystallographic orientations. Before classification, the experimental diffraction patterns were corrected for elliptical distortion, aligned according to the position of the direct beam, and denoised using wavelet-based filtering. The simulated diffraction patterns were subsequently mapped onto the same reciprocal-space coordinate system, sampling scale, and detector geometry as the corrected experimental datasets. Ideal simulated patterns do not fully reproduce the complex noise, diffuse background, inelastic-scattering contributions, and intensity variations present in experimental measurements. To reduce this simulation-to-experiment domain gap, a Sim2Real style-transfer network was therefore introduced as a data-augmentation module. The network transferred experimentally representative stochastic and textural characteristics to the geometrically aligned simulations while preserving the positions and relative geometry of the underlying Bragg-diffraction features and their corresponding phase labels. The resulting domain-adapted synthetic patterns were used to train a convolutional neural-network classifier, which was subsequently applied to the corrected and denoised experimental 4D-STEM patterns for spatially resolved phase identification. For experimental patterns containing diffraction contributions from more than one crystalline phase, the classifier output was interpreted as the dominant crystalline component associated with the most prominent Bragg features. The resulting maps therefore represent dominant-phase distributions. Patterns dominated by diffuse scattering were additionally evaluated using the independent crystallinity-index analysis described in the following subsection. This separation enables dominant crystalline-phase identification to be interpreted together with an independent assessment of local structural order.

**Normalized local crystallinity-index mapping via 4D-STEM diffraction decomposition**

To evaluate spatial variations in structural order within the 4D-STEM datasets, each experimental diffraction pattern was treated as an approximate combination of a Bragg-associated crystalline component and a non-Bragg residual component. The resulting quantity is reported as a normalized crystallinity index rather than as an absolute crystalline volume fraction. The position and shape of the direct beam were first determined by fitting a two-dimensional super-Gaussian function, and a central-beam exclusion mask was then applied to prevent the intense low-angle signal from biasing the subsequent fitting and residual calculations. Noise reduction was further achieved by subtracting an average background pattern derived from empty frames. Following these corrections, Bragg-disk coordinates were identified at each scan position using template-based peak detection. The Bragg-associated crystalline signal was reconstructed by summing parametric kernels centered at the detected Bragg-disk positions, with their amplitudes sampled directly from the corresponding background-corrected experimental intensities. The non-Bragg residual was defined as the non-negative intensity remaining after subtraction of the reconstructed crystalline component from the corrected experimental pattern. This residual may include contributions from amorphous or highly disordered scattering, diffuse scattering, inelastic scattering, and measurement noise.

The normalized local crystallinity index, $C$, was calculated as

$$C = \frac{I_{cryst}}{I_{cryst} + I_{resid}}$$

where $I_{cryst}$ is the integrated intensity of the reconstructed Bragg-associated crystalline component and $I_{resid}$ is the integrated non-Bragg residual intensity. Both quantities were evaluated outside the central-beam exclusion mask. The residual intensity was constrained to be non-negative to avoid numerical artifacts. The index ranges from 0 to 1, with higher values indicating a greater relative contribution from Bragg-associated scattering and therefore a higher degree of local structural order under the applied acquisition and processing conditions. The index was used primarily for relative spatial and temporal comparisons between datasets acquired under comparable conditions.

**Prototype cell assembly and test**

All assembly steps were conducted in an argon-filled glove box with oxygen and water levels maintained below 0.1 ppm. Full cells were fabricated using a custom mold consisting of two conductive steel rods and a polyether ether ketone cylinder with an inner diameter of 10 mm. First, 90 mg of Cl1.0/Cl1.5 solid electrolyte was cold-pressed at 80 MPa for 30 s to form a pellet with a diameter of 10 mm. Composite cathode powder with a mass loading of 26.75 mg $cm^{-2}$ was then placed on one side of the pellet and pressed at 360 MPa for 10 min. On the opposite side, a 10 mm Li–In foil disk, prepared by pressing a 100 μm-thick indium foil together with a 50 μm-thick lithium foil at 80 MPa for 30 s, was then pressed at 125 MPa for 30 s. Pressure was maintained by tightening bolts at three corners between two steel plates. Galvanostatic charge–discharge measurements were performed on the assembled mold cells using a Neware battery tester (CT-4008Q, China) at 30 ± 0.5 °C inside an environmental chamber. The voltage window was set between 2.0 and 3.7 V *vs.* $Li^+$/Li–In. The cells were initially conditioned with three cycles at 0.1 C, and then cycled at 0.33 C.

## Data availability

All processed data supporting the findings of this study are available in the main text and Supplementary Information. Raw 4D-STEM datasets, DFT input files and additional microscopy data are available from the corresponding authors upon reasonable request. Materials are available from the corresponding authors under a material transfer agreement.

## Code availability

Analysis scripts used for the 4D-STEM data processing are available from the corresponding authors upon reasonable request.

## Funding

This work was supported by the National Natural Science Foundation of China (12474186, 52401159, and 52425101), the Shanghai Jiao Tong University AI for Engineering Initiative, and the Shanghai Magnolia Talent Plan Pujiang Project

(25PJD063).

## Author contributions

Z.N. and Y.X. conceived the study and designed the experiments. X.H. prepared the samples and performed the spectroscopic and electrochemical measurements. H.Z. developed the 4D-STEM data-analysis algorithms. H.L. performed the theoretical calculations. X.H., H.Z., and X.Y. performed the principal microscopy experiments, analyzed the data, and prepared the figures. Z.M., S.C., J.W. and M.L. contributed to experimental measurements, data acquisition, validation, and analysis. L.C., Z.Y., X.L., C.O., Z.N., and Y.X. supervised the project. X.H., H.Z., and X.Y. wrote the original draft. All authors reviewed and edited the manuscript.

## Competing interests

J.W., M.L., Z.Y., C.O., and Z.N. are employees of Contemporary Amperex Technology Co., Ltd. The authors declare no other competing interests.

Supplementary Information

# Visualizing surface-reaction-driven chloride migration and cascading failure in sulfide solid electrolytes

Xiaoya He[1,2†], Haoran Zhang[2†], Xinyang Yue[3†], Hang Liu[1], Zian Mao[1], Jialong Wang[4], Min Liu[4], Shufen Chu[1], Liwei Chen[3]*, Zhaoxin Yu[4]*, Xi Liu[3,5]*, Chuying Ouyang[4]*, Ziyang Ning[4]*, Yujun Xie[1]*

[1]Global Institute of Future Technology, Shanghai Jiao Tong University, Shanghai 200240, P.R. China

[2]Shanghai Jiao Tong University Global College, Shanghai Jiao Tong University, Shanghai 200240, P.R. China

[3]School of Chemistry and Chemical Engineering, In-situ Center for Physical Science, Frontiers Science Center for Transformative Molecules, Shanghai Jiao Tong University, Shanghai 200240, P.R. China

[4]Contemporary Amperex Technology Co., Ltd., Ningde, P.R. China

[5]School of Chemistry and Chemical Engineering, Ningxia University, Yinchuan 750021, P.R. China

[†]These authors contributed equally to this work.

*Corresponding author. Email: lwchen2018@sjtu.edu.cn; yuzx@catl-21c.com; liuxi@sjtu.edu.cn; ouyangcy@catl.com; ningzy02@catl-21c.com; yujun.xie@sjtu.edu.cn

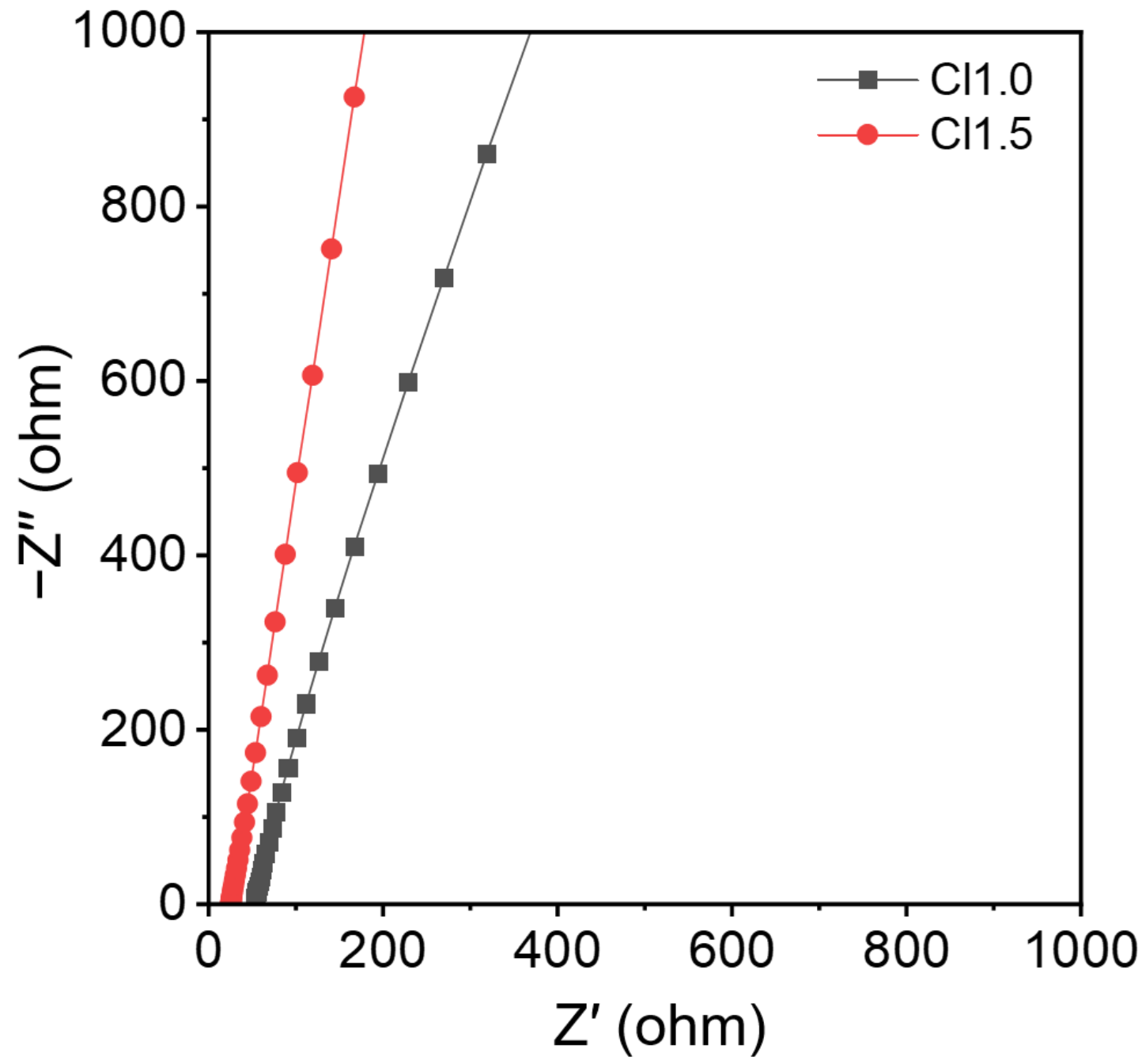

**Supplementary Fig. 1 |** Ionic conductivity enhancement in Cl-rich argyrodite. At room temperature, Cl1.5 shows a higher ionic conductivity of approximately 4.5 mS cm$^{-1}$ than Cl1.0, which exhibits approximately 2.3 mS cm$^{-1}$, indicating that Cl enrichment enhances $Li^{+}$ transport in $Li_{7-x}PS_{6-x}Cl_x$ argyrodites

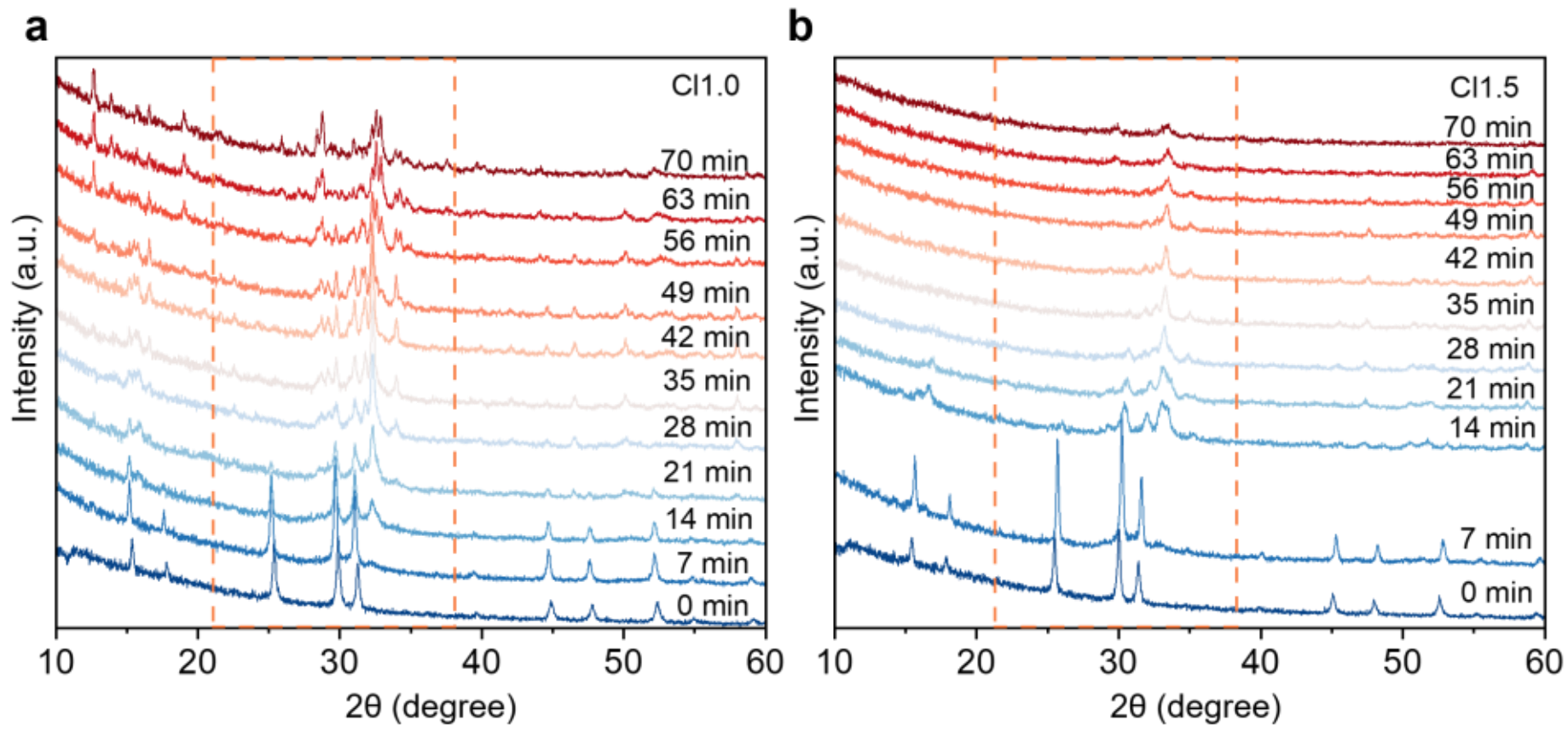

**Supplementary Fig. 2 |** In situ X-ray diffraction analysis of structural evolution. Time-resolved X-ray diffraction patterns of (a) Cl1.0 and (b) Cl1.5 recorded under a 50% relative humidity atmosphere. The 2θ range of 23°–37° is identified as the most sensitive region to investigate argyrodite framework degradation.

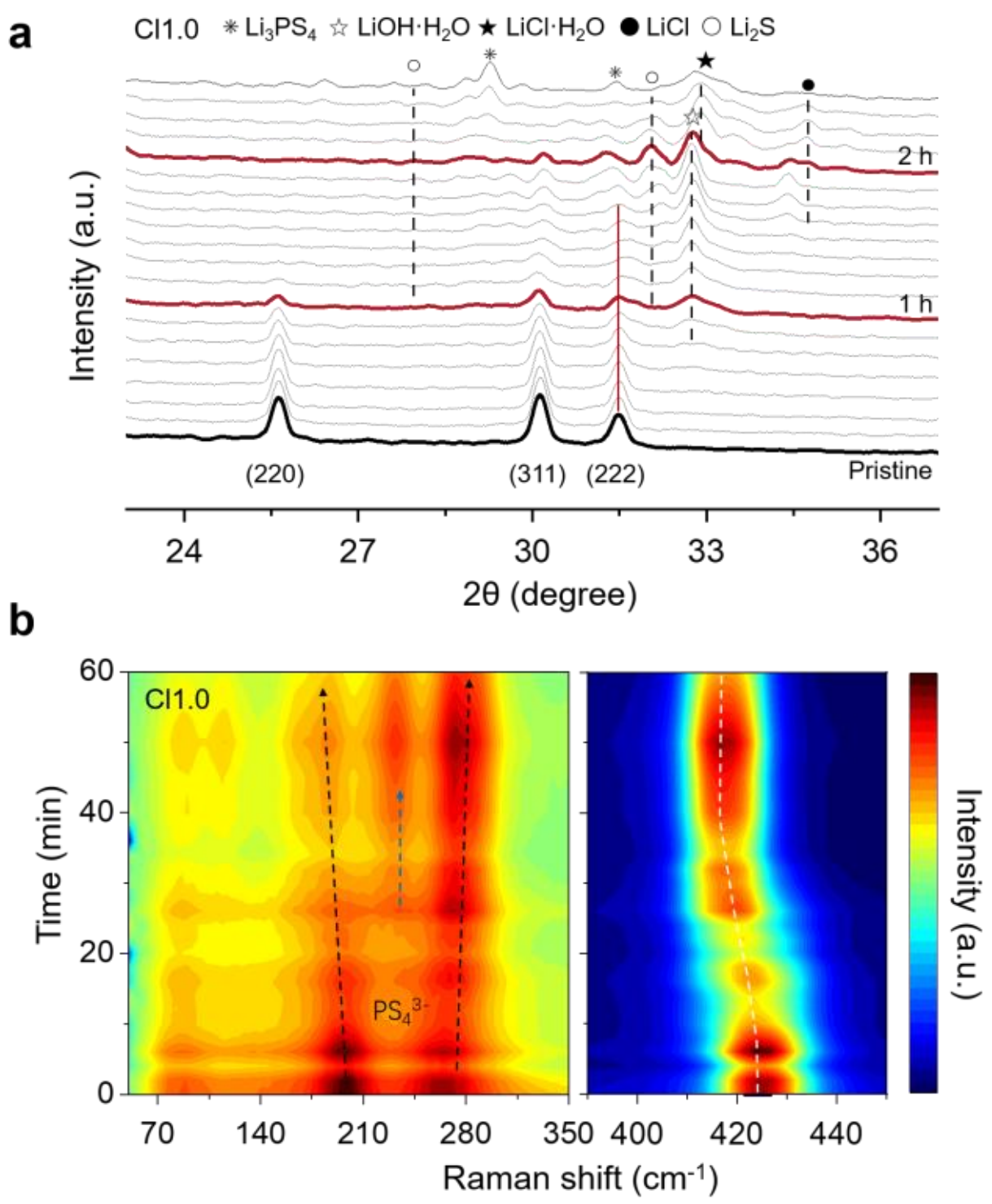

**Supplementary Fig. 3 |** In situ macroscopic characterization of the $Li_6PS_5Cl$ (Cl1.0) degradation. (a) In situ XRD patterns and (b) in situ Raman spectra of Cl1.0 under 1000 ppm $H_2O$/Ar atmosphere.

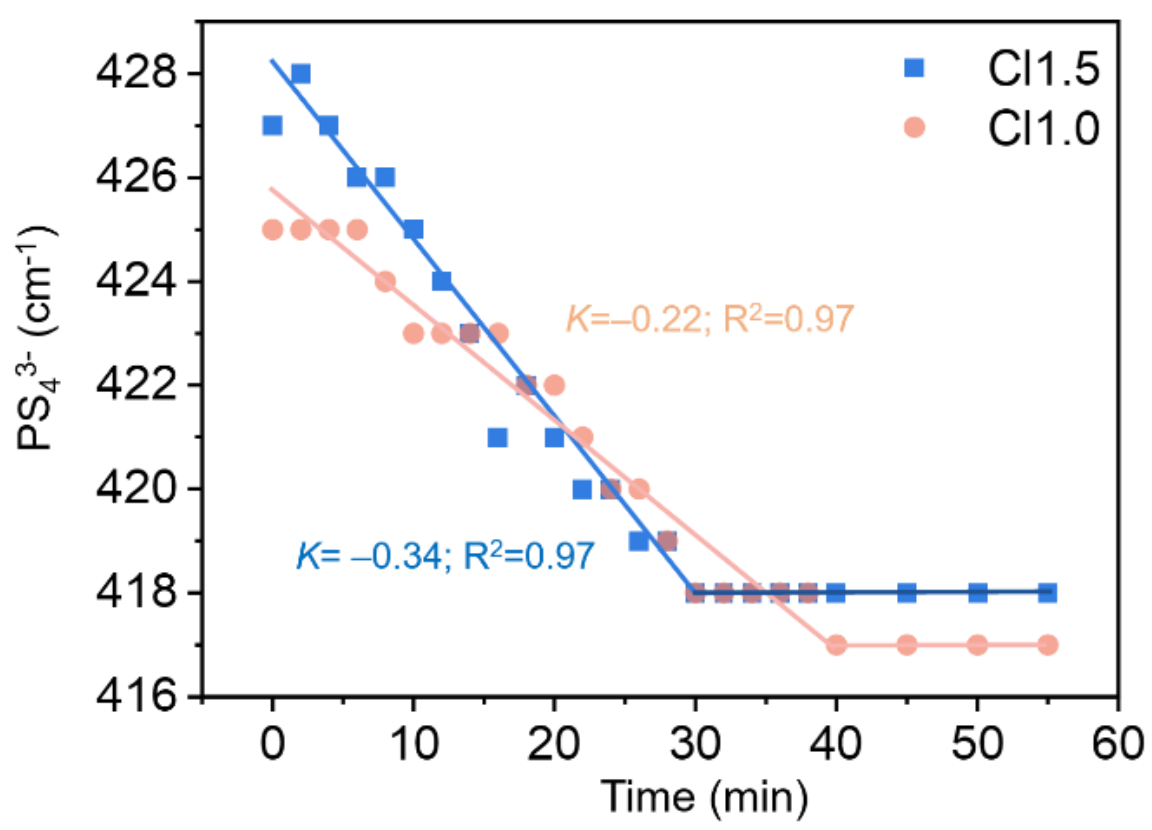

**Supplementary Fig. 4 |** Time-dependent Raman shift of $PS_4^{3-}$ of Cl1.0 and Cl1.5. The rapid red shift of the $PS_4^{3-}$ Raman peak of Cl1.5 over exposure time under a 1000 ppm $H_2O$/Ar atmosphere shows rapid perturbation/reorganization of the $PS_4^{3-}$ environment.

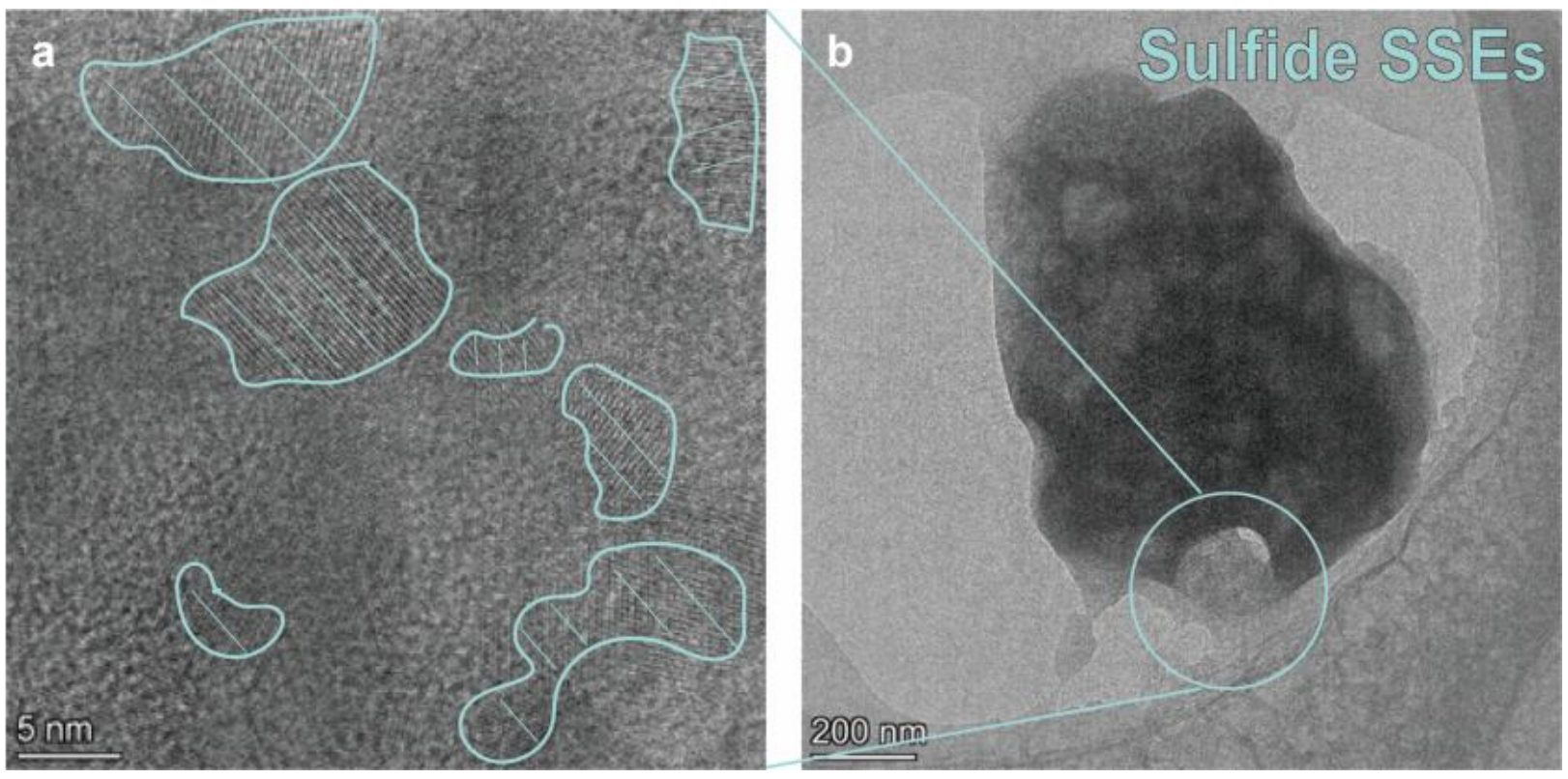

**Supplementary Fig. 5 |** Electron beam-induced amorphization in sulfide SSEs. Fragmented lattice regions (outlined in light blue) are surrounded by amorphous regions from (a, b) TEM images indicating electron beam-induced amorphization of sulfide SSEs.

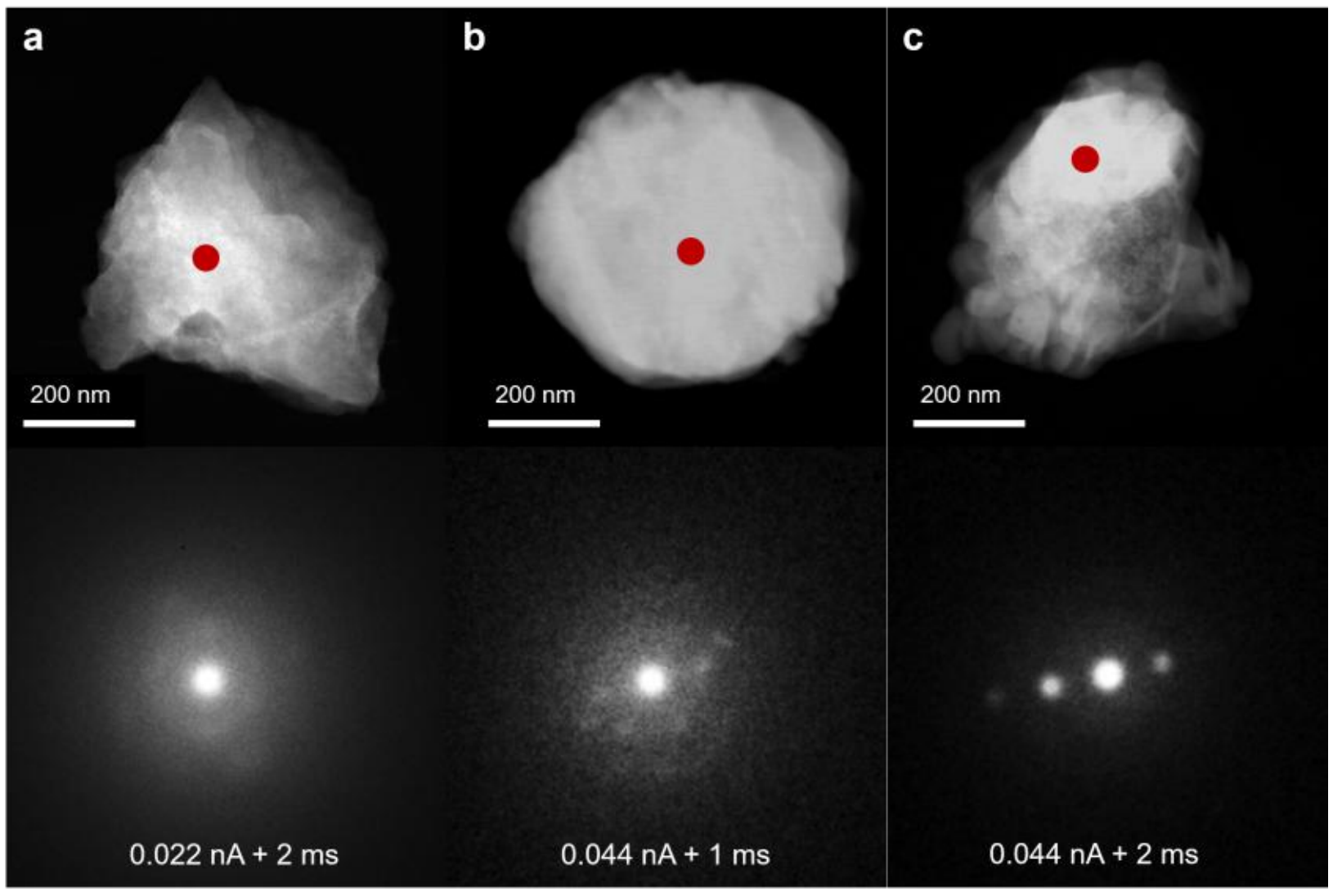

**Supplementary Fig. 6 |** Optimization of electron beam parameters for cryogenic 4D STEM. (a–c) STEM images (upper panel) and corresponding nanobeam electron diffraction patterns (bottom panel) acquired under varying beam currents and exposure times. Labels indicate specific conditions of 0.022 nA with 2 ms in a, 0.044 nA with 1 ms in b, and 0.044 nA with 2 ms in c. Bottom panels show the improvement in diffraction signal-to-noise ratio as the electron dose increases. Optimized parameters in c allow for clear diffraction spots suitable for phase identification while preserving the crystalline integrity of the electrolyte particles.

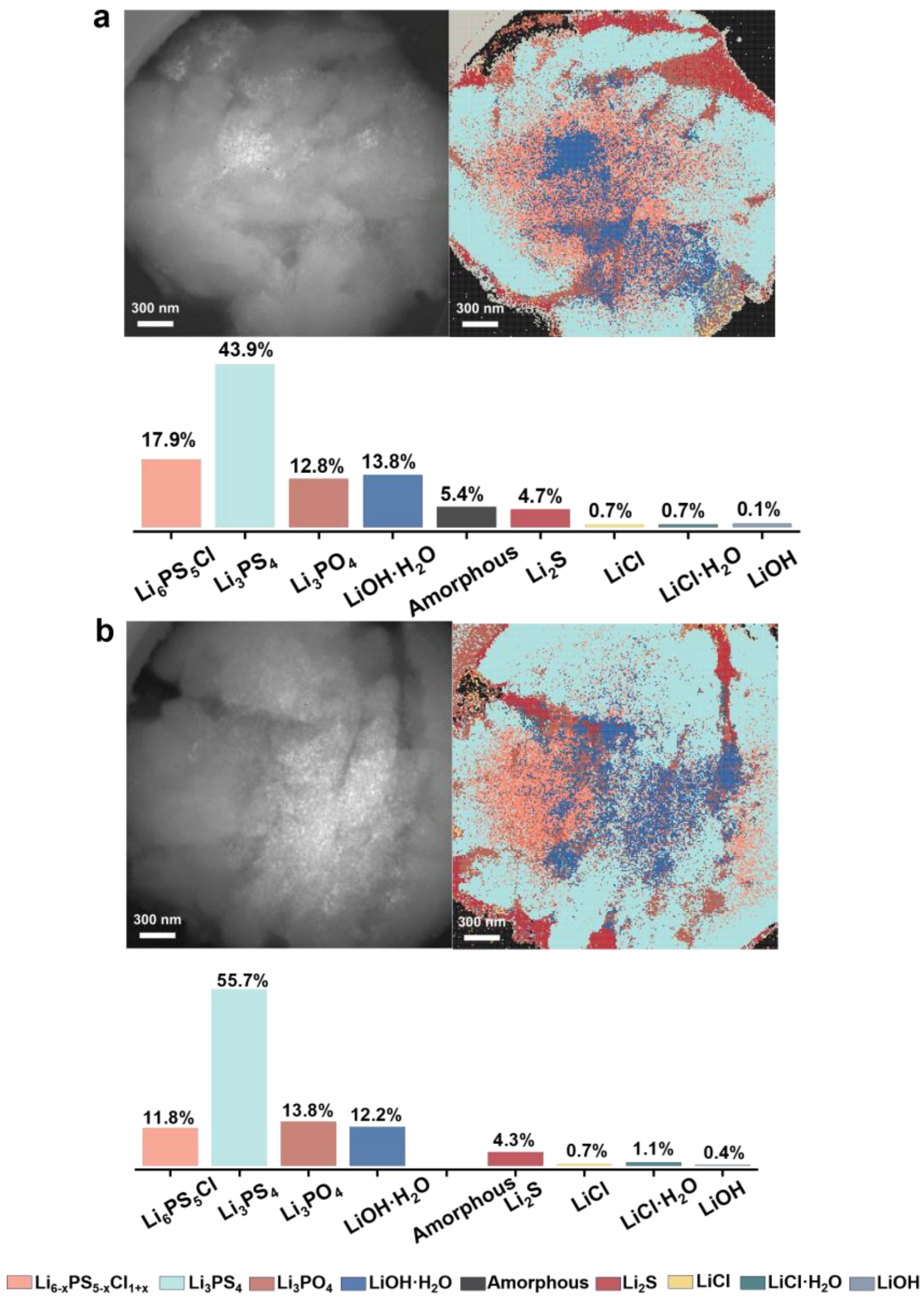

**Supplementary Fig. 7 |** Additional statistical phase distribution and quantitative analysis. (a, b) HAADF images and corresponding 4D STEM phase maps of individual particles from the deeply hydrolyzed Cl1.5 sample. Color-coded maps show the spatial distribution of the residual Cl1.5 phase alongside hydrolysis products including $Li_3PS_4$, $Li_3PO_4$, and $LiOH{\cdot}H_2O$. The bottom bar charts show the quantitative area fraction for each identified phase.

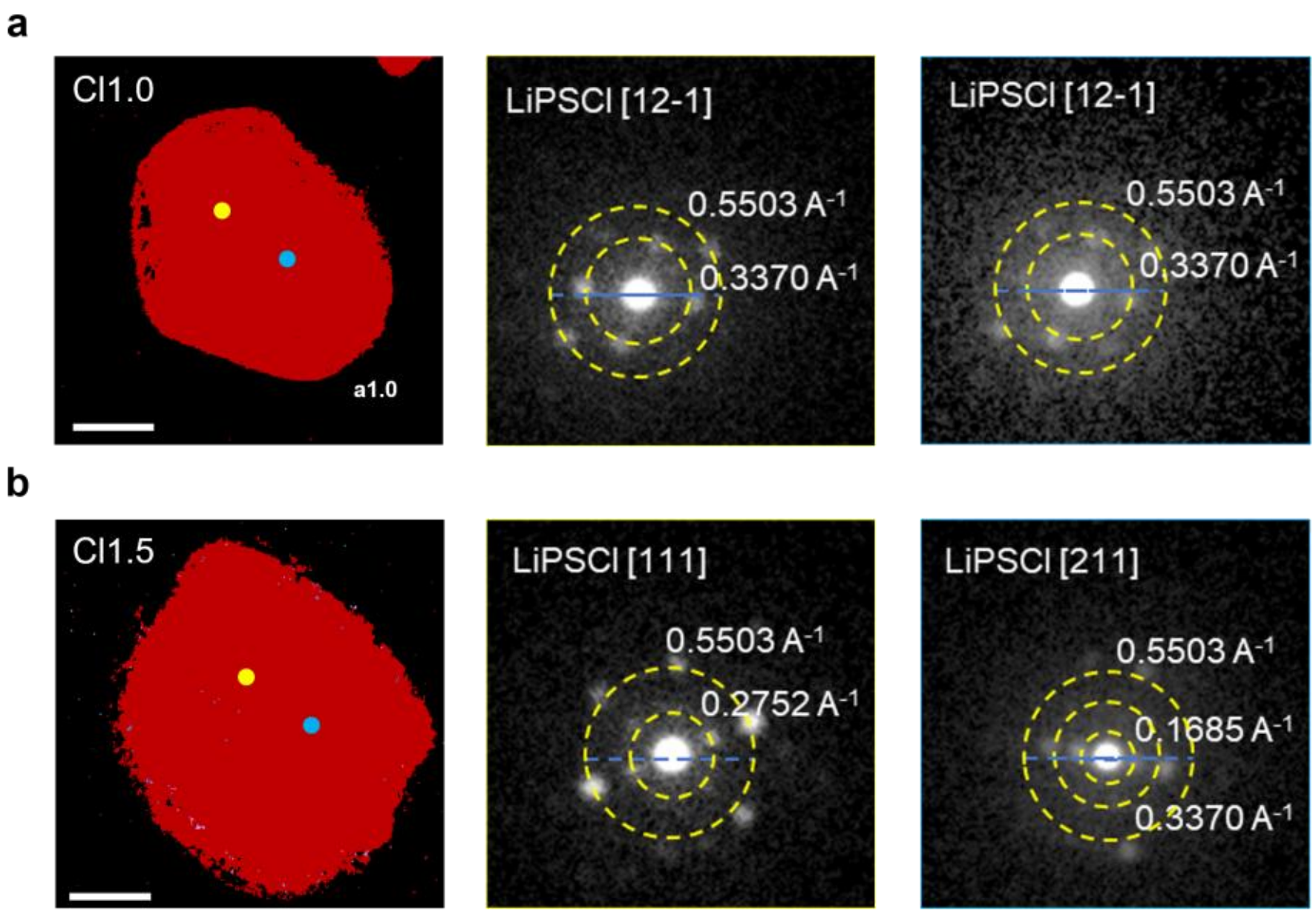

**Supplementary Fig. 8 |** Microstructure and diffraction analysis of $Li_{7-x}PS_{6-x}Cl_x$ electrolyte particles. (a) Uncycled Cl1.0 electrolyte particle. (Left) Overlap map of the Cl1.0 particle, with two selected positions marked by yellow and blue dots. (Middle and right) Corresponding electron diffraction patterns acquired at the marked positions. (b) Uncycled Cl1.5 electrolyte particle. Scale bars, 200 nm.

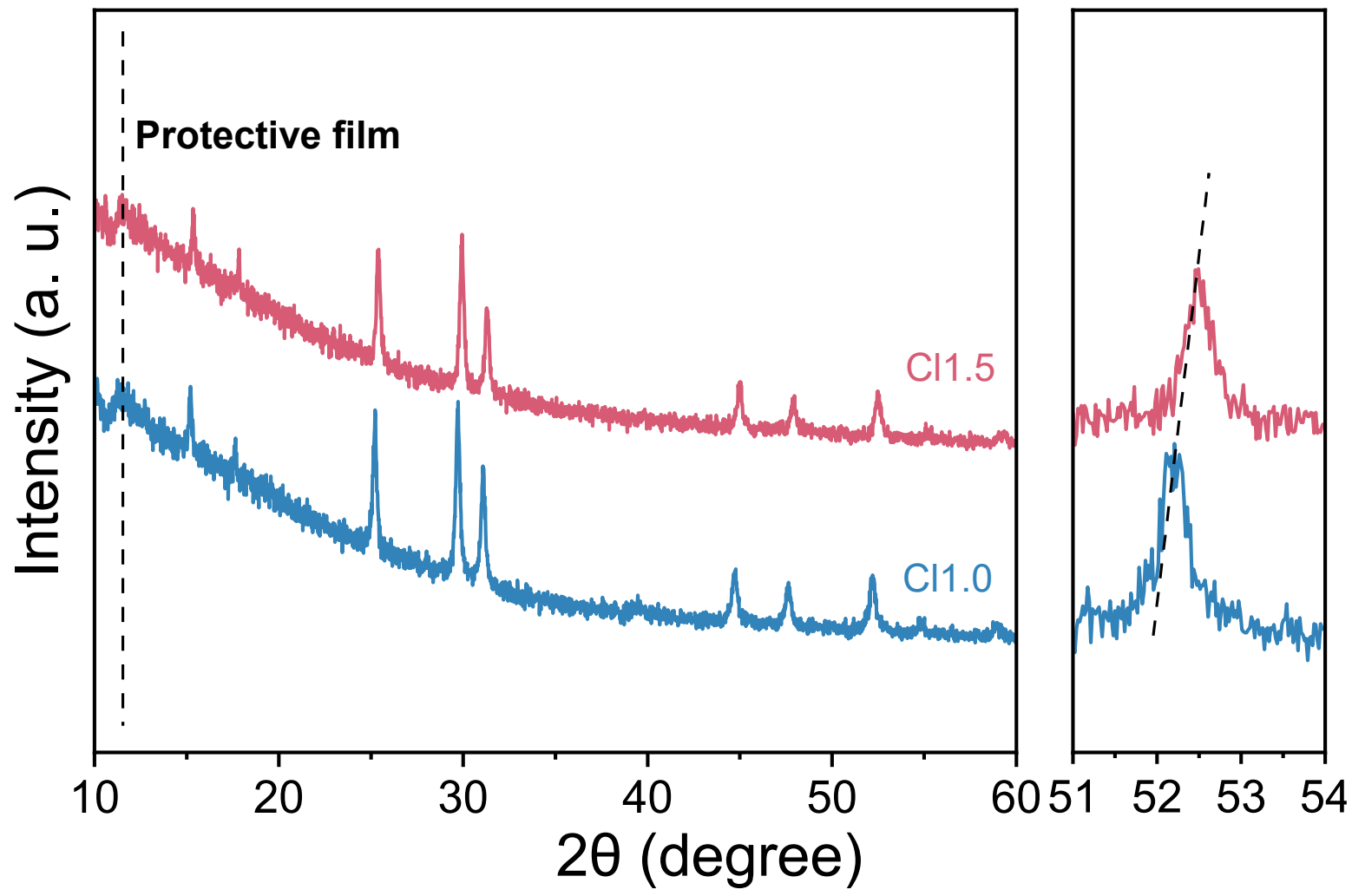

**Supplementary Fig. 9 |** Phase characterization of pristine sulfide electrolytes. X-ray diffraction patterns for Cl1.0 and Cl1.5, exhibiting a single cubic argyrodite phase without detectable impurities. The vertical dashed line at low angles can be ascribed to the polyimide protective film used during measurement. A magnified view of the diffraction peaks highlights the lattice shrinkage of Cl1.5 due to the varied chlorine content**.**

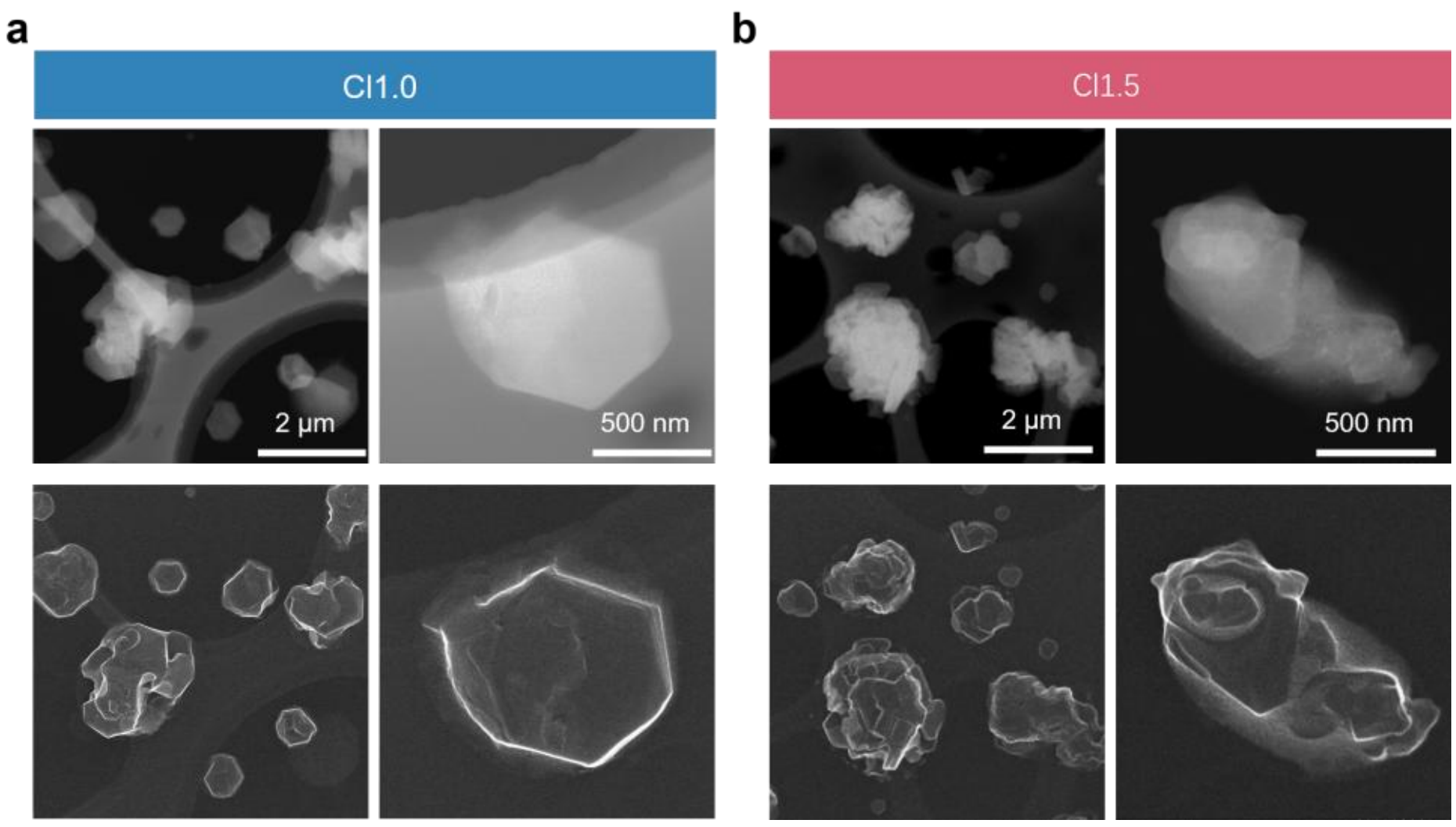

**Supplementary Fig. 10 |** Morphological characterization of pristine electrolyte particles. HAADF-STEM and secondary electron images of (a) pristine Cl1.0 and (b) pristine Cl1.5 at various magnifications.

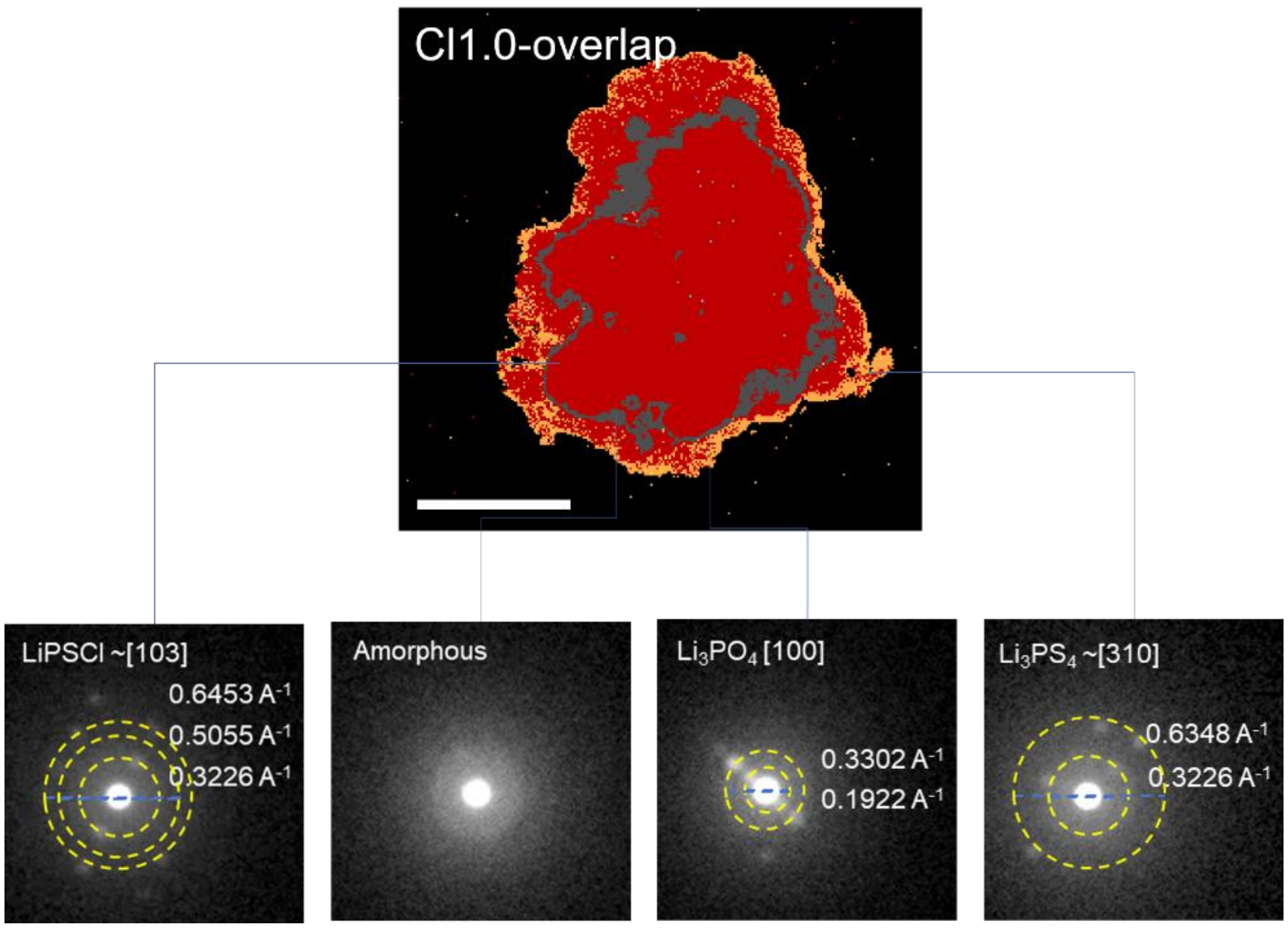

**Supplementary Fig. 11 |** Phase distribution and corresponding electron diffraction pattern of Cl1.0 electrolyte particle after 30 min exposure. Scale bar, 200 nm.

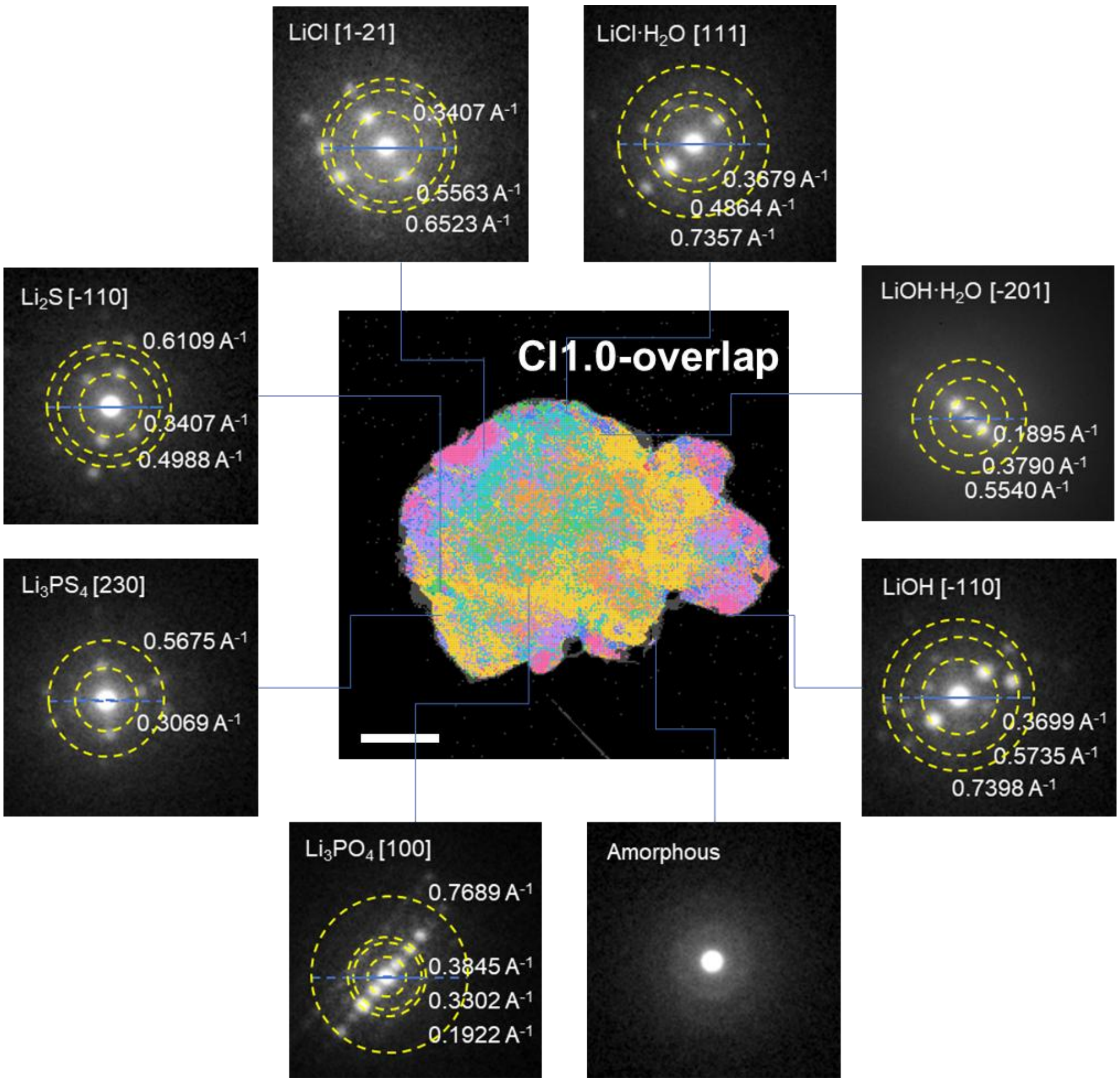

**Supplementary Fig. 12 |** Phase distribution and corresponding electron diffraction pattern of Cl1.0 electrolyte particle after 3 h exposure. Scale bar, 200 nm.

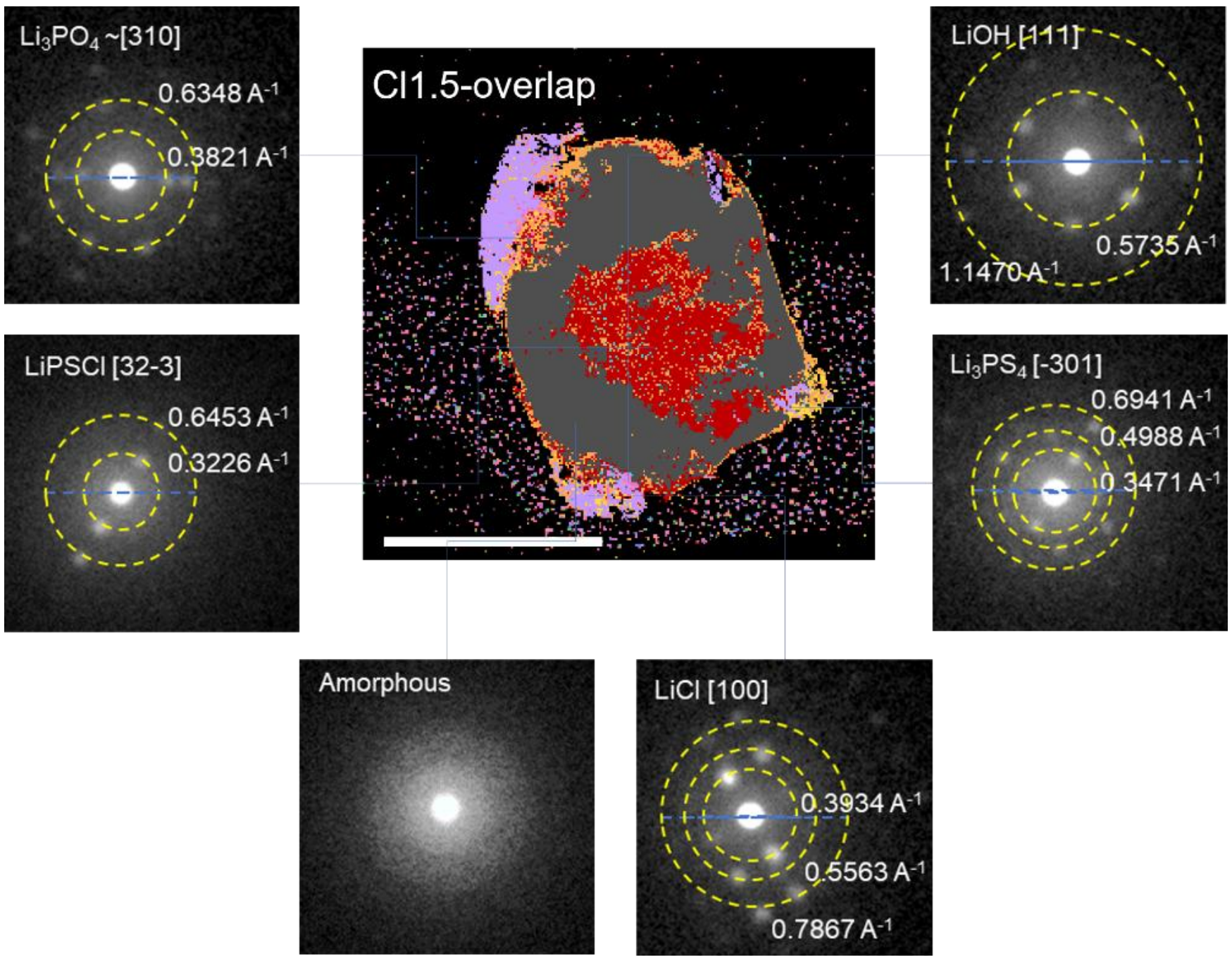

**Supplementary Fig. 13 |** Phase distribution and corresponding electron diffraction pattern of Cl1.5 electrolyte particle after 30 min exposure. Scale bar, 200 nm.

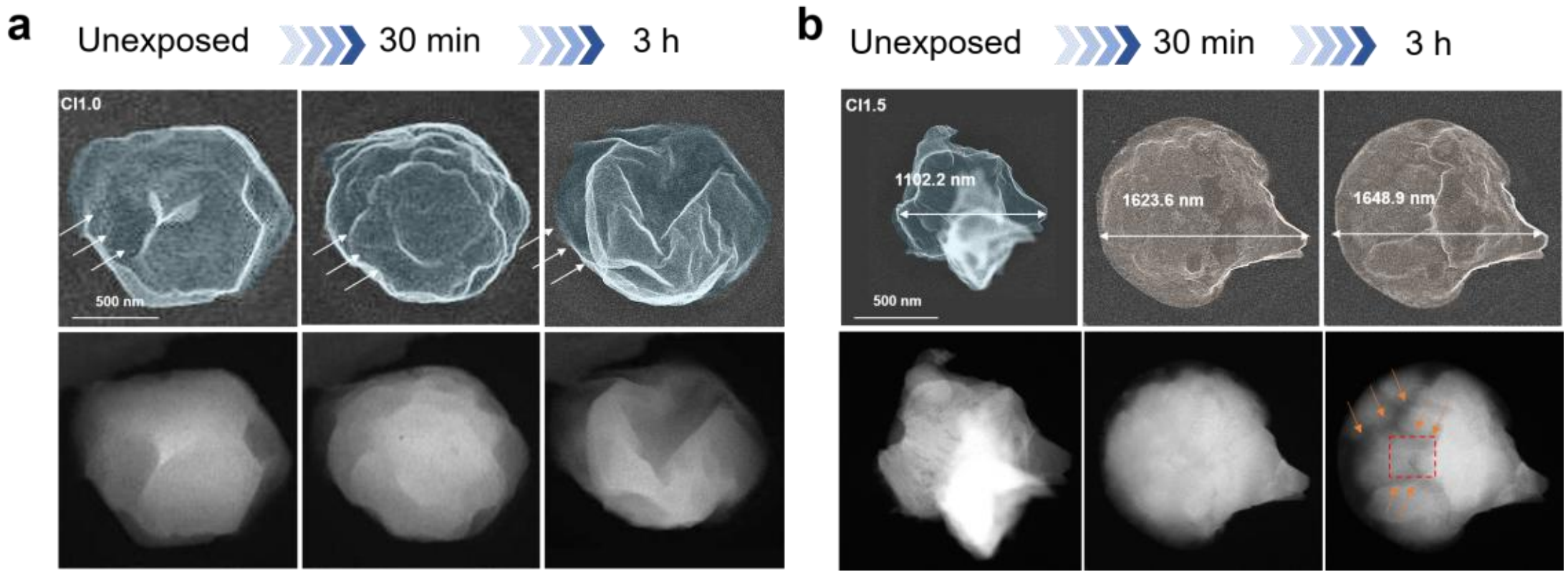

**Supplementary Fig. 14 |** HAADF-STEM and secondary-electron images of morphological evolution and structural collapse. Time-resolved STEM images showing the morphological changes of (a) Cl1.0 and (b) Cl1.5 particles during moisture exposure for up to 3 h. (a) White arrows highlight the expansion of the particle edges and the loss of crystalline facets in the Cl1.0 sample. (b) A representative Cl1.5 particle exhibits apparent expansion and structural collapse, with its projected maximum dimension increasing from 1102.2 to 1648.9 nm. Orange arrows and the dashed red box in the 3-h panel indicate the formation of internal voids and structural fragmentation.

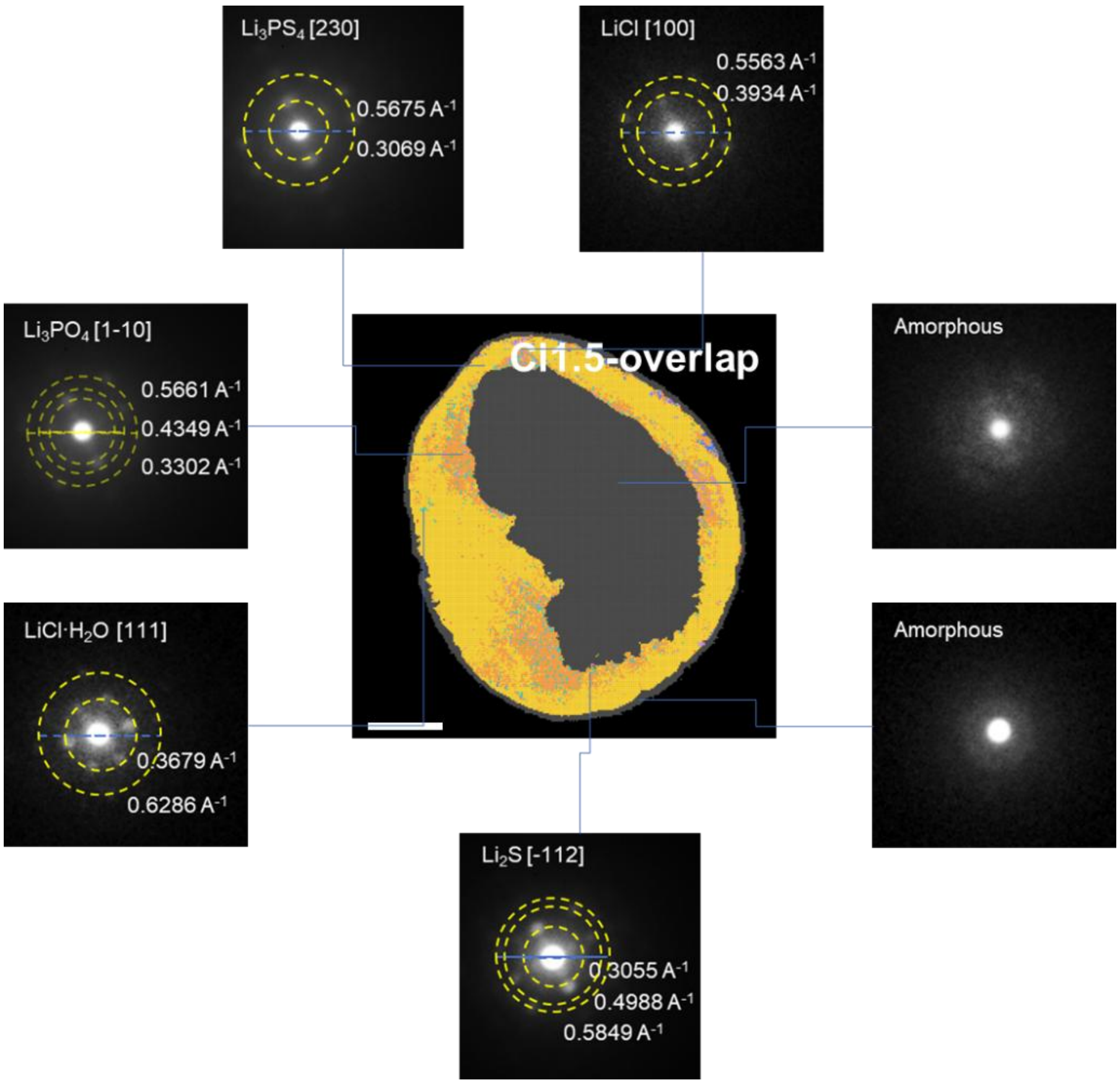

**Supplementary Fig. 15 |** Phase distribution and corresponding electron diffraction pattern of Cl1.5 electrolyte particle after 3 h exposure. Scale bar, 200 nm.

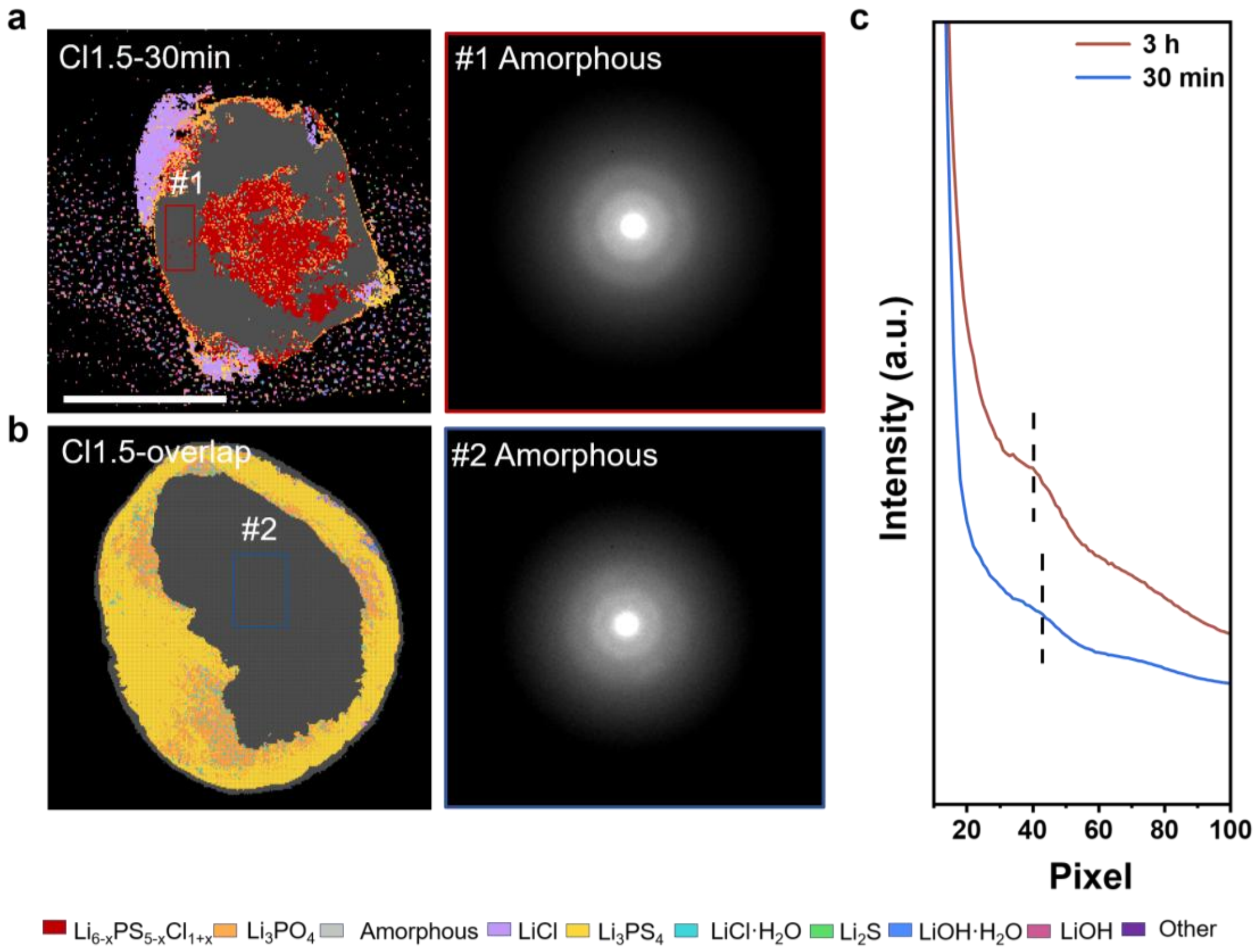

**Supplementary Fig. 16 |** Evolution of local structural order within amorphous regions. 4D STEM phase maps and representative electron diffraction patterns for the Cl1.5 electrolyte after (a) 30 minutes and (b) 3 hours of moisture exposure. Grey areas in the phase maps represent amorphous regions identified by diffuse rings in diffraction patterns 1 and 2. (c) Comparison of radially integrated intensity profiles for the amorphous regions at different exposure times. The profile for the 3-hour sample shows more distinct features than the 30-minute sample, indicating an increase in local structural order.

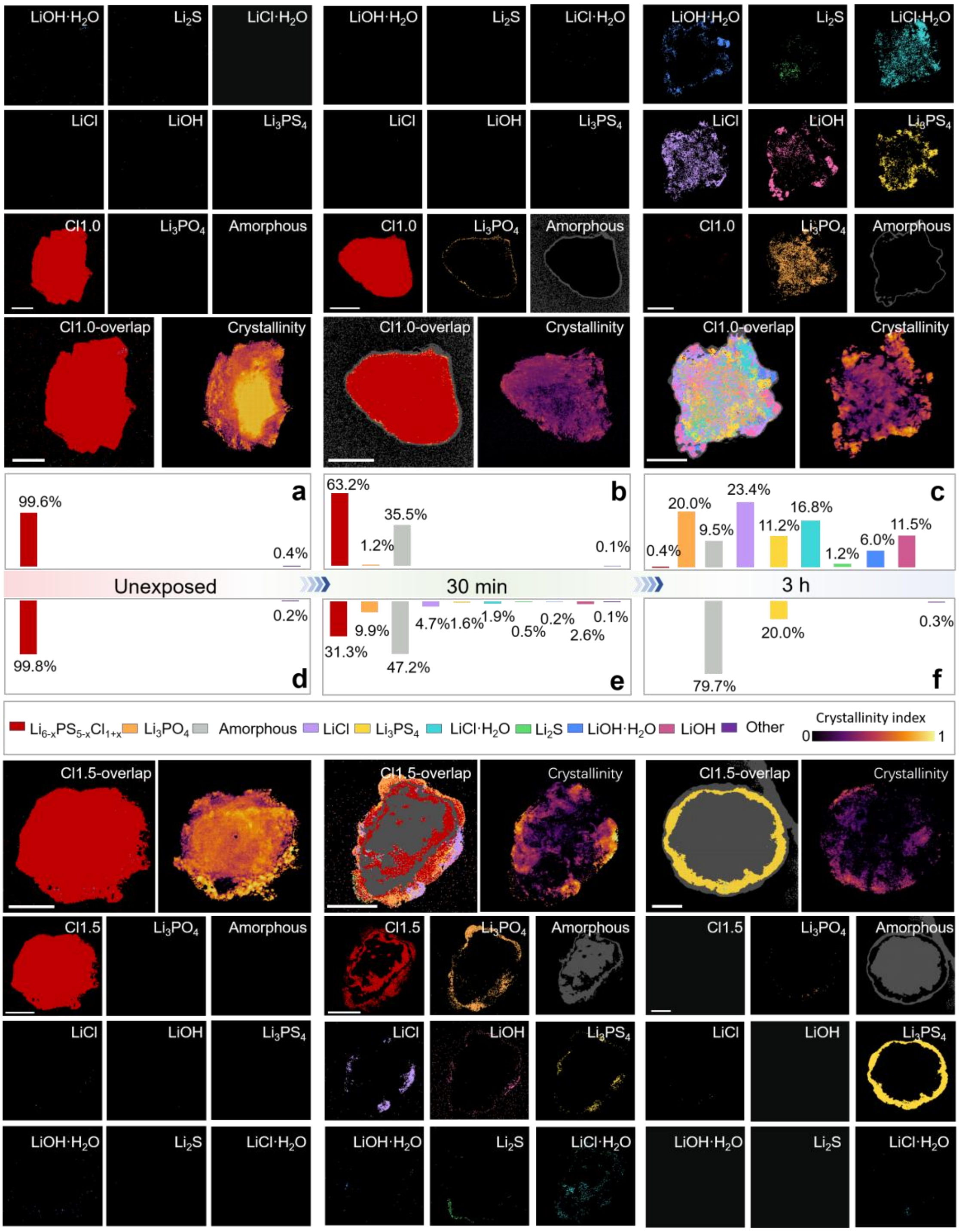

**Supplementary Fig. 17 |** Additional spatiotemporal observation of the degradation process in Cl1.0 and Cl1.5 solid electrolytes. Multi-panel figures combined with phase fraction histogram, crystallinity index (blue-red color scaled), color-coded composites mapping distributions (with phases marked in corresponding panels) of (a, c, e) Cl1.0 and (b, d, f) Cl1.5 after 1000 ppm $H_2O$/Ar atmosphere exposure of (a, b) 0 min, (c, d) 30 min, and (e, f) 3 hour. Electron diffraction analysis of multi-component phases in the particle is shown in Supplementary Figs. 18–22.

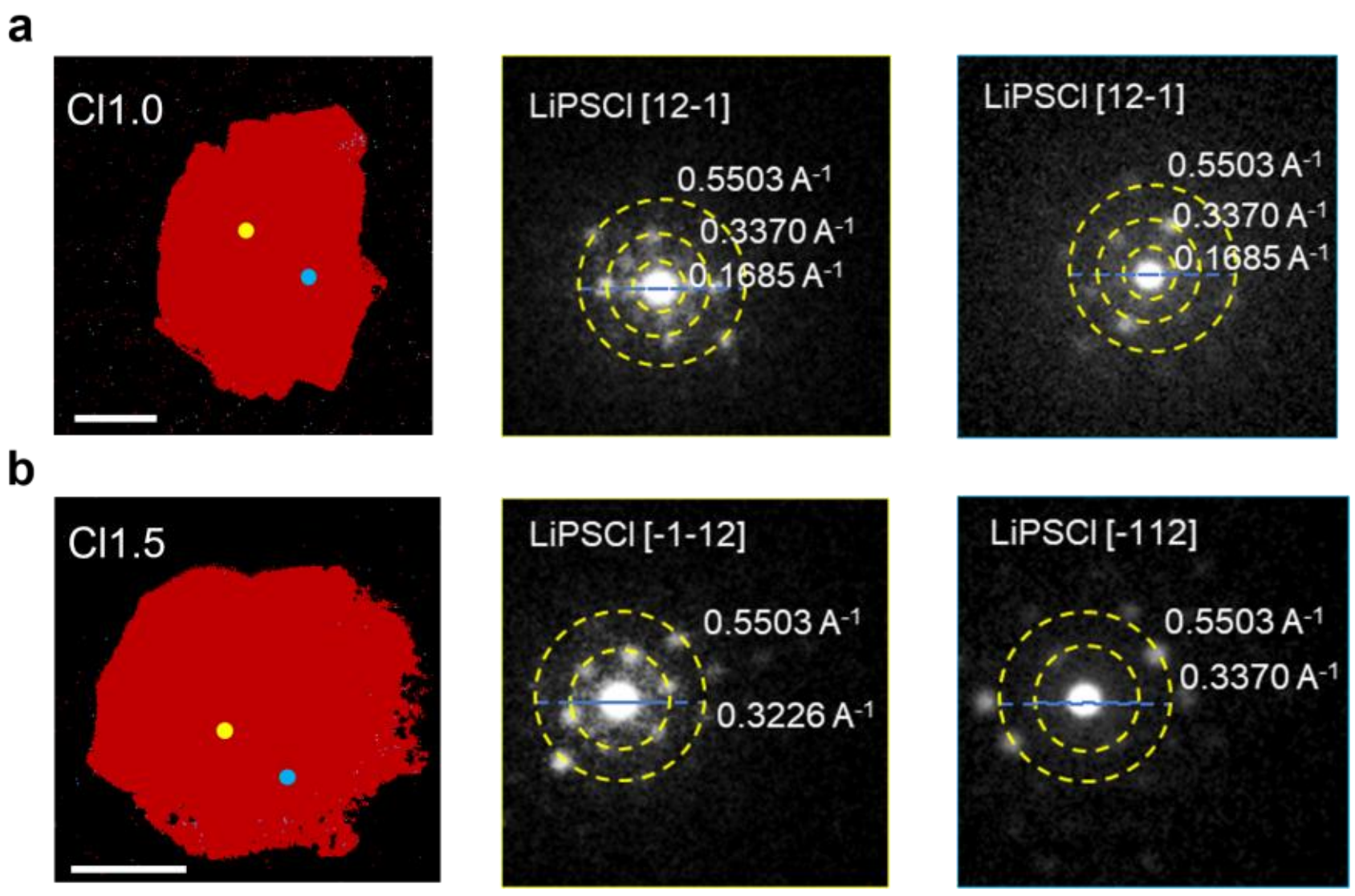

**Supplementary Fig. 18 |** Phase distribution and diffraction analysis of $Li_{7-x}PS_{6-x}Cl_x$ electrolyte particles. (a) Unexposed Cl1.0 electrolyte particle. (Left) Overlap map of the Cl1.0 particle, with two selected positions marked by yellow and blue dots. (Middle and right) Corresponding electron diffraction patterns acquired at the marked positions. (b) Unexposed Cl1.5 electrolyte particle. Scale bars, 200 nm.

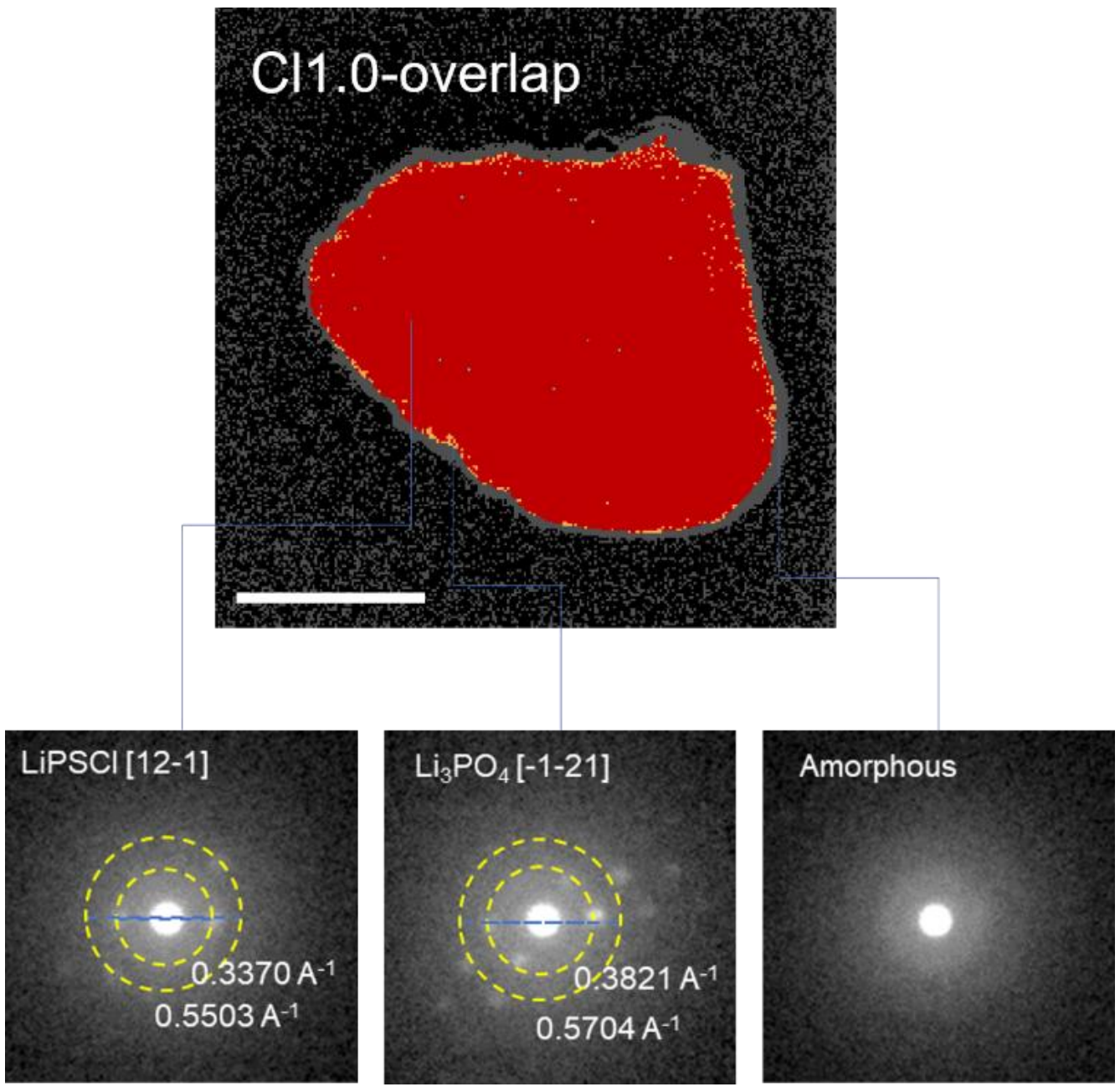

**Supplementary Fig. 19 |** Phase distribution and corresponding electron diffraction pattern of Cl1.0 electrolyte particle after 30 min exposure. Scale bar, 200 nm.

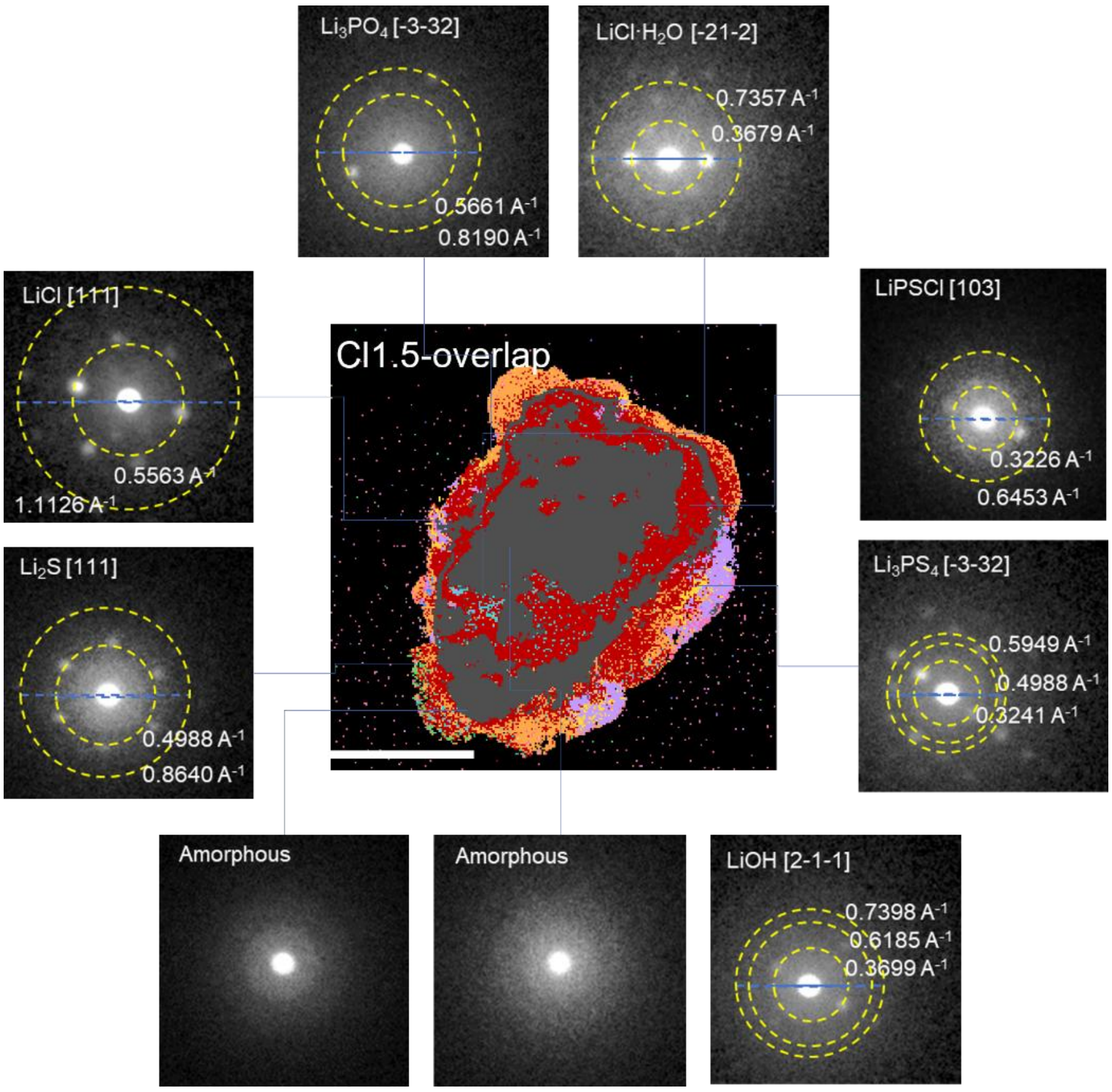

**Supplementary Fig. 20 |** Phase distribution and corresponding electron diffraction pattern of Cl1.5 electrolyte particle after 30 min exposure. Scale bar, 200 nm.

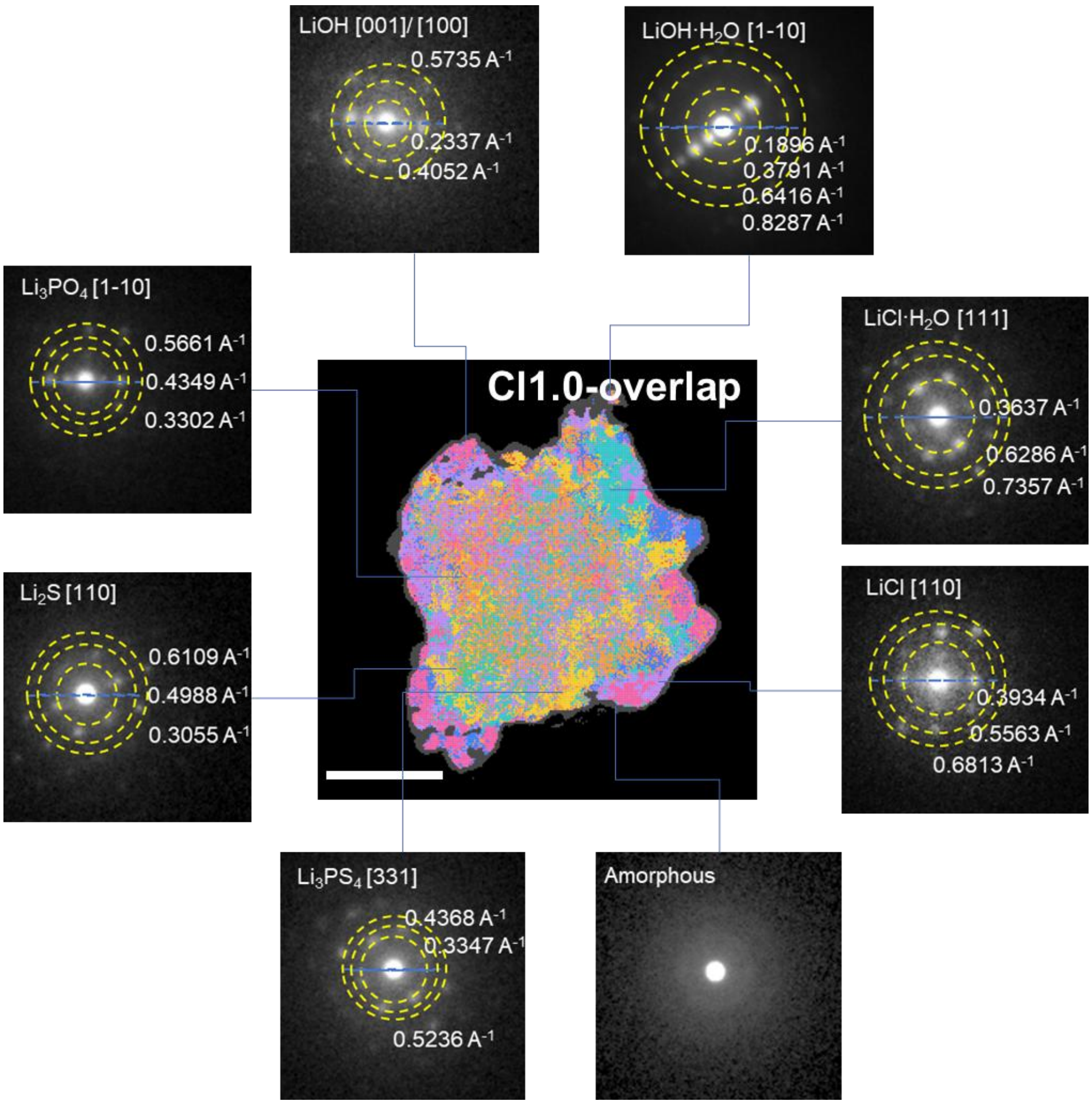

**Supplementary Fig. 21 |** Phase distribution and corresponding electron diffraction pattern of Cl1.0 electrolyte particle after 3 h exposure. Scale bar, 200 nm.

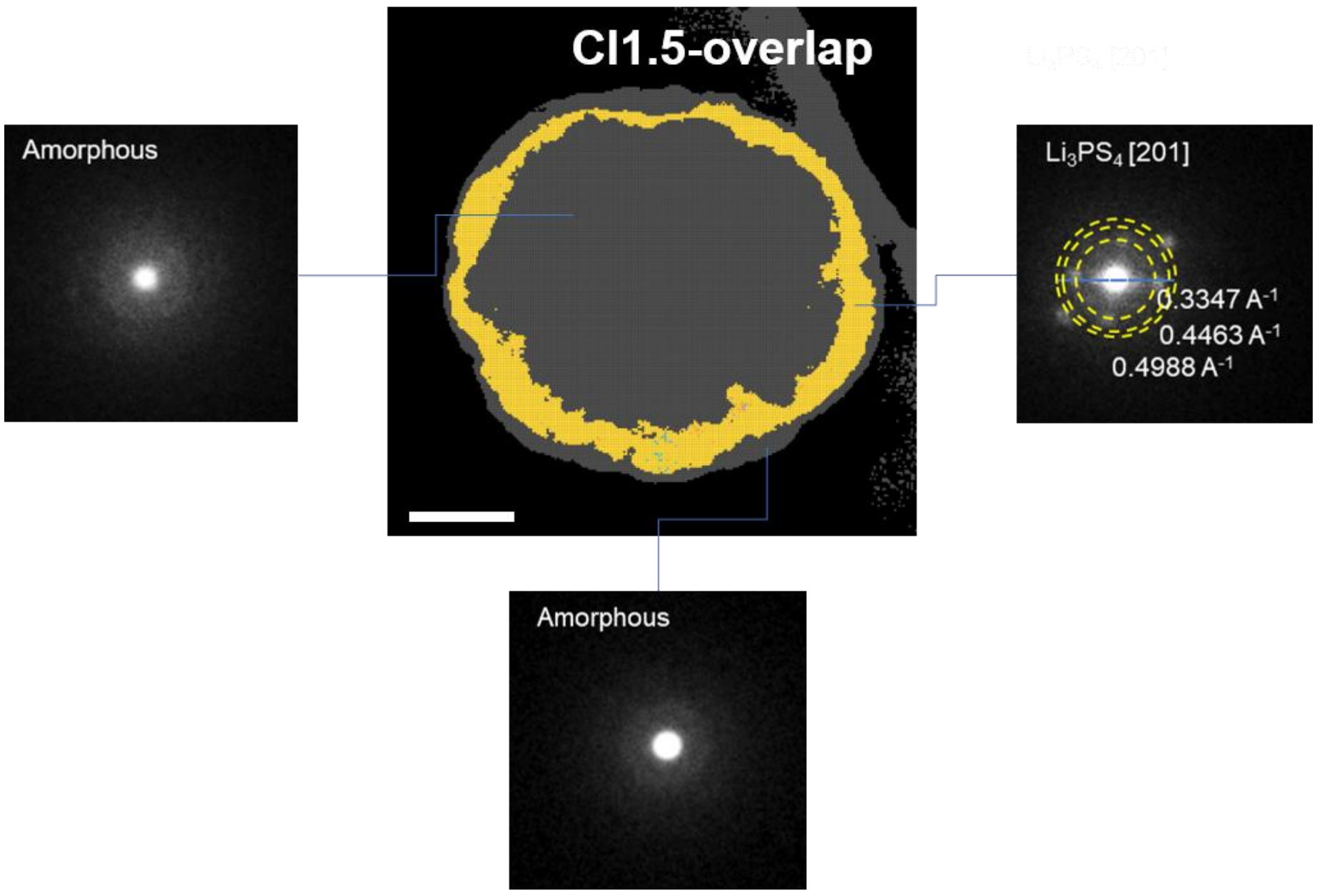

**Supplementary Fig. 22 |** Phase distribution and corresponding electron diffraction pattern of Cl1.5 electrolyte particle after 3 h exposure. Scale bar, 200 nm.

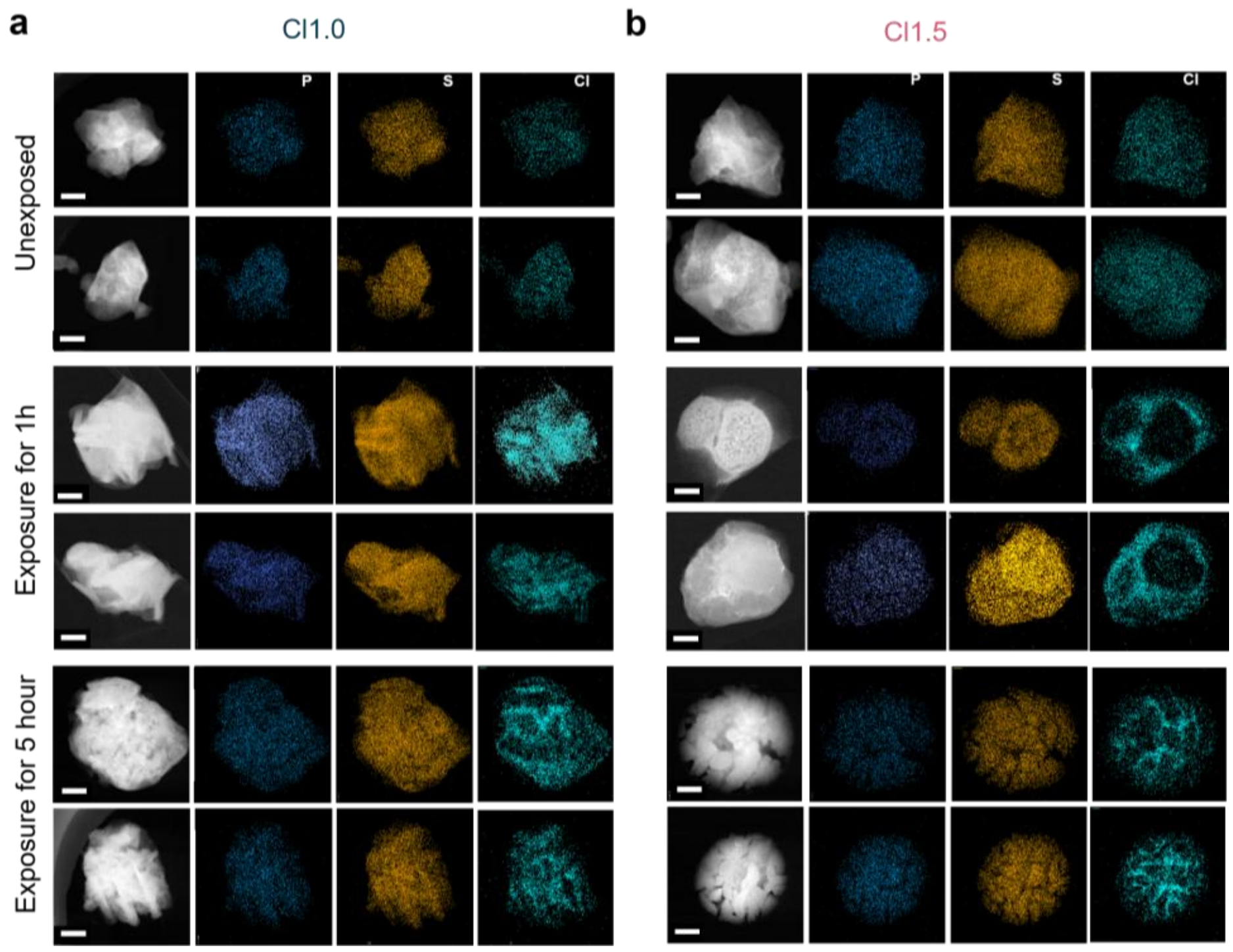

**Supplementary Fig. 23 |** Elemental distribution during hydrolysis. Energy dispersive X-ray spectroscopy mapping for (a) Cl1.0 and (b) Cl1.5 particles over an exposure duration from 0 to 5 hours. The STEM images and corresponding elemental maps of phosphorus, sulfur, and chlorine illustrate the spatial distribution of these elements as hydrolysis proceeds. Scale bars, 200 nm.

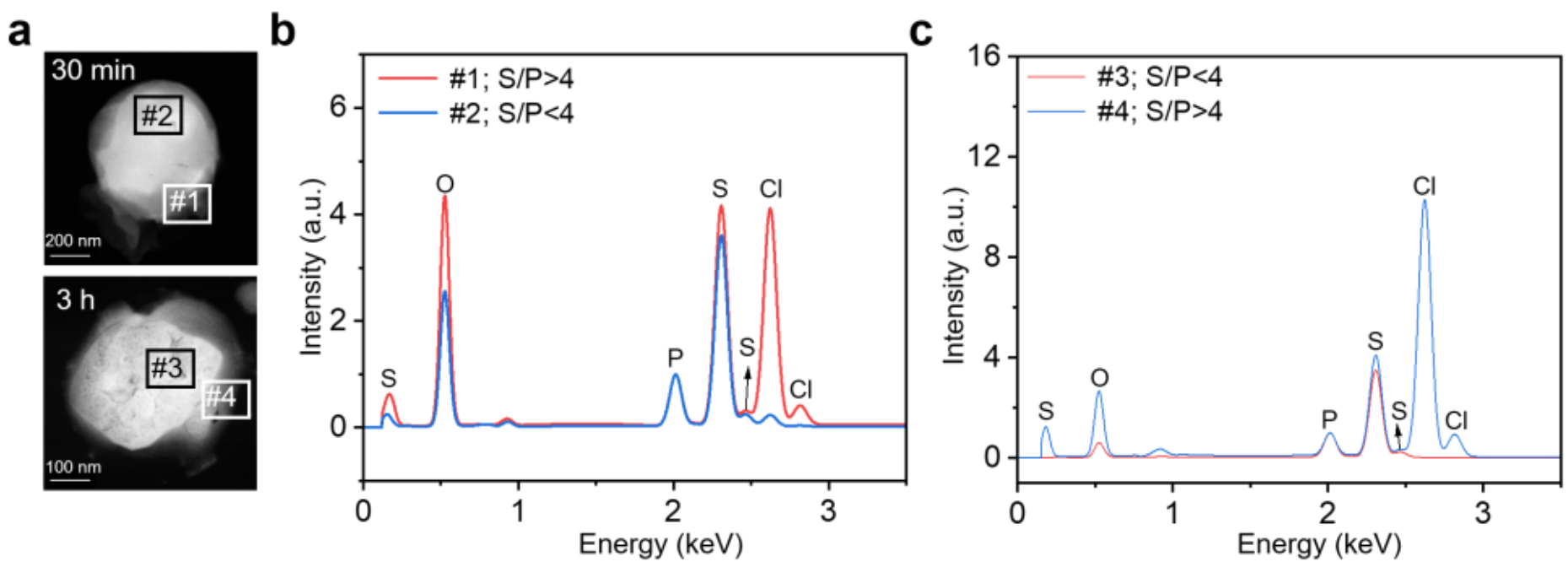

**Supplementary Fig. 24 |** Elemental composition and energy dispersive X-ray spectroscopy analysis of hydrolysis regions. (a) HAADF-STEM images of electrolyte particles after 30 min and 3 h of moisture exposure, with labeled regions 1–4 for local chemical analysis. (b, c) EDX spectra on the indicated regions at different exposure times.

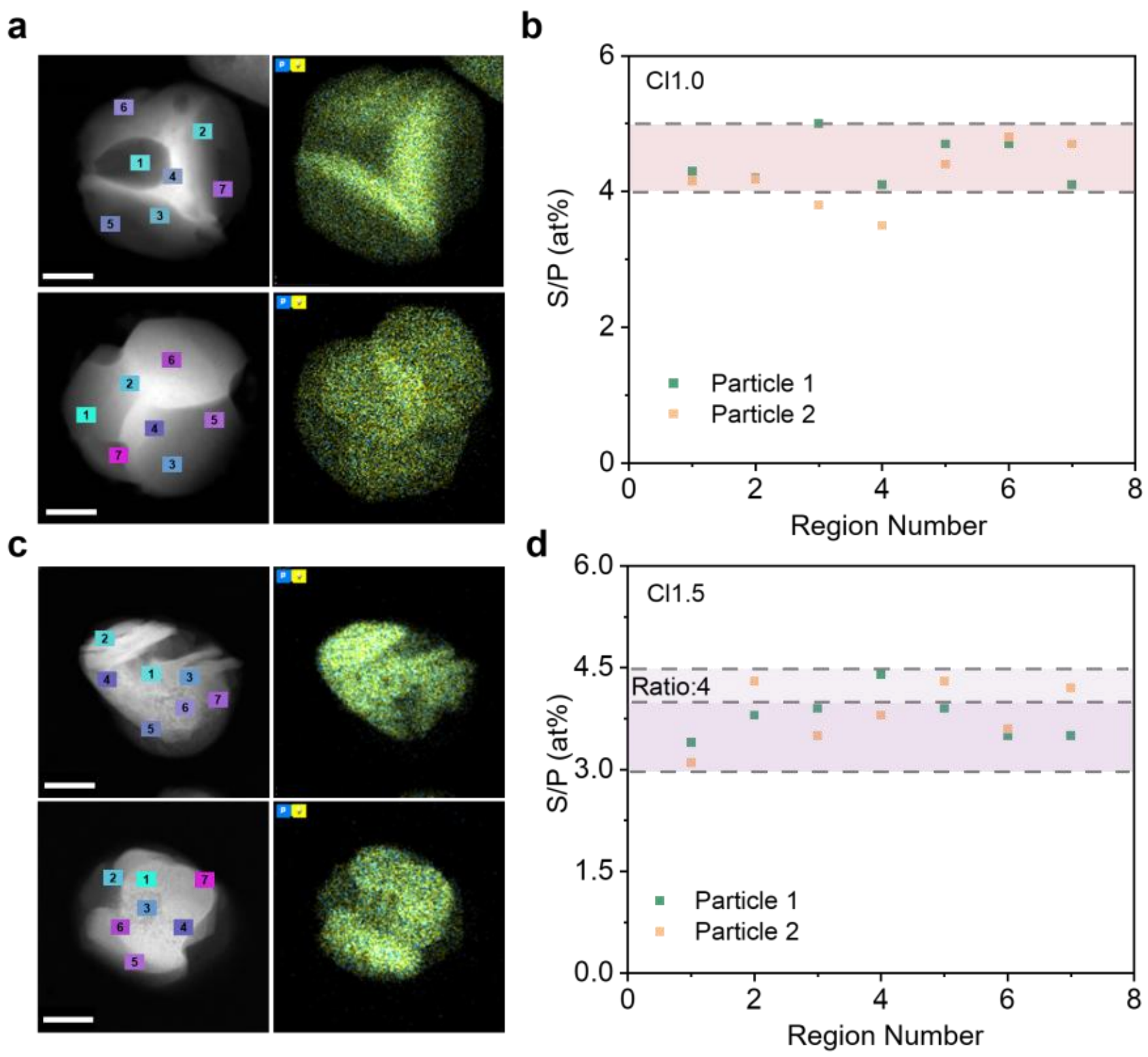

**Supplementary Fig. 25 |** Local elemental analysis and sulfur to phosphorus ratio quantification. HAADF-STEM images and corresponding EDS maps of (a) degraded Cl1.0 and (c) Cl1.5 particles after storage in a 1000 ppm $H_2O$/Ar atmosphere for 20 h. Labels 1 to 7 indicate specific local regions selected for quantitative chemical analysis on the particle interior and surface. (b, d) Calculated S/P atomic ratios for two representative particles of each composition. Scale bars, 200 nm.

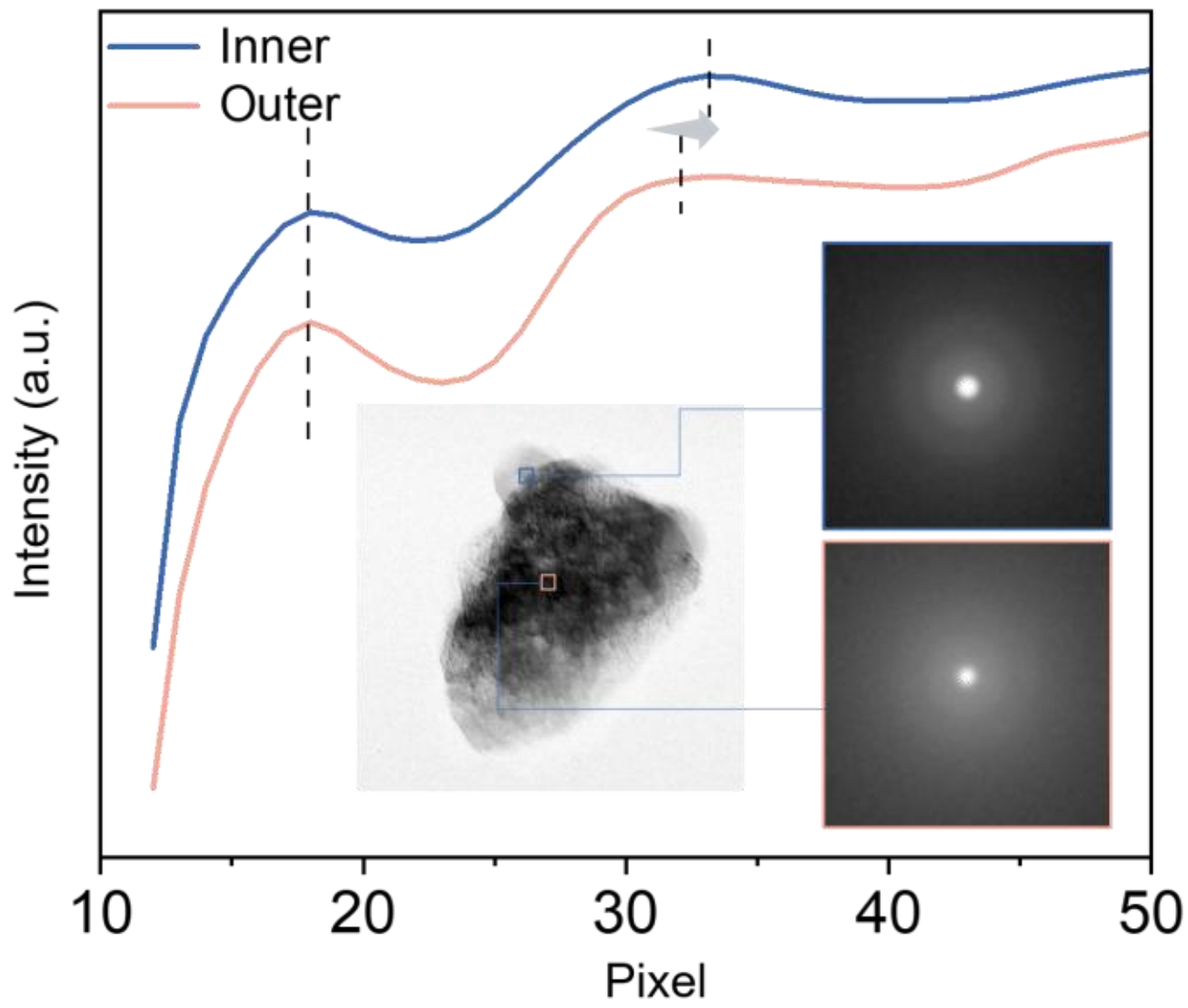

**Supplementary Fig. 26 |** Structural and intensity comparison between the edge and interior amorphous regions of the degraded argyrodite particle. The line intensity profiles (blue: outer region, pink: inner region) show the variations in signal intensity across the particle, highlighting differences between the edge and interior areas. The corresponding electron diffraction patterns collected from the outer (blue frame) and inner (pink frame) amorphous regions both exhibit diffuse halos, confirming the amorphous nature of the degraded phases at both locations. The shift in intensity peaks indicates structural or compositional variations between the edge and interior amorphous domains.

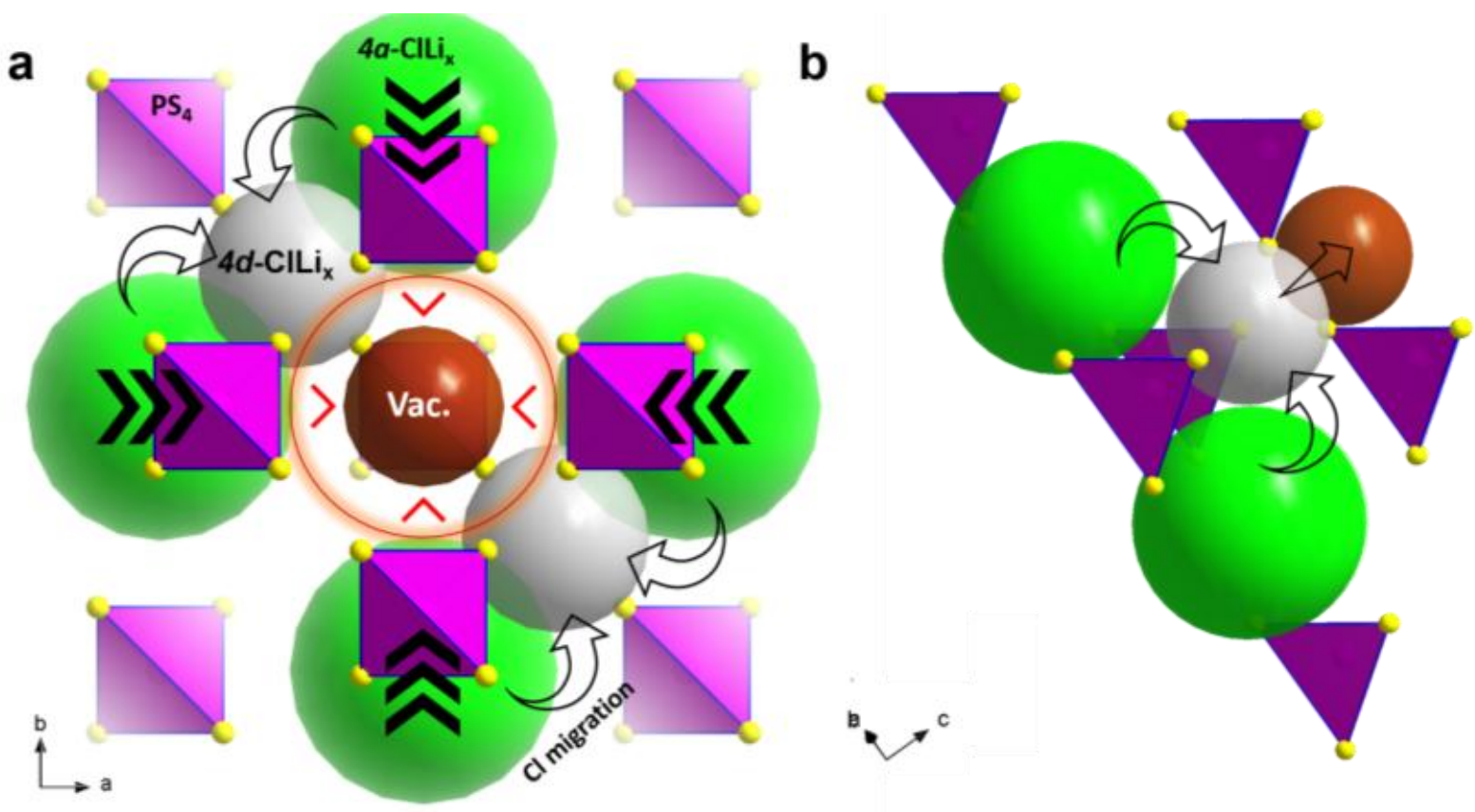

**Supplementary Fig. 27 |** Structural response of Cl1.5 to modeled LiCl removal and its relationship to the proposed chloride-redistribution pathway. Cl-deficient Cl1.5 structures viewed along the (a) [001] and (b) [−110] directions. The green and grey polyhedra represent the $Li_xCl$ coordination environments at the *4a* and *4d* sites, respectively. Removal of a LiCl unit generates local Li/Cl vacancies, indicated by the red sphere, and is accompanied by distortion of the adjacent $PS_4^{3-}$ tetrahedra shown in purple. This local structural response may modify the energetics of chloride redistribution along the proposed *4a* → *4d* → *4a* pathway but does not independently establish the migration rate.

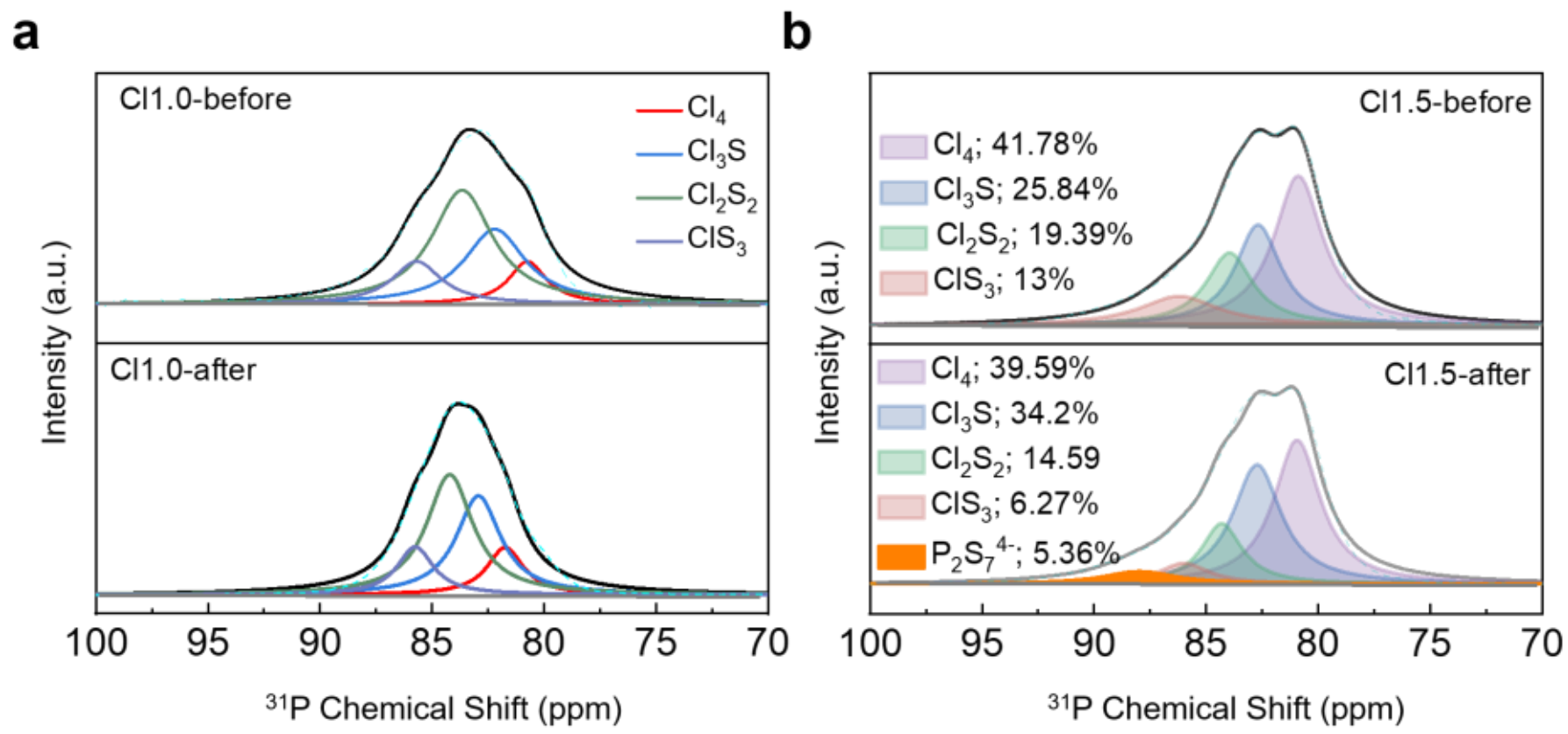

**Supplementary Fig. 28 |** Molecular framework evolution analyzed by nuclear magnetic resonance spectroscopy. (a) Solid-state $^{31}P$ magic-angle spinning nuclear magnetic resonance spectra for the Cl1.0 before and after hydrolysis. (b) $^{31}P$ MAS NMR spectra for the Cl1.5 before and after hydrolysis.

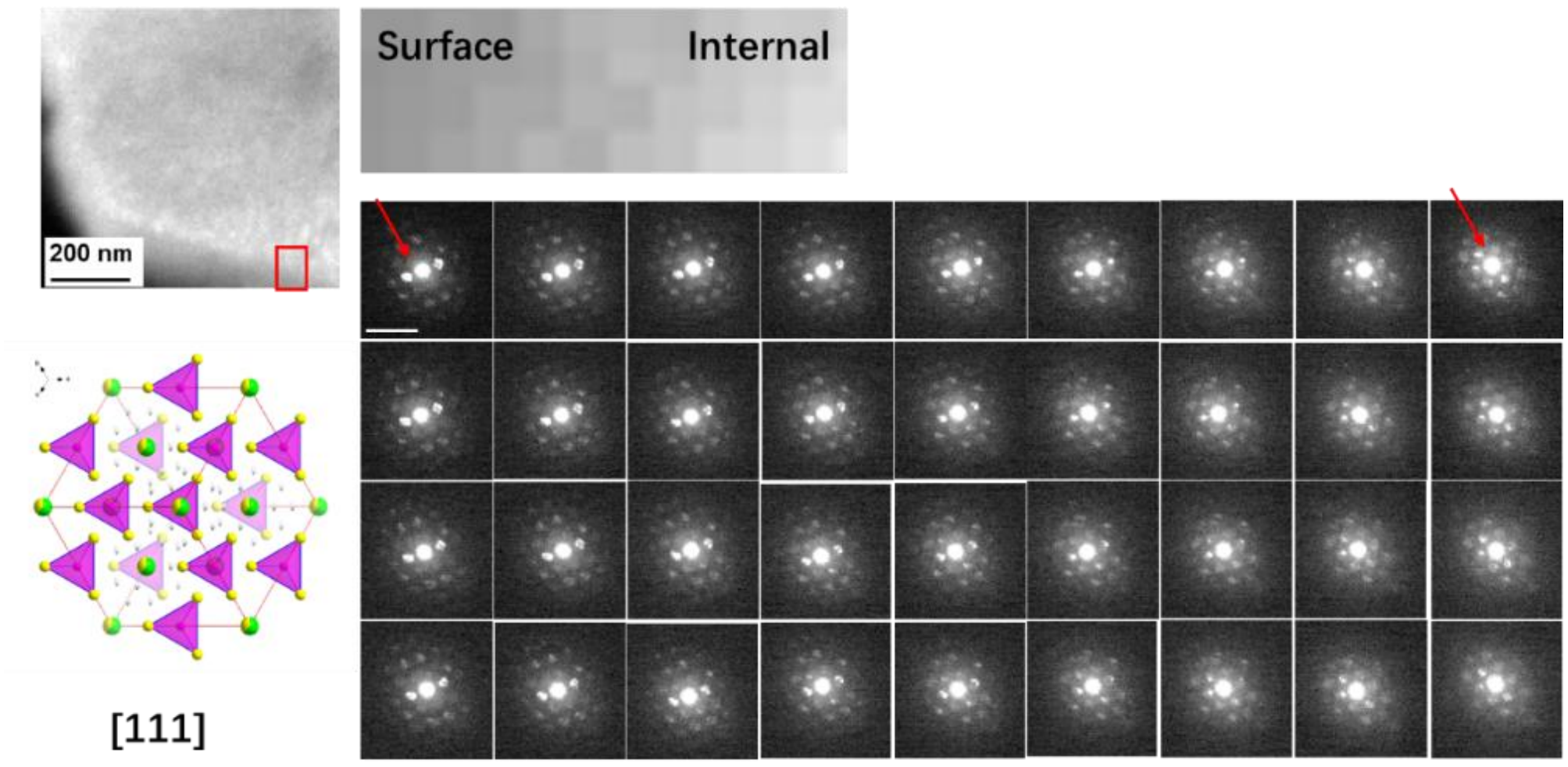

**Supplementary Fig. 29 |** Microstructural and phase evolution of the Cl1.5 argyrodite electrolyte after initial hydrolysis under a 1000 ppm $H_2O$/Ar atmosphere. A high-angle annular dark-field scanning transmission electron microscopy image shows the morphology of a Cl1.5 particle, with the red box marking the region of interest for local analysis. The schematic illustrates the argyrodite crystal structure, viewed from the [111] direction. Nanobeam electron diffraction patterns acquired along a line from the particle surface to the interior reveal a gradient in crystallinity and the presence of additional or diffuse diffraction features induced by moisture exposure. Red arrows indicate these additional or diffuse diffraction features. Scale bar, 10 1/nm.

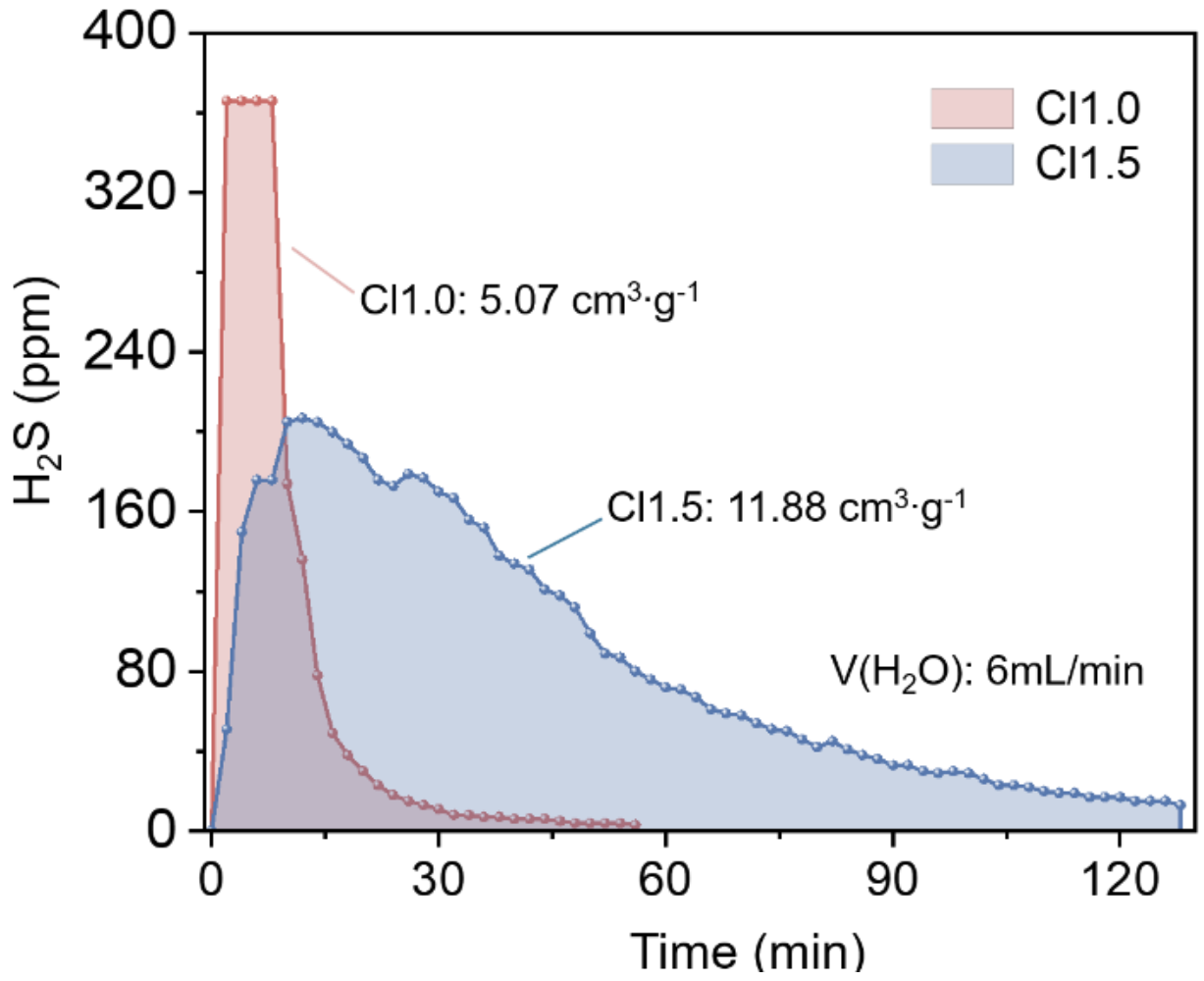

**Supplementary Fig. 30 |** In situ quantification of $H_2S$ gas evolution for Cl1.5 and Cl1.0 during moisture exposure. Real-time monitoring of $H_2S$ concentration (ppm) over time for the Cl1.0 (red) and Cl1.5 (blue) samples. The moisture exposure was conducted using a 1000 ppm $H_2O$/Ar stream at a flow rate of 6 mL $min^{-1}$. The shaded regions represent the integrated $H_2S$ volume released per gram of material, calculated from the $H_2S$ concentration–time profiles.

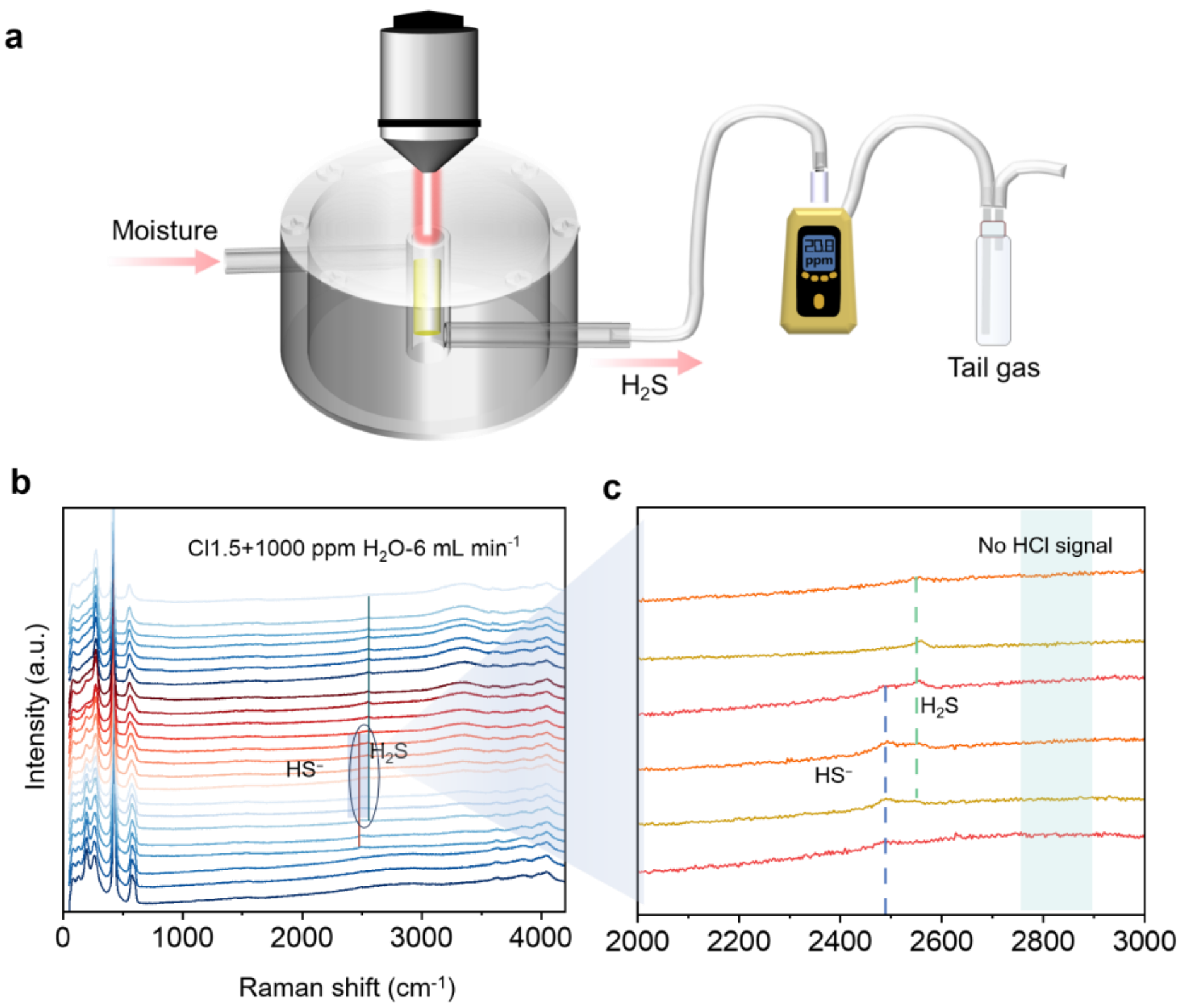

**Supplementary Fig. 31 |** Surface hydrolysis and gas evolution in argyrodite electrolytes. (a) Schematic illustration showing the integrated experimental setup for simultaneous in situ Raman spectroscopy and hydrogen sulfide gas detection. (b) Time-resolved Raman spectra of the Cl1.5 electrolyte recorded during exposure to a 1000 ppm $H_2O$/Ar stream at a flow rate of 6 mL min$^{-1}$. (c) The magnified view of the Raman spectra reveals vibrational signals corresponding to $HS^-$ and $H_2S$.

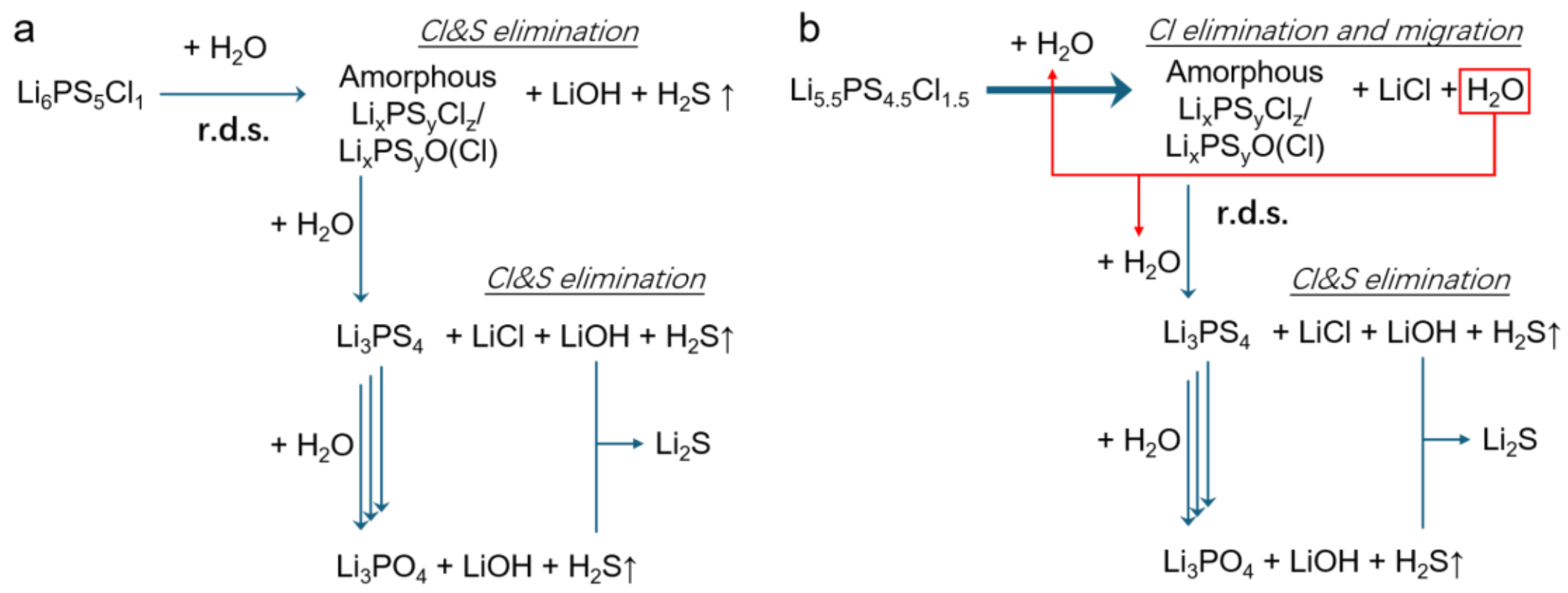

**Supplementary Fig. 32 |** Proposed moisture-degradation pathways of Cl1.0 and Cl1.5 inferred from the combined structural, chemical, and gas-evolution measurements.

Cl1.0 (a) exhibits predominantly surface-localized degradation during the early stage of moisture exposure, while the particle interior largely retains its crystalline structure. This spatial pattern is consistent with an inward-progressing reaction front whose propagation may be limited by moisture transport into the particle and by water consumption during near-surface hydrolysis. The pronounced early $H_2S$ signal, together with the formation of LiOH-containing surface products, indicates substantial sulfur removal and hydrolysis near the particle exterior. Continued exposure results in deeper hydrolysis and the development of a spatially layered distribution of crystalline and amorphous degradation products.

In Cl1.5 (b), LiCl-containing phases accumulate near the particle surface, while Cl-depleted amorphous regions extend into the particle interior. The relatively weak initial $H_2S$ signal and the spatial separation between oxygen-rich surface products and internal crystallinity loss are consistent with chloride redistribution contributing to structural disorder before extensive hydrolysis of the particle bulk. Continued exposure is accompanied by further chloride redistribution, amorphization, void formation, and the formation of $Li_3PS_4$-containing degradation products. The sequence illustrated here represents a mechanistic interpretation of the spatially and temporally resolved measurements rather than a directly observed elementary reaction pathway.

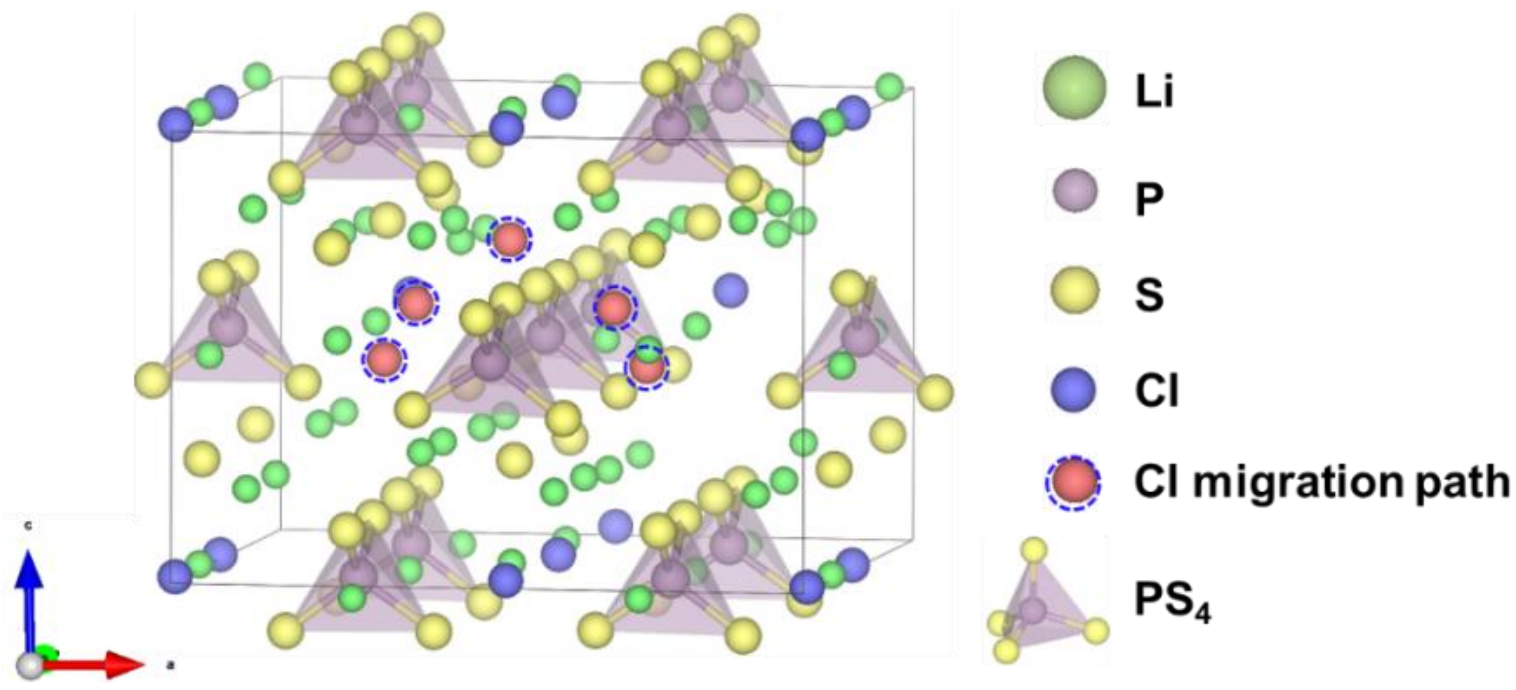

**Supplementary Fig. 33 |** Theoretical investigation of chloride ion migration pathways. Calculated chloride ion migration path within the Cl1.0 cell. The red spheres highlighted with dashed circles represent the sequential positions of a chlorine ion as it moves through the lattice sites.

Because the calculated barriers were obtained from idealized structural models, they should be interpreted primarily as comparative descriptors of pathway accessibility rather than as direct predictions of absolute room-temperature diffusion rates. The experimental material contains moisture-reacted surfaces, Li and Cl vacancies, structural disorder, grain boundaries, and dynamically evolving interfaces that are not fully represented in the idealized migration models. Water coordination, defect interactions, and local lattice softening during the early stages of degradation may further modify the effective migration barrier. Nevertheless, the calculations establish that the anion configuration of Cl1.5 offers a substantially more favorable kinetic pathway than that of Cl1.0, consistent with the experimentally observed difference in chloride redistribution.

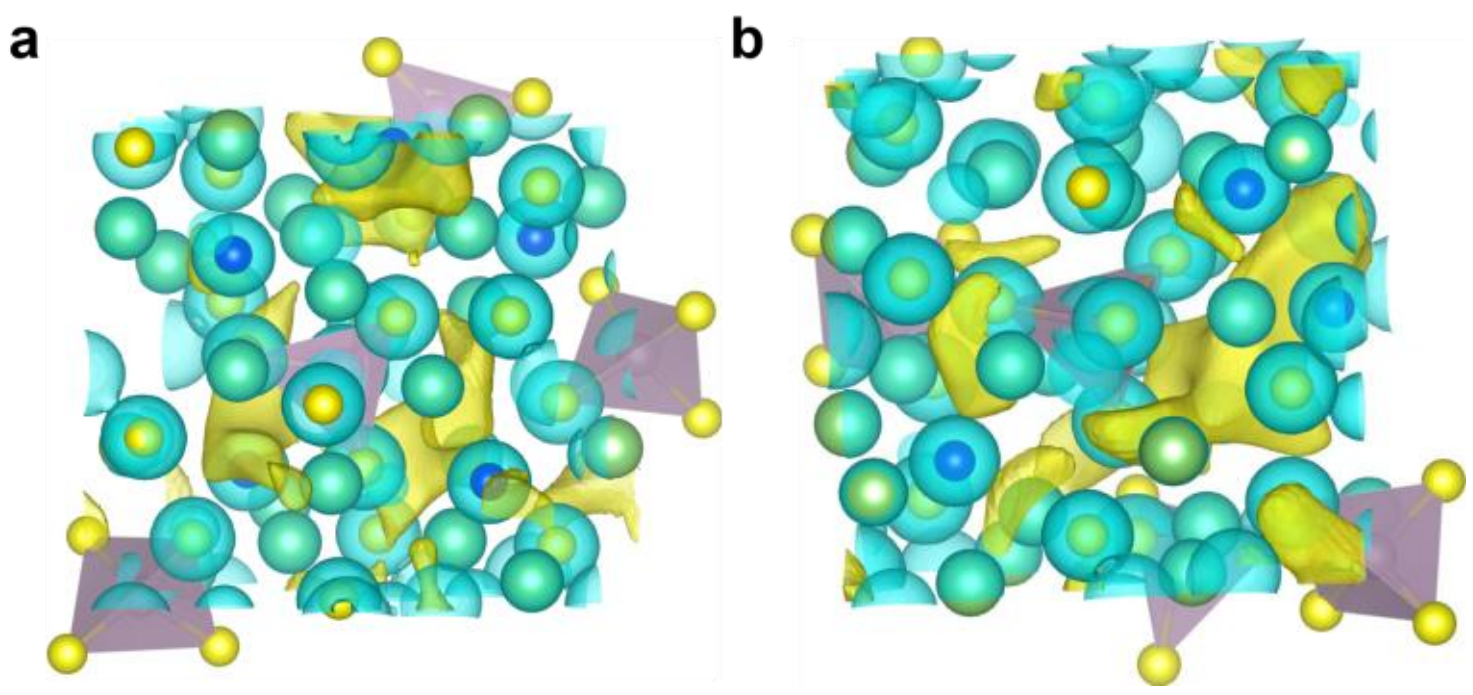

**Supplementary Fig. 34 |** Calculated electrostatic-potential distributions of the Cl1.5 surface model (a) before and (b) after modeled chloride removal and surface LiCl formation. Yellow and cyan iso-surfaces denote positive and negative electrostatic potentials, respectively. The Cl-deficient model exhibits a localized increase in positive potential near the surface. This change is consistent with an electrostatic contribution that may favor chloride redistribution toward the surface, although the electrostatic-potential analysis alone does not determine the operative migration pathway or rate.

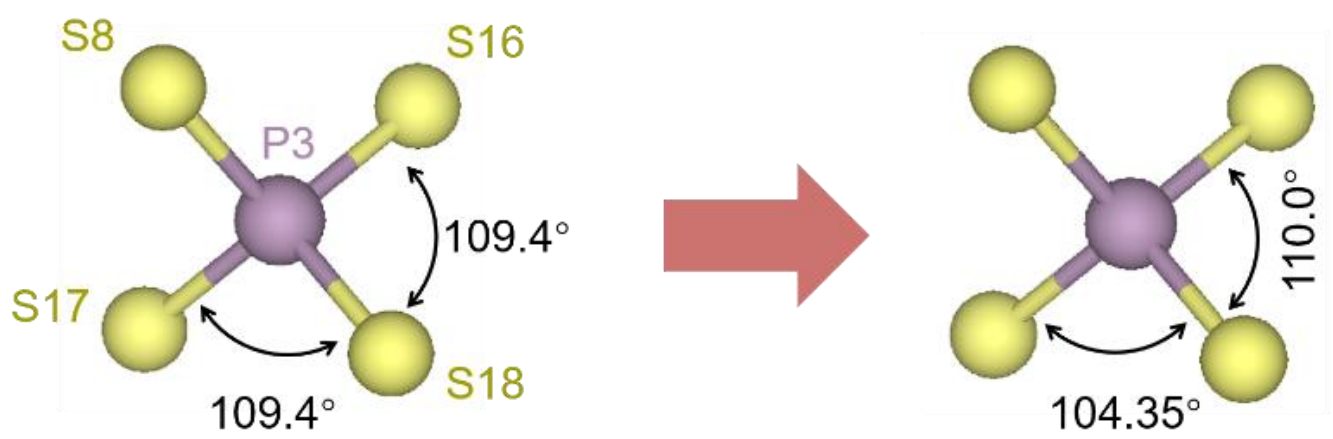

**Supplementary Fig. 35 |** Evolution of representative S–P–S bond angles in a $PS_4^{3-}$ tetrahedron before and after modeled chloride removal and surface LiCl formation. Chloride removal produces changes in the S–P–S bond angles, indicating local distortion of the thiophosphate framework that may contribute to structural destabilization.

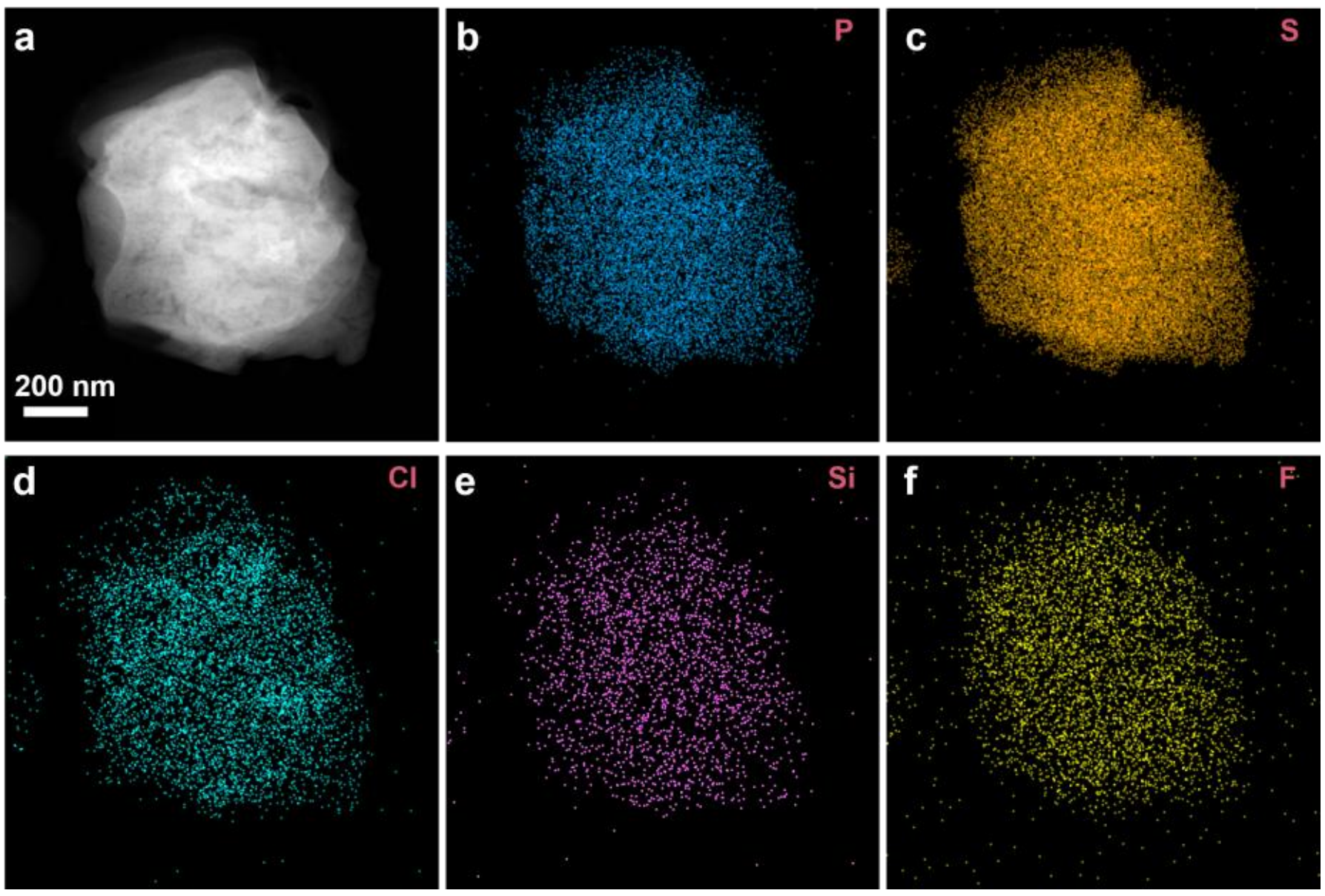

**Supplementary Fig. 36 |** Elemental distribution under controlled environmental exposure. (a) STEM image of the modified electrolyte particle after exposure to a dry room environment with a dew point of −40 °C for 7 hours. (b–f) Energy dispersive X-ray spectroscopy mapping showing the spatial distribution of phosphorus, sulfur, chlorine, silicon, and fluorine.

EDS mapping showed broadly distributed Si and F signals across the Cl1.5 particle surface, consistent with DFMS-derived surface modification.

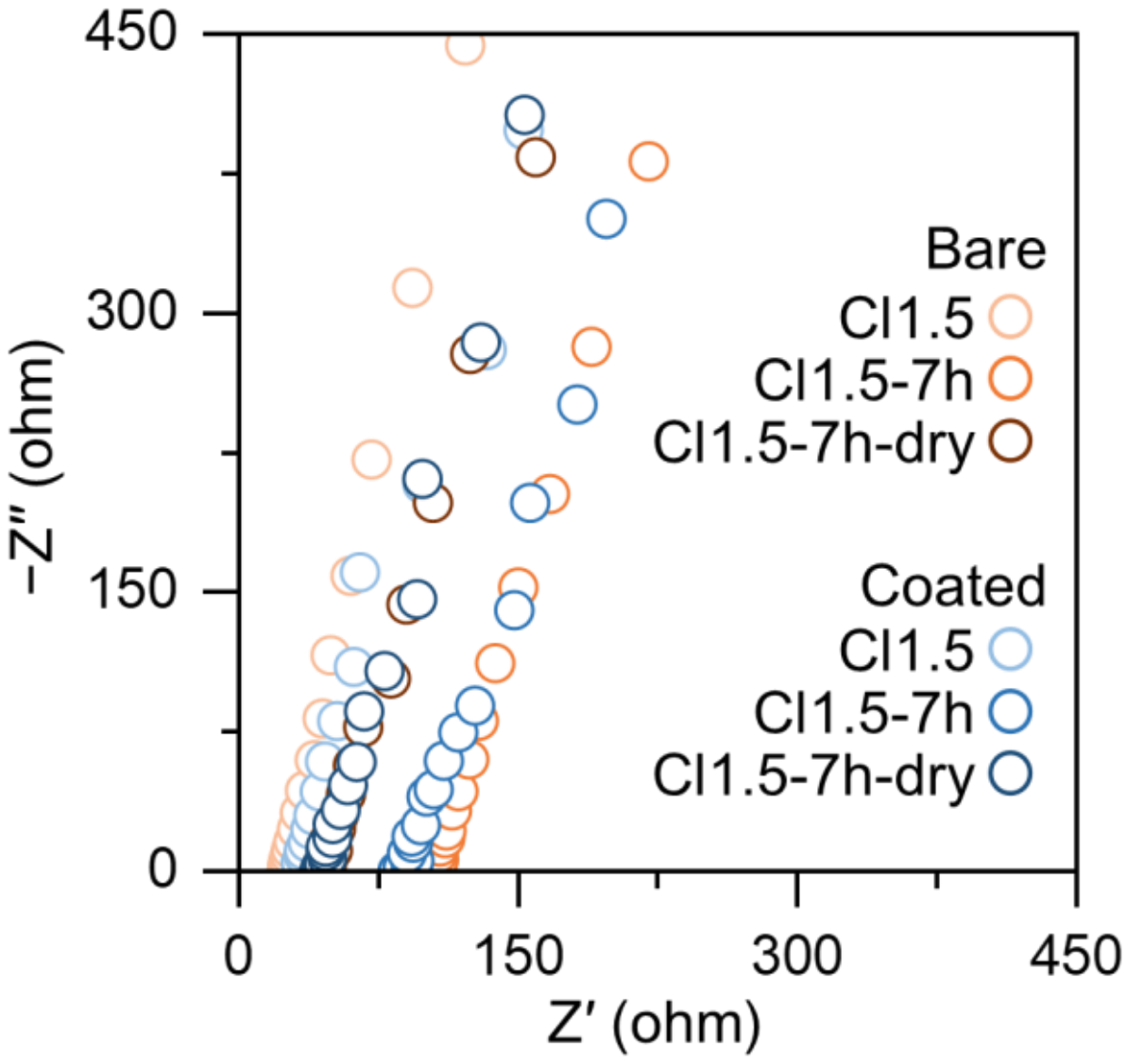

**Supplementary Fig. 37 |** Interfacial modification of Cl1.5. Electrochemical impedance spectroscopy (EIS) of bare Cl1.5 and DFMS-coated Cl1.5, and moisture-exposed Cl1.5, with and without subsequent drying.

DFMS-coated Cl1.5 exhibited improved conductivity retention from 21% to 36% after identical moisture exposure and enhanced thermal recovery from 53% to 64%, despite a slight reduction in initial conductivity (from 4.50 to 3.34 mS·cm$^{-1}$).

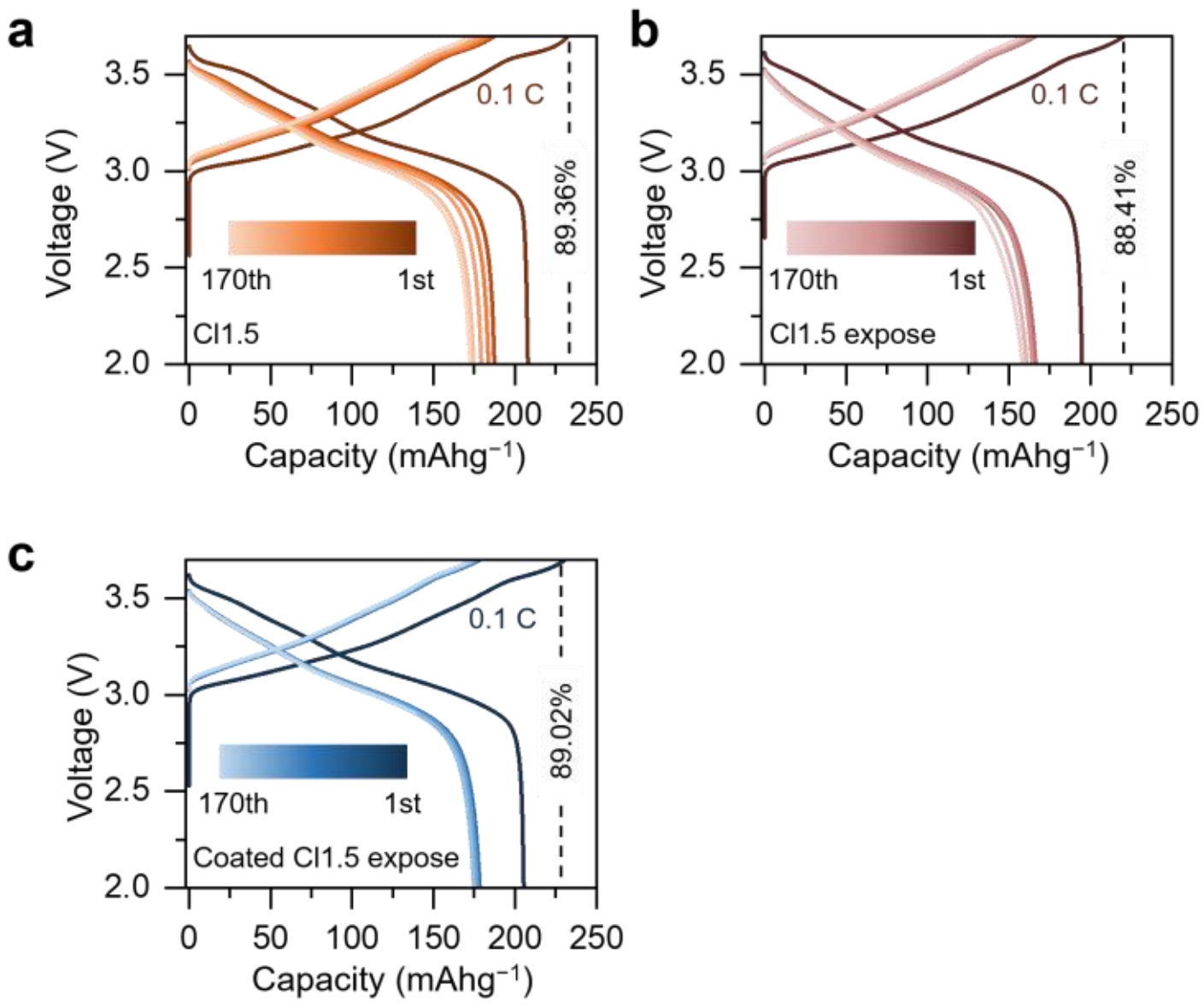

**Supplementary Fig. 38 |** Interfacial modification of Cl1.5. Voltage-Capacity profiles of Li–In || NCM811 cells using (a) bare Cl1.5, (b) exposed Cl1.5, and (c) exposed DFMS-coated Cl1.5.

**Supplementary Table 1 |** Atomic ratio analysis of as-synthesized $Li_6PS_5Cl$ and $Li_{5.5}PS_{4.5}Cl_{1.5}$ measured by EDS.

| **Targeted** | **Measurement** | **Cl/P ratio** | **Avg. ratio (Cl/P)** | **S/P ratio** | **Avg. ratio (S/P)** | **(S/P) At%** |
|---|---|---|---|---|---|---|
| $Li_6PS_5Cl$-0min | M1 | 0.94 | 0.95 | 4.98 | 4.97 | 17.35<br>16.33 |
| | M2 | 0.95 | | 4.96 | | 17.53<br>16.61 |
| $Li_{5.5}PS_{4.5}Cl_{1.5}$-0min | M1 | 1.47 | 1.44 | 4.44 | 4.48 | 15.91<br>23.45 |
| | M2 | 1.40 | | 4.52 | | 16.26<br>22.75 |

**Supplementary Table 2 |** Ionic conductivity (mS cm$^{-1}$) and stability of Cl1.5 and DFMS-treated Cl1.5 after 7 h of exposure to a dry-room environment (dew point, −40 °C).

| **Sample** | **Before exposure** | **After 7 h of dry-room exposure** | **Conductivity retention** | **After drying** | **Recovery rate** |
|---|---|---|---|---|---|
| Bare Cl1.5 | 4.50 | 0.93 | 21% | 2.4 | 53.33% |
| Coated Cl1.5 | 3.34 | 1.19 | 36% | 2.1 | 63.52% |